\documentclass[aps,prd,preprint,superscriptaddress,nopacs,nofootinbib,longbibliography]{revtex4-1}

\usepackage{amsmath,amssymb,latexsym,amsthm}
\usepackage{graphicx}
\usepackage[dvipdfm,hyperfootnotes=false]{hyperref}
\usepackage{txfonts}
\usepackage{mathrsfs}
\usepackage{color}
\DeclareSymbolFont{symbols}{OMS}{cmsy}{m}{n}
\DeclareSymbolFont{largesymbols}{OMX}{cmex}{m}{n}

\begin{document}

\title{
Out-of-time-ordered correlators of the Hubbard model:
SYK strange metal in the spin freezing crossover region
}

\author{Naoto Tsuji}
\affiliation{RIKEN Center for Emergent Matter Science (CEMS), Wako 351-0198, Japan}
\author{Philipp Werner}
\affiliation{Department of Physics, University of Fribourg, 1700 Fribourg, Switzerland}

\begin{abstract}
The Sachdev-Ye-Kitaev (SYK) model describes a strange metal that shows peculiar non-Fermi liquid properties without quasiparticles.
It exhibits a maximally chaotic behavior characterized by out-of-time-ordered correlators (OTOCs),
and is expected to be a holographic dual to black holes. 
While a faithful realization of the SYK model in condensed matter systems may be involved,
a striking similarity between the SYK model and the Hund-coupling induced spin-freezing crossover 
in multi-orbital Hubbard models has recently been pointed out. 
To further explore this connection, 
we study OTOCs for fermionic single-orbital and multi-orbital Hubbard models, which are prototypical models for strongly correlated electrons in solids. 
We introduce an imaginary-time four-point correlation function with an appropriate time ordering, which by means of the spectral representation and the out-of-time-order fluctuation-dissipation theorem can be analytically continued to real-time OTOCs.
Based on this approach, we numerically evaluate real-time OTOCs for Hubbard models in the thermodynamic limit, using the dynamical mean-field theory in combination with a numerically exact continuous-time Monte Carlo impurity solver. 
The results for the single-orbital model show that a certain spin-related OTOC 
captures local moment formation in the vicinity of the metal-insulator transition, 
while the self-energy does not show SYK-like non-Fermi liquid behavior.  
On the other hand, for the two- and three-orbital models with nonzero Hund coupling
we find that the OTOC exhibits a rapid damping at short times
and an approximate power-law decay at longer times in the spin-freezing crossover regime characterized by fluctuating local moments and a non-Fermi liquid self-energy $\Sigma(\omega) \sim \sqrt{\omega}$. 
These results are in a good agreement with the behavior of the SYK model,
providing firm evidence for the close relation between the spin-freezing crossover physics of multi-orbital Hubbard models and the SYK strange metal.

\end{abstract}


\date{\today}


\maketitle

\section{Introduction}

The strange-metal state found by Sachdev and Ye \cite{SachdevYe1993} 
in a random all-to-all interacting Heisenberg spin model 
shows a peculiar non-Fermi liquid behavior without quasiparticle excitations,
which is reminiscent of the unconventional electronic and magnetic properties 
of high-temperature superconductors above the superconducting dome.
The model exhibits various unusual properties
\cite{SachdevYe1993,GeorgesParcolletSachdev2000,GeorgesParcolletSachdev2001}, 
including the absence of magnetic ordering 
(in the fermionic representation)
and a residual entropy down to zero temperature, and an approximate scale invariance at low energies.
In particular, the resulting self-energy has
a characteristic frequency dependence of $\Sigma(\omega)\propto \sqrt{\omega}$ at low frequencies. 
The Sachdev-Ye model has been generalized to a $t$-$J$-like model 
with spins coupled to fermions \cite{ParcolletGeorgesKotliarSengupta1998, ParcolletGeorges1999},
which exhibits a wide 
doping range with similar properties.

Recently, a related fermionic model 
with random all-to-all interactions and no single-particle hopping has been introduced by Kitaev
\cite{Kitaev2015, Sachdev2015, PolchinskiRosenhaus2016, MaldacenaStanford2016}.
This model, dubbed Sachdev-Ye-Kitaev (SYK) model, 
not only retains the non-Fermi liquid properties of the Sachdev-Ye model
but also allows for semi-analytic calculations of out-of-time-ordered correlators (OTOCs) \cite{LarkinOvchinnikov1969}.
OTOCs are a novel type of four-point correlation functions 
such as $\langle \hat A(t)\hat B(0)\hat A(t)\hat B(0)\rangle$
that ignore the usual time-ordering rule.
They are used to describe quantum chaotic properties and information scrambling of quantum many-body systems
\cite{ShenkerStanford2014a, ShenkerStanford2014b, ShenkerStanford2015, MaldacenaShenkerStanford2016, Hosur2016, Swingle2016}, 
and allow for experimental observations \cite{Garttner2017,Li2017,Meier2017}.
The SYK model has been shown to be maximally chaotic in the large-$N$ limit
\cite{Kitaev2015, MaldacenaStanford2016}, 
in the sense that OTOCs 
grow exponentially ($\sim c_0-c_1 e^{\lambda t}$) at early time with the growth rate $\lambda=2\pi k_BT/\hbar$
\cite{MaldacenaShenkerStanford2016, TsujiShitaraUeda2018b}
($k_B$ is the Boltzmann constant, $T$ is the temperature, and $\hbar$ is the Planck constant). 
This property is shared with black holes in Einstein gravity \cite{ShenkerStanford2014a, ShenkerStanford2014b, ShenkerStanford2015, MaldacenaShenkerStanford2016},
which supports expectations  
that the SYK model may be the holographic dual to gravitational theories.
As shown in Ref.~\cite{Bagrets2017}, the initial exponential growth of OTOCs in the SYK model crosses over to an exponential decay at intermediate times, and eventually to a power-law relaxation in the long-time limit.
In the mean time, the SYK model has been generalized to lattice models in which each site represents an SYK atom \cite{Gu2017, Davison2017, Song2017, Chowdhury2018}.

\begin{figure}[t]
\begin{center}
\includegraphics[width=0.98\columnwidth]{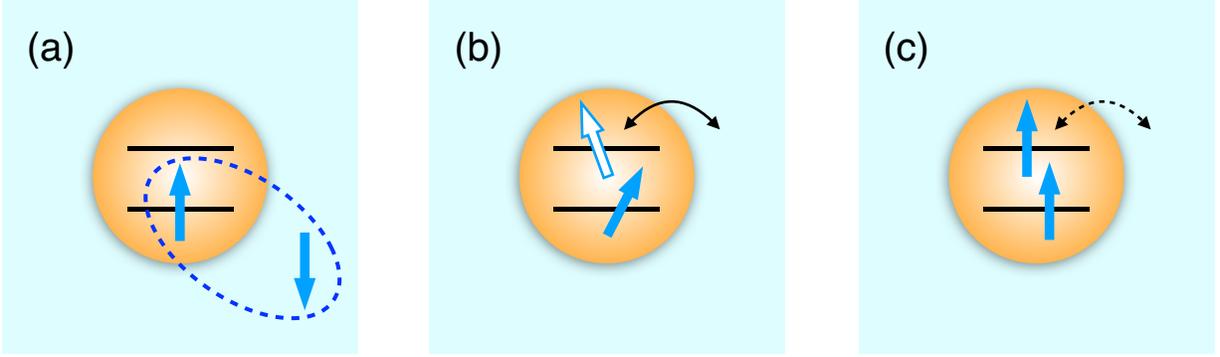}
\caption{Schematic illustration of the spin freezing crossover from (a) a Fermi liquid state with a Kondo singlet via (b) a fluctuating moment state to (c) a frozen moment state in multi-orbital systems with nonzero Hund coupling. Here we focus on a single lattice site (orange) and represent the rest of the lattice by an electron bath (light blue).
}
\label{fig:spin freezing}
\end{center}
\end{figure}

\begin{figure}[t]
\begin{center}
\includegraphics[angle=0, width=0.49\columnwidth]{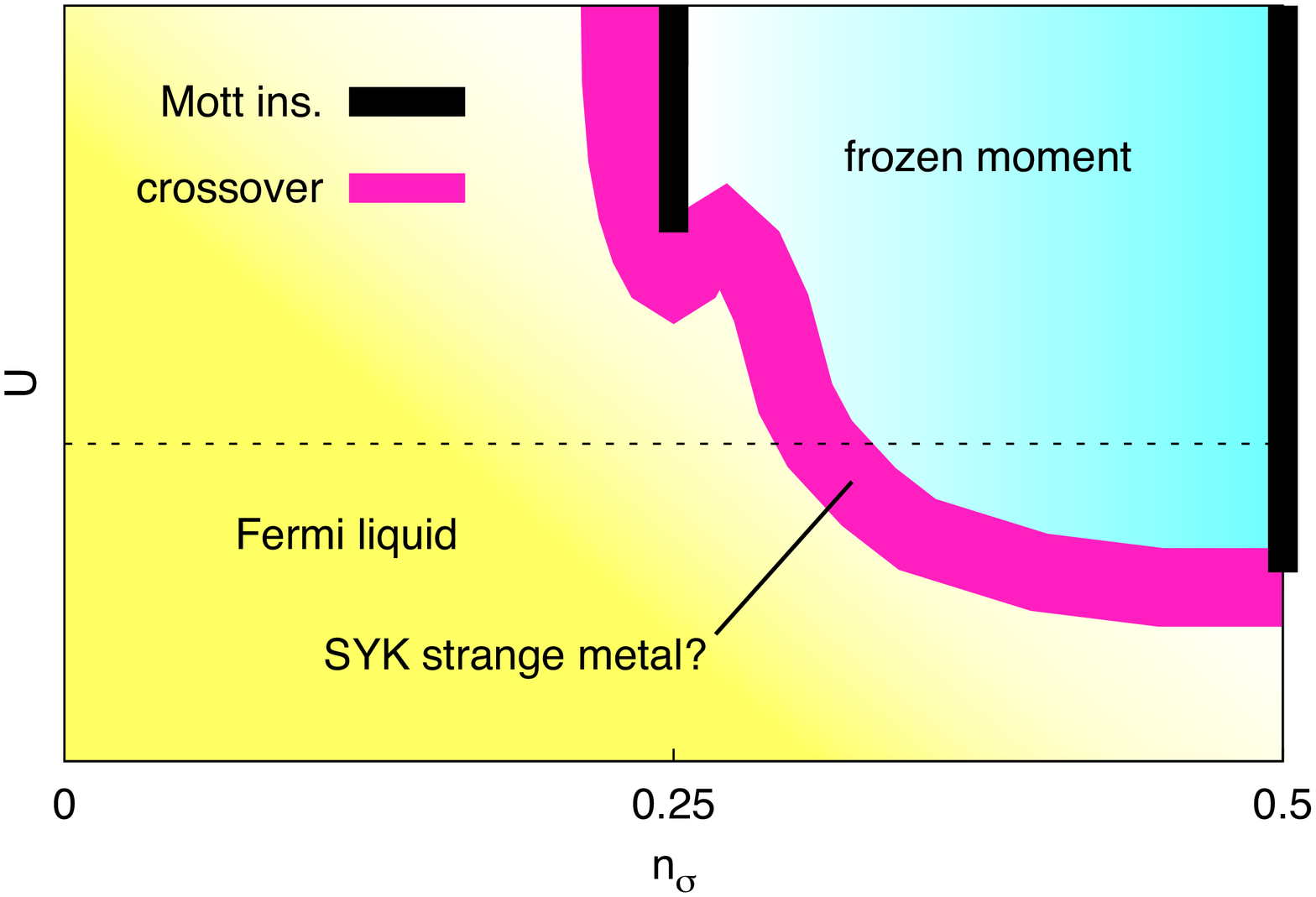}\hfill
\includegraphics[angle=0, width=0.462\columnwidth]{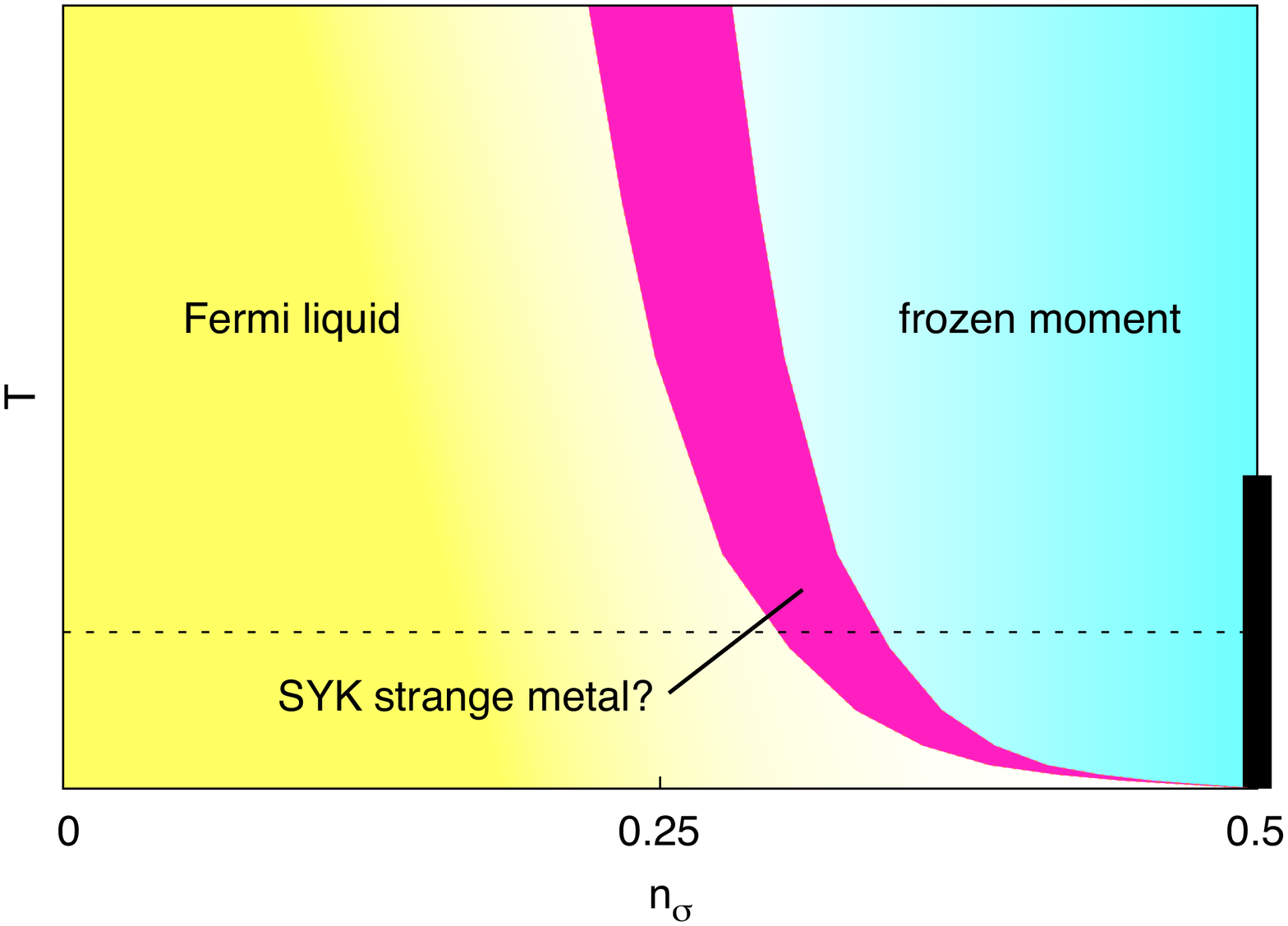}
\caption{
Sketch of the phase diagram of the two-orbital Hubbard model in the space of the interaction $U$ and filling per orbital and spin $n_\sigma$ for 
a fixed ratio $J/U>0$ and temperature of the order of 1/100 of the bandwidth (left panel), and in the space of the temperature $T$ and $n_\sigma$ at fixed $U$ (right panel). The bold black lines indicate Mott insulating solutions appearing at integer total fillings. Doping of the half-filled Mott insulator ($n_\sigma=0.5$) results in an incoherent metal state with frozen magnetic moments (blue shaded region). At lower fillings, there is a crossover to a Fermi liquid metal. The crossover region (pink) is characterized by the non-Fermi liquid self-energy $\Sigma(\omega)\sim \sqrt{\omega}$. The dashed lines show the parameter region that we explore in Sec.~\ref{numerical results}.
}
\label{fig:illustration}
\end{center}
\end{figure}

While the SYK model shows intriguing universal properties, 
a question that is of particular interest here is: 
where can we find a concrete realization of the SYK strange-metal state in condensed matter systems? 
A faithful implementation of the random all-to-all interaction without single-particle hopping in fermion systems may be involved.
Despite its difficulty, there have been
several proposals for the realization of the SYK model in condensed-matter systems \cite{Danshita2017, Garcia-Alvarez2017, PikulinFranz2017, Chen2018}.
On the other hand, as pointed out in Ref.~\cite{Werner2018}, there is a striking similarity between the SYK strange metal 
and the spin-freezing crossover regime of multi-orbital Hubbard models with nonzero Hund coupling \cite{Werner2008, IshidaLiebsch2010, Medici2011, Hoshino2015}, which are prototypical models of strongly correlated electron materials.
In these multi-orbital lattice systems, there is a competition between the Hund effect that favors the formation of local magnetic moments and the Kondo effect that screens local magnetic moments. When the two effects are balanced, there emerges a fluctuating moment state with non-Fermi liquid properties (Fig.~\ref{fig:spin freezing}). This so-called spin-freezing crossover regime separates a Fermi liquid metal from a spin-moment-frozen metal. 
A schematic phase diagram of the two-orbital Hubbard model with Hund coupling in the space of
the Coulomb interaction $U$ and 
the filling per orbital and spin $n_\sigma$
is shown in Fig.~\ref{fig:illustration}, where the ratio between
the Hund coupling $J$ and $U$ is fixed. The crossover from the Fermi liquid to an incoherent metal state with frozen
magnetic moments occurs in the region of the doped half-filled Mott insulator. Near the crossover line (red curve), the self-energy shows
a non-Fermi-liquid frequency dependence $\Sigma(\omega)\sim \sqrt{\omega}$ over a significant energy range, and a spin-spin correlation function which decays on the imaginary-time axis as $\langle \hat S_z(\tau)\hat S_z(0)\rangle \sim 1/\tau$ \cite{Werner2008}. This is exactly the same non-Fermi liquid behavior as realized in the Sachdev-Ye and SYK models. At low enough temperature, there is a crossover to the Fermi liquid scaling (${\rm Im}\,\Sigma(\omega)\sim \omega^2$) \cite{Stadler2015}, similar to what is found in lattice generalizations of the SYK model \cite{Chowdhury2018}.

There are similarities and differences at the level of the Hamiltonian.
The local interaction term of multi-orbital Hubbard models with nonzero Hund coupling can be written in the form 
$\sum_{\alpha\beta\gamma\delta} U_{\alpha\beta\gamma\delta}c_\alpha^\dagger c_\beta^\dagger c_\gamma c_\delta$, where $U_{\alpha\beta\gamma\delta}$ denotes the interaction matrix element, 
$\alpha, \beta, \cdots$ labels the spin and orbital, and $c^\dagger$ ($c$) is the electron creation (annihilation) operator. 
Although the pattern of $U_{\alpha\beta\gamma\delta}$ in realistic systems may be complicated,
$U_{\alpha\beta\gamma\delta}$ does not in general resemble a Gaussian random distribution. 
Furthermore, there are additional single-particle hopping terms in the Hubbard models, which can be a relevant or irrelevant perturbation depending on the strength \cite{Garcia-Garcia2018}. Hence it is a nontrivial question whether or not the spin-freezing crossover regime in multi-orbital Hubbard models can be regarded as a realization of the SYK strange metal.

In this work, we calculate out-of-time-ordered correlators for single-orbital and multi-orbital Hubbard models in the thermodynamic limit. 
Since OTOCs are dynamical
(real-time) four-point correlation
functions with an unusual ordering of operator sequences, which often requires
a formidable effort of summing all the relevant diagrams and
solving a Bethe-Salpeter equation,
it is numerically challenging
to evaluate OTOCs for correlated many-body systems.
Previously, the exponential growth of OTOCs has been calculated analytically 
by summing ladder diagrams for the SYK and several related models \cite{Kitaev2015, MaldacenaStanford2016, Banerjee2017, Garcia-Garcia2018}. 
For single-particle problems, OTOCs have been numerically calculated
for the quantum kicked rotor model \cite{Rozenbaum2017} and the quantum stadium billiard \cite{Rozenbaum2018}.
For many-body problems, OTOCs have been studied by using field theoretic approaches \cite{Aleiner2016, Stanford2016, PatelSachdev2017, Patel2017, ChowdhurySwingle2017, LiaoGalitski2018}. 
OTOCs have also been numerically evaluated for relatively small-size systems using
exact diagonalization \cite{Chen2016b, Fan2016, FuSachdev2016, He2016, Huang2016, Shen2016, Yao2016, Bohrdt2017, Dora2017a, Dora2017b}. 
For interacting fermion lattice models, which are typically studied 
in condensed matter physics, a useful approach is provided by
the dynamical mean-field theory (DMFT) \cite{GeorgesKotliarKrauthRozenberg1996}, which can be directly applied to
the thermodynamic limit, and becomes an exact treatment in the limit of large lattice dimension \cite{MetznerVollhardt1989}. 
It involves a self-consistent mapping of the lattice model to an effective quantum impurity model. 
DMFT has been generalized to
deal with nonequilibrium states by switching from the imaginary-time Matsubara formalism
to the real-time Kadanoff-Baym or Keldysh formalism 
defined on the singly-folded time contour \cite{Aoki2014}.
To calculate OTOCs, nonequilibrium DMFT can be further extended to a doubly folded time-contour
formalism, with which OTOCs have been evaluated for the Falicov-Kimball model \cite{TsujiWernerUeda2017},
an integrable lattice fermion model that 
can be exactly solved within DMFT  
\cite{FreericksZlatic2003}.
The application of this method to Hubbard models, however, is difficult due to the lack of 
reliable and versatile impurity solvers that can be used within the doubly folded time-contour formalism.

As an alternative approach, we develop a general method to derive real-time OTOCs from
imaginary-time four-point correlation functions through an analytic continuation
and the use of the recently formulated out-of-time-order fluctuation-dissipation theorem \cite{TsujiShitaraUeda2018a},
in an analogous way as the retarded Green's function is obtained by 
analytic continuation from the imaginary-time Matsubara Green's function.
The advantage of this method is that the imaginary-time four-point function
can be accurately evaluated numerically for general quantum many-body systems
by using an appropriate continuous-time quantum Monte Carlo (QMC) method \cite{Gull2011}.
This is in contrast to the direct application of QMC methods to the calculation
of real-time correlation functions, which usually suffers from a sign problem \cite{WernerOkaMillis2009},
that is exacerbated in the OTOC case by the doubling of the real-time contour. 
We use the proposed method to evaluate OTOCs of single-orbital and multi-orbital Hubbard models 
within the framework of DMFT using a QMC impurity solver. 

The results show that accurate real-time OTOCs can be obtained up to a few hopping times, and also the general
long-time behavior can be approximately captured, while the details of the oscillations at intermediate and long
times cannot be resolved. 
For the half-filled single-orbital Hubbard model, we evaluate several different types of OTOCs in the vicinity of the
metal-insulator transition. We show that although these functions capture nontrivial correlations in the strongly interacting metallic regime, 
there is no evidence for the non-Fermi liquid behavior with SYK-like exponents
due to the dominance of the Kondo effect. 
We then turn to the doped Mott insulating phase of the two- and three-orbital Hubbard models with nonzero Hund coupling,
for which we focus on a spin-related OTOC that has a counterpart in the SYK model and is sensitive to
fluctuating magnetic moments.
We find that the OTOC 
in the spin-freezing crossover regime 
damps quickly (roughly exponentially) at short times, and decays as a power law at longer times. 
We confirm that this behavior agrees qualitatively with that of the OTOC for the SYK model with a finite number of orbitals obtained from exact diagonalization. The power-law exponents agree almost quantitatively if we identify the variance of the SYK interaction with the square of the Hund coupling, as suggested in Ref.~\cite{Werner2018}.  
Our analysis of OTOCs thus establishes a close connection between
the spin-freezing crossover regime of multi-orbital Hubbard models and the SYK strange metal.

The paper is organized as follows: In Sec.~\ref{imaginary-time formalism},
we define the imaginary-time four-point correlation functions and discuss their
general properties. 
In Sec.~\ref{real-time OTOC}, we prove that these imaginary-time four-point functions
can be analytically continued to real-time OTOCs by introducing the spectral representation
and using the out-of-time-order fluctuation-dissipation theorem.
In Sec.~\ref{numerical results}, we present numerical results for OTOCs
of the single-, two- and three-orbital Hubbard models, and compare them with the SYK model. 
Section~\ref{sec:conclusions} contains a summary and conclusions. 

\section{Imaginary-time four-point functions}
\label{imaginary-time formalism}

In this section, we define imaginary-time four-point correlation functions,
which we prove in the next section to be analytically continuable to real-time OTOCs,
and discuss their general properties.
For arbitrary operators $\hat A$ and $\hat B$, we define
\begin{align}
C_{(AB)^2}^M(\tau)
&\equiv
\begin{cases}
-\langle \hat A(\tau+\tfrac{\beta\hbar}{2})\hat B(\tfrac{\beta\hbar}{2})\hat A(\tau)\hat B(0) \rangle
& 0\le \tau\le \frac{\beta\hbar}{2},
\\
-\langle \hat B(\tfrac{\beta\hbar}{2})\hat A(\tau+\tfrac{\beta\hbar}{2})\hat B(0)\hat A(\tau) \rangle
& -\tfrac{\beta\hbar}{2} \le \tau < 0.
\end{cases}
\label{imaginary-time OTOC}
\end{align}
Here $\beta=(k_BT)^{-1}$ is the inverse temperature,
$\langle \cdots \rangle\equiv {\rm Tr}(e^{-\beta\hat H}\cdots)/Z$ represents the statistical average,
$\hat H$ is the Hamiltonian of the system, $Z\equiv {\rm Tr}(e^{-\beta\hat H})$ is the partition function,
and $\hat A(\tau)=e^{\frac{\tau}{\hbar}\hat H}\hat A e^{-\frac{\tau}{\hbar}\hat H}$ is the Heisenberg representation
for the imaginary-time evolution.
We use the label `$M$' for the imaginary-time four-point function because of the analogy with
the Matsubara Green's function defined by
\begin{align}
C_{AB}^M(\tau)
&=
\begin{cases}
-\langle \hat A(\tau)\hat B(0) \rangle & 0\le\tau\le\beta\hbar,
\\
\mp\langle \hat B(0)\hat A(\tau) \rangle & -\beta\hbar\le\tau<0,
\end{cases}
\label{imaginary-time TOC}
\end{align}
where the sign $+$ is taken when both $\hat A$ and $\hat B$ are fermionic (i.e., Grassmann odd)
and the sign $-$ is taken when either $\hat A$ or $\hat B$ is bosonic (i.e., Grassmann even). 
In Eq.~(\ref{imaginary-time TOC}) the function $C_{AB}^M(\tau)$ is defined 
for $-\beta\hbar\le\tau\le\beta\hbar$, while in Eq.~(\ref{imaginary-time OTOC}) the function $C_{(AB)^2}^M(\tau)$
is defined for $-\frac{\beta\hbar}{2}\le\tau\le\frac{\beta\hbar}{2}$.
Note that the definition (\ref{imaginary-time OTOC}) does not have a sign change,
independent of the statistical nature (bosonic or fermionic) of $\hat A$ and $\hat B$,
since the order of $\hat A$ and $\hat B$ is exchanged twice for $-\frac{\beta\hbar}{2}\le\tau<0$
in Eq.~(\ref{imaginary-time OTOC}).

An important property of the imaginary-time four-point function $C_{(AB)^2}^M(\tau)$ is its time periodicity,
\begin{align}
C_{(AB)^2}^M(\tau+\tfrac{\beta\hbar}{2})
&=
C_{(AB)^2}^M(\tau)
\quad
(-\tfrac{\beta\hbar}{2}\le\tau< 0).
\label{OTOC periodicity}
\end{align}
The proof for (\ref{OTOC periodicity}) follows straightforwardly from the definition (\ref{imaginary-time OTOC}):
\begin{align}
C_{(AB)^2}^M(\tau+\tfrac{\beta\hbar}{2})
&=
-\langle \hat A(\tau+\beta\hbar)\hat B(\tfrac{\beta\hbar}{2})\hat A(\tau+\tfrac{\beta\hbar}{2})\hat B(0)\rangle
\notag
\\
&=
-\frac{1}{Z}{\rm Tr}[e^{-\beta\hat H} \hat A(\tau+\beta\hbar)\hat B(\tfrac{\beta\hbar}{2})\hat A(\tau+\tfrac{\beta\hbar}{2})\hat B(0)]
\notag
\\
&=
-\frac{1}{Z}{\rm Tr}[\hat A(\tau)e^{-\beta\hat H}\hat B(\tfrac{\beta\hbar}{2})\hat A(\tau+\tfrac{\beta\hbar}{2})\hat B(0)]
\notag
\\
&=
-\langle \hat B(\tfrac{\beta\hbar}{2})\hat A(\tau+\tfrac{\beta\hbar}{2})\hat B(0)\hat A(\tau)\rangle
=
C_{(AB)^2}^M(\tau)
\quad
(-\tfrac{\beta\hbar}{2}\le \tau <0).
\end{align}
We can extend the region of the definition of $C_{(AB)^2}^M(\tau)$ from $-\frac{\beta\hbar}{2}\le\tau\le\frac{\beta\hbar}{2}$
to $-\infty<\tau<\infty$ by repeatedly applying (\ref{OTOC periodicity}) for $n\frac{\beta\hbar}{2}\le\tau<(n+1)\frac{\beta\hbar}{2}$
($n=0, \pm 1, \pm 2, \dots$).
In this way, $C_{(AB)^2}^M(\tau)$ can be considered as a periodic function of $\tau$ with period $\frac{\beta\hbar}{2}$.
This allows us to Fourier transform $C_{(AB)^2}^M(\tau)$ into
\begin{align}
C_{(AB)^2}^M(i\varpi_n)
&=
\int_0^{\frac{\beta\hbar}{2}}d\tau\,
e^{i\varpi_n\tau}C_{(AB)^2}^M(\tau),
\label{matsubara OTOC}
\end{align}
where $\varpi_n$ 
takes the discrete values 
\begin{align}
\varpi_n
&=
\frac{4n\pi}{\beta\hbar}
\quad
(n\in\mathbb Z).
\end{align}
Let us compare this situation with the one for the usual imaginary-time two-point function $C_{AB}^M(\tau)$,
which is (anti)periodic,
\begin{align}
C_{AB}^M(\tau+\beta\hbar)
&=
\pm C_{AB}^M(\tau),
\label{TOC periodicity}
\end{align}
with period $\beta\hbar$. In Eq.~(\ref{TOC periodicity}), the minus sign is taken
when both $\hat A$ and $\hat B$ are fermionic, and plus otherwise.
Due to the (anti)periodicity, one can Fourier transform $C_{AB}^M(\tau)$ into
\begin{align}
C_{AB}^M(i\omega_n)
&=
\int_0^{\beta\hbar}d\tau\, e^{i\omega_n\tau}C_{AB}^M(\tau)
\label{matsubara TOC}
\end{align}
with the Matsubara frequency given by
\begin{align}
\omega_n
&=
\begin{cases}
\displaystyle
\frac{2n\pi}{\beta\hbar} & \mbox{either $\hat A$ or $\hat B$ is bosonic},
\\
\displaystyle
\frac{(2n+1)\pi}{\beta\hbar} & \mbox{both $\hat A$ and $\hat B$ are fermionic}.
\end{cases}
\end{align}
The period for $C_{(AB)^2}^M(\tau)$ (\ref{imaginary-time OTOC}) is half of that for $C_{AB}^M(\tau)$ (\ref{imaginary-time TOC}).
Due to this difference, the frequency step changes between $C_{(AB)^2}^M(i\varpi_n)$ ($\ref{matsubara OTOC}$)
and $C_{AB}^M(i\omega_n)$ (\ref{matsubara TOC}).

\section{Analytic continuation to real-time OTOCs}
\label{real-time OTOC}

In the previous section, we have introduced the imaginary-frequency four-point function $C_{(AB)^2}^M(i\varpi_n)$ (\ref{matsubara OTOC}). Let us recall that the Matsubara Green's function $C_{AB}^M(i\omega_n)$ (\ref{matsubara TOC})
can be analytically continued to the retarded Green's function $C_{AB}^R(\omega)$
via the replacement $i\omega_n\to \omega+i\delta$. 
Here $C_{AB}^R(\omega)=\int_{-\infty}^{\infty}dt\, e^{i\omega t}C_{AB}^R(t,0)$ is the Fourier transform of
\begin{align}
C_{AB}^R(t,t')
&\equiv
-i\theta(t-t')\langle [\hat A(t), \hat B(t')]_\mp \rangle
\label{retarded Green}
\end{align}
with $[,]_\mp$ representing the anticommutator ($\{,\}$) when both $\hat A$ and $\hat B$ are fermionic
and the commutator ($[,]$) otherwise.
It is thus natural to ask what kind of function corresponds to the analytic continuation
of $C_{(AB)^2}^M(i\varpi_n)$.

Below we show that the analytic continuation of $C_{(AB)^2}^M(i\varpi_n)$ through $i\varpi_n\to \omega+i\delta$
is given by what we call {\it the retarded OTOC} $C_{(AB)^2}^R(\omega)$,
which is defined by the Fourier transform $C_{(AB)^2}^R(\omega)=\int_{-\infty}^{\infty} dt\, e^{i\omega t}C_{(AB)^2}^R(t,0)$
of
\begin{align}
C_{(AB)^2}^R(t,t')
&\equiv
-i\theta(t-t')[\langle \hat A(t)\hat B(t'), \hat A(t)\hat B(t')\rangle
-\langle \hat B(t')\hat A(t), \hat B(t')\hat A(t)\rangle].
\label{retarded OTOC}
\end{align}
Here $\theta(t)$ is the step function defined by $\theta(t)=1$ ($t\ge 0$) and $=0$ ($t<0$), and we used the notation of the bipartite statistical average
\begin{align}
\langle \hat X, \hat Y\rangle
&\equiv
{\rm Tr}(\hat\rho^{\frac{1}{2}}\hat X\hat\rho^{\frac{1}{2}}\hat Y)
\end{align}
(with $\hat\rho=e^{-\beta\hat H}/Z$ being the density matrix),
which has previously appeared in the study of OTOCs \cite{MaldacenaShenkerStanford2016, Yao2016,
PatelSachdev2017, Patel2017, TsujiShitaraUeda2018a, LiaoGalitski2018}. 
In terms of the bipartite statistical average,
the imaginary-time four-point function introduced in the previous section can be written as
\begin{align}
C_{(AB)^2}^M(\tau)
&=
\begin{cases}
-\langle \hat A(\tau)\hat B(0), \hat A(\tau)\hat B(0) \rangle
& 0\le \tau\le \frac{\beta\hbar}{2},
\\
-\langle \hat B(0)\hat A(\tau), \hat B(0)\hat A(\tau) \rangle
& -\tfrac{\beta\hbar}{2} \le \tau < 0.
\end{cases}
\end{align}
If we introduce the commutator-anticommutator representation of OTOCs,
\begin{align}
C_{[A,B]_{\alpha_1}[A,B]_{\alpha_2}}(t,t')
&\equiv
\langle [\hat A(t), \hat B(t')]_{\alpha_1}, [\hat A(t), \hat B(t')]_{\alpha_2} \rangle
\quad
(\alpha_1, \alpha_2=\pm),
\end{align}
$C_{(AB)^2}^R(t,t')$ can be written in the form 
\begin{align}
C_{(AB)^2}^R(t,t')
&=
-i\theta(t-t')C_{\{A,B\}[A,B]}(t,t').
\label{retarded OTOC 2}
\end{align}

The original motivation to employ this form
was that the squared commutator $\langle [\hat A(t), \hat B(t')]^2\rangle$ might be ill-defined
in the context of quantum field theory, because two operators can approach each other arbitrarily close in time,
which may cause divergences. In this situation, one usually needs to regularize the squared commutator.
One prescription to regularize it is to take the bipartite statistical average,
$\langle [\hat A(t), \hat B(t')], [\hat A(t), \hat B(t')] \rangle$, with which the two commutators are separated
in the imaginary-time direction \cite{MaldacenaShenkerStanford2016}. 
There is an information-theoretic meaning 
of the difference between the usual and bipartite statistical averages, which is given by
the Wigner-Yanase (WY) skew information \cite{WignerYanase1963},
\begin{align}
I_{\frac{1}{2}}(\hat\rho,\hat O)
&\equiv
-\frac{1}{2}{\rm Tr}([\hat\rho^{\frac{1}{2}}, \hat O]^2)
=
\langle \hat O^2\rangle-\langle \hat O, \hat O \rangle,
\end{align}
for a quantum state $\hat\rho$ and an observable $\hat O$ (which is a hermitian operator).
It represents the information content of quantum fluctuations of the observable $\hat O$
contained in the quantum state $\hat\rho$ (for further details on the WY skew information in the present context,
we refer to Refs.~\cite{TsujiShitaraUeda2018a, Luo2005}).
If quantum fluctuations are suppressed (e.g., in the semiclassical regime), 
one expects that OTOCs in the form of the usual and bipartite statistical averages would share
common semiclassical features such as the chaotic exponential growth in the short-time regime (butterfly effect).
It has also recently been pointed out that OTOCs in the form of the usual statistical average
may involve scattering processes that contribute to the exponential growth but are not relevant to
many-body chaos, while OTOCs with the bipartite statistical average correctly capture 
chaotic properties \cite{LiaoGalitski2018}.
Hereafter we focus on OTOCs in the form of the bipartite statistical average.

The relation between $C_{(AB)^2}^M(i\varpi_n)$ and $C_{(AB)^2}^R(\omega)$ is most clearly seen in the spectral
representation. To obtain the spectral representation for $C_{(AB)^2}^M(i\varpi_n)$, we expand it 
in the basis of eigenstates of $\hat H$ denoted by $|n\rangle$ with eigenenergies $E_n$,
\begin{align}
C_{(AB)^2}^M(\tau)
&=
-\frac{1}{Z}\sum_{klmn} e^{-\frac{\beta}{2}(E_k+E_m)}
e^{\frac{1}{\hbar}(E_k-E_l+E_m-E_n)\tau}
\langle k|\hat A|l\rangle \langle l|\hat B|m\rangle
\langle m|\hat A|n\rangle \langle n|\hat B|k\rangle
\notag
\\
&=
-\frac{1}{Z}\int_{-\infty}^{\infty}d\omega'\, e^{-\omega'\tau}\sum_{klmn} e^{-\frac{\beta}{2}(E_k+E_m)}
\delta(\omega'+\tfrac{1}{\hbar}(E_k-E_l+E_m-E_n))
\notag
\\
&\quad\times
\langle k|\hat A|l\rangle \langle l|\hat B|m\rangle
\langle m|\hat A|n\rangle \langle n|\hat B|k\rangle.
\end{align}
From the first to the second line, we inserted 
$1=\int_{-\infty}^{\infty} d\omega'\, \delta(\omega'+\frac{1}{\hbar}(E_k-E_l+E_m-E_n))$,
where $\delta(\omega)$ is the delta function. By Fourier transforming $C_{(AB)^2}^M(\tau)$, we obtain
\begin{align}
C_{(AB)^2}^M(i\varpi_n)
&=
\frac{1}{Z}\int_{-\infty}^{\infty}d\omega'\, \frac{1-e^{-\frac{\beta\hbar\omega'}{2}}}{i\varpi_n-\omega'}\sum_{klmn} e^{-\frac{\beta}{2}(E_k+E_m)}
\delta(\omega'+\tfrac{1}{\hbar}(E_k-E_l+E_m-E_n))
\notag
\\
&\quad\times
\langle k|\hat A|l\rangle \langle l|\hat B|m\rangle
\langle m|\hat A|n\rangle \langle n|\hat B|k\rangle
\notag
\\
&=
\frac{1}{Z}\int_{-\infty}^{\infty}d\omega'\, \frac{1}{i\varpi_n-\omega'}\sum_{klmn} 
(e^{-\frac{\beta}{2}(E_k+E_m)}-e^{-\frac{\beta}{2}(E_l+E_n)})
\delta(\omega'+\tfrac{1}{\hbar}(E_k-E_l+E_m-E_n))
\notag
\\
&\quad\times
\langle k|\hat A|l\rangle \langle l|\hat B|m\rangle
\langle m|\hat A|n\rangle \langle n|\hat B|k\rangle.
\end{align}
Motivated by the above expression, let us define the spectral function for the OTOC by
\begin{align}
\mathscr A_{(AB)^2}(\omega)
&\equiv
\frac{1}{Z}\sum_{klmn} 
(e^{-\frac{\beta}{2}(E_k+E_m)}-e^{-\frac{\beta}{2}(E_l+E_n)})
\delta(\omega+\tfrac{1}{\hbar}(E_k-E_l+E_m-E_n))
\notag
\\
&\quad\times
\langle k|\hat A|l\rangle \langle l|\hat B|m\rangle
\langle m|\hat A|n\rangle \langle n|\hat B|k\rangle.
\label{OTOC spectral function}
\end{align}
Note that $\mathscr A_{(AB)^2}(\omega)$ takes real values 
when $\hat B=\hat A^\dagger$, since $\mathscr A_{(AB)^2}(\omega)^\ast=\mathscr A_{(B^\dagger A^\dagger)^2}(\omega)$.
However, in this case $\mathscr A_{(AA^\dagger)^2}(\omega)$ is not necessarily positive semidefinite for $\omega\ge 0$. 
One exception is the low-temperature limit, where $\mathscr A_{(AA^\dagger)^2}(\omega)$ becomes positive semidefinite for $\omega\ge 0$. To see this, let us denote the ground state as $|g\rangle$
with the eigenenergy $E_g$. In the zero-temperature limit, the spectral function approaches
\begin{align}
\mathscr A_{(AA^\dagger)^2}(\omega)
&\to
\frac{1}{Z}\sum_{ln} 
e^{-\beta E_g}
\delta(\omega+\tfrac{1}{\hbar}(2E_g-E_l-E_n))
\langle g|\hat A|l\rangle \langle l|\hat A^\dagger|g\rangle
\langle g|\hat A|n\rangle \langle n|\hat A^\dagger|g\rangle
\notag
\\
&\quad
-\frac{1}{Z}\sum_{km} 
e^{-\beta E_g}
\delta(\omega+\tfrac{1}{\hbar}(E_k+E_m-2E_g))
\langle k|\hat A|g\rangle \langle g|\hat A^\dagger|m\rangle
\langle m|\hat A|g\rangle \langle g|\hat A^\dagger|k\rangle
\notag
\\
&=
\sum_{km} 
[\delta(\omega-\tfrac{1}{\hbar}(E_k+E_m-2E_g))
-\delta(\omega+\tfrac{1}{\hbar}(E_k+E_m-2E_g))]
|\langle g|\hat A|k\rangle|^2
|\langle g|\hat A|m\rangle|^2
\notag
\\
&\ge
0
\quad
(\omega\ge 0).
\end{align}
The spectral sum is given by
\begin{align}
\int_{-\infty}^{\infty} d\omega\, \mathscr A_{(AB)^2}(\omega)
&=
\langle \{\hat A, \hat B\}, [\hat A, \hat B] \rangle
=:
c_{AB}.
\end{align}
Using the spectral function $\mathscr A_{(AB)^2}(\omega)$, 
the imaginary-frequency function $C_{(AB)^2}^M(i\varpi_n)$ can be written as
\begin{align}
C_{(AB)^2}^M(i\varpi_n)
&=
\int_{-\infty}^{\infty} d\omega'\, \frac{\mathscr A_{(AB)^2}(\omega')}{i\varpi_n-\omega'}.
\label{matsubara OTOC Lehmann}
\end{align}
This is analogous to the Lehmann representation for the Matsubara Green's function,
\begin{align}
C_{AB}^M(i\omega_n)
&=
\int_{-\infty}^{\infty} d\omega' \frac{\mathscr A_{AB}(\omega')}{i\omega_n-\omega'},
\end{align}
where $\mathscr A_{AB}(\omega)$ is the spectral function for the Matsubara Green's function defined by
\begin{align}
\mathscr A_{AB}(\omega)
&\equiv
\frac{1}{Z}\sum_{kl} (e^{-\beta E_k}\mp e^{-\beta E_l})\delta(\omega+\tfrac{1}{\hbar}(E_k-E_l))
\langle k|\hat A|l\rangle \langle l|\hat B|k\rangle.
\end{align}
Here the sign $+$ is taken when both $\hat A$ and $\hat B$ are fermionic and the sign $-$ is taken otherwise.

In a similar manner, we can obtain the spectral representation of the retarded OTOC,
which is expanded in the eigenbasis of the Hamiltonian as
\begin{align}
C_{(AB)^2}^R(t,t')
&=
-i\theta(t-t')\frac{1}{Z}\sum_{klmn}
e^{-\frac{\beta}{2}(E_k+E_m)}
\Big[e^{\frac{i}{\hbar}(E_k-E_l+E_m-E_n)(t-t')}
\langle k|\hat A|l\rangle \langle l|\hat B|m\rangle
\langle m|\hat A|n\rangle \langle n|\hat B|k\rangle
\notag
\\
&\quad
-e^{-\frac{i}{\hbar}(E_k-E_l+E_m-E_n)(t-t')}
\langle k|\hat B|l\rangle \langle l|\hat A|m\rangle
\langle m|\hat B|n\rangle \langle n|\hat A|k\rangle\Big].
\label{retarded OTOC 3}
\end{align}
We permute the summation labels for the second term in Eq.~(\ref{retarded OTOC 3}) 
as $k\to l\to m\to n\to k$ to obtain
\begin{align}
C_{(AB)^2}^R(t,t')
&=
-i\theta(t-t')\frac{1}{Z}\sum_{klmn}
\big[e^{-\frac{\beta}{2}(E_k+E_m)}-e^{-\frac{\beta}{2}(E_l+E_n)}\big]
e^{\frac{i}{\hbar}(E_k-E_l+E_m-E_n)(t-t')}
\notag
\\
&\quad\times
\langle k|\hat A|l\rangle \langle l|\hat B|m\rangle
\langle m|\hat A|n\rangle \langle n|\hat B|k\rangle.
\end{align}
By using the expression for the Fourier transformation of the step function
\begin{align}
\theta(t)
&=
\frac{i}{2\pi}\int_{-\infty}^{\infty} d\omega' \frac{e^{-i\omega't}}{\omega'+i\delta}
\end{align}
with a positive infinitesimal constant $\delta$,
we can Fourier transform the retarded OTOC as
\begin{align}
C_{(AB)^2}^R(\omega)
&=
\int_{-\infty}^{\infty} d\omega' \frac{1}{\omega'+i\delta}
\frac{1}{Z}\sum_{klmn}
\big[e^{-\frac{\beta}{2}(E_k+E_m)}-e^{-\frac{\beta}{2}(E_l+E_n)}\big]
\notag
\\
&\quad\times
\delta(\omega-\omega'+\tfrac{1}{\hbar}(E_k-E_l+E_m-E_n))
\langle k|\hat A|l\rangle \langle l|\hat B|m\rangle
\langle m|\hat A|n\rangle \langle n|\hat B|k\rangle.
\end{align}
One notices that the same form of the spectral function $\mathscr A_{(AB)^2}(\omega)$ (\ref{OTOC spectral function})
has appeared in the above expression. Thus, we find that the retarded OTOC has a spectral representation 
\begin{align}
C_{(AB)^2}^R(\omega)
&=
\int_{-\infty}^{\infty} d\omega' \frac{\mathscr A_{(AB)^2}(\omega')}{\omega-\omega'+i\delta}.
\label{retarded OTOC Lehmann}
\end{align}
One can see that $C_{(AB)^2}^R(\omega)$ is analytic in the upper half of the complex plane.
In the limit of $\omega\to\infty$, it behaves as
\begin{align}
C_{(AB)^2}^R(\omega)
&\sim
\frac{c_{AB}}{\omega}.
\label{1/omega}
\end{align}
By comparing Eq.~(\ref{matsubara OTOC Lehmann}) and (\ref{retarded OTOC Lehmann}),
we prove that the imaginary-frequency function $C_{(AB)^2}^M(i\varpi_n)$
can be analytically continued to the retarded OTOC $C_{(AB)^2}^R(\omega)$
through $i\varpi_n\to\omega+i\delta$,
\begin{align}
C_{(AB)^2}^M(i\varpi_n)
&\xrightarrow{i\varpi_n\to\omega+i\delta}
C_{(AB)^2}^R(\omega).
\end{align}

Since $C_{(AB)^2}^R(\omega)$ is analytic in the upper half plane and uniformly decays to zero as in Eq.~(\ref{1/omega})
for $\omega\to\infty$, it should satisfy the Kramers-Kronig relation,
\begin{align}
{\rm Re}\, C_{(AB)^2}^R(\omega)
&=
-\frac{1}{\pi}\mathcal P\int_{-\infty}^{\infty} d\omega' 
\frac{{\rm Im}\, C_{(AB)^2}^R(\omega')}{\omega-\omega'},
\\
{\rm Im}\, C_{(AB)^2}^R(\omega)
&=
\frac{1}{\pi}\mathcal P\int_{-\infty}^{\infty} d\omega' 
\frac{{\rm Re}\, C_{(AB)^2}^R(\omega')}{\omega-\omega'}.
\end{align}

We also define the advanced OTOC as
\begin{align}
C_{(AB)^2}^A(t,t')
&\equiv
i\theta(t'-t)C_{\{A,B\},[A,B]}(t,t')
\notag
\\
&=
i\theta(t'-t)[\langle \hat A(t)\hat B(t'), \hat A(t)\hat B(t')\rangle
-\langle \hat B(t')\hat A(t), \hat B(t')\hat A(t)\rangle].
\label{advanced OTOC}
\end{align}
In the same way as for the retarded OTOC, the advanced OTOC has the spectral representation 
\begin{align}
C_{(AB)^2}^A(\omega)
&=
\int_{-\infty}^{\infty} d\omega' \frac{\mathscr A_{(AB)^2}(\omega')}{\omega-\omega'-i\delta}.
\label{advanced OTOC Lehmann}
\end{align}
Hence the advanced OTOC $C_{(AB)^2}^A(\omega)$ is analytic in the lower half plane.
By comparing Eq.~(\ref{matsubara OTOC Lehmann}) and Eq.~(\ref{advanced OTOC Lehmann}), we can see that
$C_{(AB)^2}^A(\omega)$ is obtained by analytic continuation from $C_{(AB)^2}^M(i\varpi_n)$ via $i\varpi_n\to \omega-i\delta$.
The retarded and advanced OTOCs are related via
\begin{align}
C_{(AB)^2}^R(\omega)^\ast
&=
C_{(B^\dagger A^\dagger)^2}^A(\omega).
\end{align}
In the case of $\hat B=\hat A^\dagger$, the spectral function $\mathscr A_{(AB)^2}(\omega)$
(which is real in this case) is given by the imaginary part of the retarded OTOC,
\begin{align}
\mathscr A_{(AA^\dagger)^2}(\omega)
&=
-\frac{1}{\pi}{\rm Im}\, C_{(AA^\dagger)^2}^R(\omega).
\end{align}

So far, we have explained how to obtain the retarded and advanced OTOCs
by analytic continuation of the imaginary-time four-point function $C_{(AB)^2}^M(i\varpi_n)$.
This allows us to access $\langle \hat A(t)\hat B(t'), \hat A(t)\hat B(t')\rangle-\langle \hat B(t')\hat A(t), \hat B(t')\hat A(t)\rangle$ [see Eqs.~(\ref{retarded OTOC}) and (\ref{advanced OTOC})].
In order to get the full information on OTOCs, we also need to calculate 
the complementary part,
$\langle \hat A(t)\hat B(t'), \hat A(t)\hat B(t')\rangle+\langle \hat B(t')\hat A(t), \hat B(t')\hat A(t)\rangle$. This can be done by using the out-of-time-order fluctuation-dissipation theorem,
which is the out-of-time-order extension of the conventional fluctuation-dissipation theorem, 
expressed as  
\begin{align}
C_{AB}^K(\omega)
&=
\begin{cases}
\displaystyle
\coth\left(\frac{\beta\hbar\omega}{2}\right)[C_{AB}^R(\omega)-C_{AB}^A(\omega)] & \mbox{either $\hat A$ or $\hat B$ is bosonic},
\\
\displaystyle
\tanh\left(\frac{\beta\hbar\omega}{2}\right)[C_{AB}^R(\omega)-C_{AB}^A(\omega)] & \mbox{both $\hat A$ and $\hat B$ are fermionic}.
\label{FD}
\end{cases}
\end{align}
Here we have defined the Keldysh Green's function
\begin{align}
C_{AB}^K(\omega)
&=
-i\langle [\hat A(t), \hat B(t')]_\pm \rangle
\end{align}
with the sign $+$ taken if either $\hat A$ or $\hat B$ are bosonic
and the sign $-$ taken if both $\hat A$ and $\hat B$ are fermionic.
Following the analogy between the Green's functions and OTOCs,
let us define the ``Keldysh'' component of OTOCs as
\begin{align}
C_{(AB)^2}^K(t,t')
&\equiv
-\frac{i}{2}[C_{\{A,B\}^2}(t,t')+C_{[A,B]^2}(t,t')]
\notag
\\
&=
-i[\langle \hat A(t)\hat B(t'), \hat A(t)\hat B(t')\rangle
+\langle \hat B(t')\hat A(t), \hat B(t')\hat A(t)\rangle].
\end{align}
One can see that $C_{(AB)^2}^K(t,t')$ is exactly the complementary part that we needed
to reconstruct OTOCs from the imaginary-time data.
The out-of-time-order fluctuation-dissipation theorem has an analogous form
to the conventional one,
\begin{align}
C_{(AB)^2}^K(\omega)
&=
\coth\left(\frac{\beta\hbar\omega}{4}\right)
[C_{(AB)^2}^R(\omega)-C_{(AB)^2}^A(\omega)].
\end{align}
Note that the argument of the cotangent factor ($\frac{\beta\hbar\omega}{4}$) is just half of that
for the conventional fluctuation-dissipation theorem (\ref{FD}).
The out-of-time-order fluctuation-dissipation theorem takes the same form
for arbitrary statistics (bosonic or fermionic) for the operators $\hat A$ and $\hat B$.
In Table~\ref{table:comparison}, we list the definitions and properties of the Green's function and OTOC.
One can see a clear parallelism between the two 
types of correlation functions. 

\begin{table}
\begin{tabular}{c|c|c}
& Green's function & Out-of-time-ordered correlator (OTOC) \\
\hline
\hline
Matsubara (time) & 
$C_{AB}^M(\tau)=-\langle \hat A(\tau)\hat B(0)\rangle$
&
$C_{(AB)^2}^M(\tau)=-\langle \hat A(\tau)\hat B(0), \hat A(\tau)\hat B(0) \rangle$
\\
& ($0\le\tau\le\beta\hbar$)
& ($0\le \tau\le \frac{\beta\hbar}{2}$)
\\
& $C_{AB}^M(\tau)=\mp \langle \hat B(0)\hat A(\tau)\rangle$
& $C_{(AB)^2}^M(\tau)=-\langle \hat B(0)\hat A(\tau), \hat B(0)\hat A(\tau) \rangle$
\\
& ($-\beta\hbar\le\tau<0$) & ($-\tfrac{\beta\hbar}{2} \le \tau < 0$)
\\
periodicity & $C_{AB}^M(\tau+\beta\hbar)=\pm C_{AB}^M(\tau)$
& $C_{(AB)^2}^M(\tau+\frac{\beta\hbar}{2})=C_{(AB)^2}^M(\tau)$
\\
Matsubara (frequency) 
& $\displaystyle
C_{AB}^M(i\omega_n)=\int_0^{\beta\hbar}d\tau e^{i\omega_n\tau}C_{AB}^M(\tau)$
& $\displaystyle
C_{(AB)^2}^M(i\varpi_n)=\int_0^{\frac{\beta\hbar}{2}}d\tau e^{i\varpi_n\tau}C_{(AB)^2}^M(\tau)$
\\
& $\omega_n=
\begin{cases}
2n\pi/\beta\hbar \\
(2n+1)\pi/\beta\hbar
\end{cases}
(n\in\mathbb Z)$
& $\varpi_n=4n\pi/\beta\hbar \;\;\; (n\in \mathbb Z)$
\\
retarded 
& $C_{AB}^R(t,t')=-i\theta(t-t')\langle [\hat A(t), \hat B(t')]_\mp\rangle$
& $C_{(AB)^2}^R(t,t')=-i\theta(t-t')[\langle \hat A(t)\hat B(t'), \hat A(t)\hat B(t')\rangle$
\\
& &
\qquad $-\langle \hat B(t')\hat A(t), \hat B(t')\hat A(t)\rangle]$
\\
advanced 
& $C_{AB}^A(t,t')=i\theta(t'-t)\langle [\hat A(t), \hat B(t')]_\mp\rangle$
& $C_{(AB)^2}^A(t,t')=i\theta(t'-t)[\langle \hat A(t)\hat B(t'), \hat A(t)\hat B(t')\rangle$
\\
& &
\qquad $-\langle \hat B(t')\hat A(t), \hat B(t')\hat A(t)\rangle]$
\\
Keldysh 
& $C_{AB}^K(t,t')=-i\theta(t-t')\langle [\hat A(t), \hat B(t')]_\pm\rangle$
& $C_{(AB)^2}^K(t,t')=-i\theta(t-t')[\langle \hat A(t)\hat B(t'), \hat A(t)\hat B(t')\rangle$
\\
& &
\qquad $+\langle \hat B(t')\hat A(t), \hat B(t')\hat A(t)\rangle]$
\\
analytic continuation & $C_{AB}^M(i\omega_n)\longrightarrow C_{AB}^R(\omega)$
& $C_{(AB)^2}^M(i\varpi_n)\longrightarrow C_{(AB)^2}^R(\omega)$
\\
& ($i\omega_n\to\omega+i\delta$) & ($i\varpi_n\to\omega+i\delta$)
\\
FDT & $C_{AB}^K(\omega)=
\coth\left(\frac{\beta\hbar\omega}{2}\right)^{\pm 1}$
& $C_{(AB)^2}^K(\omega)=
\coth\left(\frac{\beta\hbar\omega}{4}\right)$
\\
& \qquad\qquad\qquad\quad $\times [C_{AB}^R(\omega)-C_{AB}^A(\omega)]$
& \qquad\qquad\qquad\qquad\qquad $\times [C_{(AB)^2}^R(\omega)-C_{(AB)^2}^A(\omega)]$
\end{tabular}
\caption{Comparison between Green's function and OTOC. 
For the Green's function, the statistical average $\langle \hat X\rangle\equiv{\rm Tr}(e^{-\beta\hat H}\hat X)/Z$ is used,
while for the OTOC, the bipartite statistical average $\langle \hat X, \hat Y\rangle\equiv {\rm Tr}(e^{-\frac{\beta}{2}\hat H}\hat X e^{-\frac{\beta}{2}\hat H}\hat Y)/Z$ is used.
In the Green's function column, the upper sign is taken when either $\hat A$ or $\hat B$ is bosonic,
and the lower sign is taken when both $\hat A$ and $\hat B$ are fermionic.}
\label{table:comparison}
\end{table}

\begin{figure}[htbp]
\includegraphics[width=10cm]{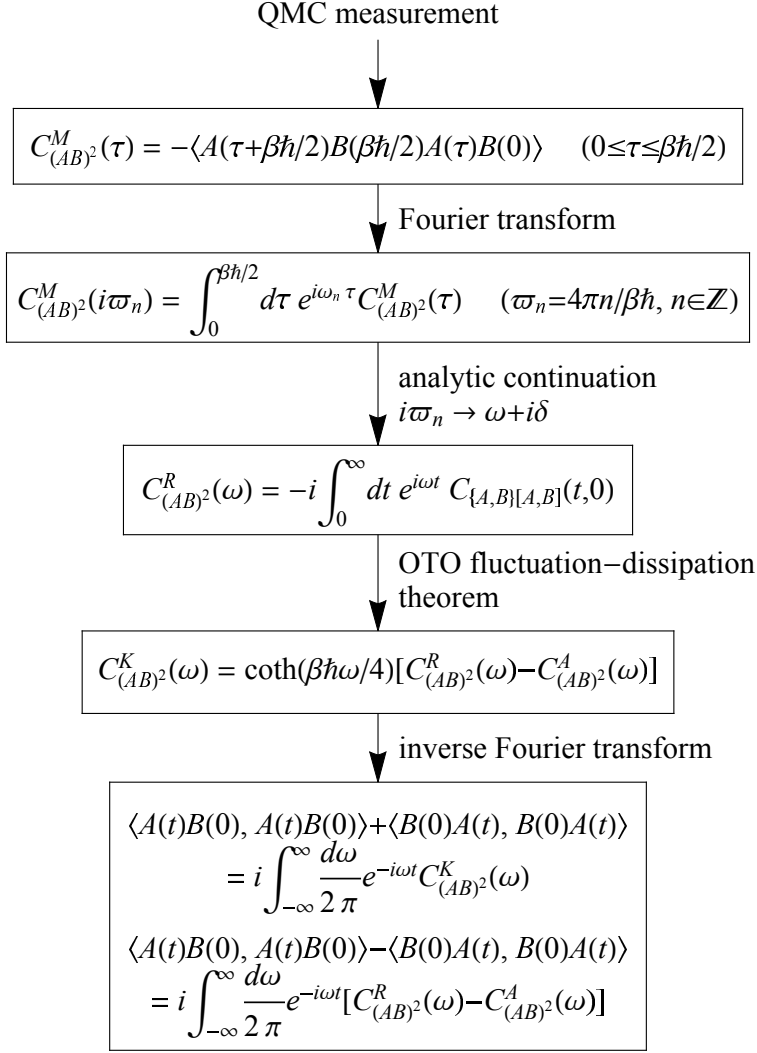}
\caption{The procedure to compute the out-of-time-ordered correlation function
$\langle A(t)B(0), A(t)B(0)\rangle \pm \langle B(0)A(t), B(0)A(t)\rangle$ from the imaginary-time data
obtained by QMC calculations.}
\label{analytic continuation}
\end{figure}

By using the out-of-time-order fluctuation-dissipation theorem, we obtain
$C_{(AB)^2}^K(\omega)$ from $C_{(AB)^2}^R(\omega)$ and $C_{(AB)^2}^A(\omega)$.
Finally, we perform the inverse Fourier transformation of $C_{(AB)^2}^K(\omega)$
and $C_{(AB)^2}^R(\omega)-C_{(AB)^2}^A(\omega)$ to derive
$\langle \hat A(t)\hat B(t'), \hat A(t)\hat B(t')\rangle \pm
\langle \hat B(t')\hat A(t), \hat B(t')\hat A(t)\rangle$.
We summarize the procedure of deriving real-time OTOCs
from the measurement of the imaginary-time four-point function 
in Fig.~\ref{analytic continuation}.

\section{Numerical results for the Hubbard model}
\label{numerical results}

\subsection{Observables and numerical procedure}

We consider the single-orbital, two-orbital, and three-orbital Hubbard models on an infinitely-connected Bethe lattice, which can be solved exactly within DMFT \cite{GeorgesKotliarKrauthRozenberg1996}. In this case, the noninteracting density of states is semicircular with bandwidth $4v_\ast$, for which there exists a simplified DMFT self-consistency condition between the hybridization function $\Delta_{\alpha\sigma}$ of the DMFT impurity problem and the local (impurity) Green's function $G_{\alpha\sigma}$: 
\begin{equation}
\Delta_{\alpha\sigma}(\tau)=v_\ast^2 G_{\alpha\sigma}(\tau), 
\label{semicirc}
\end{equation}
with $\alpha$ the orbital and $\sigma$ the spin index. We will use $v_\ast$ as the unit of energy and measure time in units of $\hbar/v_\ast$. 

Three different types of imaginary-time four-point functions $C_{(AB)^2}^M(\tau)$
of the form of Eq.~(\ref{imaginary-time OTOC})
with $(\hat A, \hat B)=(c_\sigma^\dagger, c_\sigma), (\hat n_\sigma, \hat n_\sigma)$, and $(\hat n, \hat n)$
are calculated in the interval $0\le \tau\le \frac{\beta\hbar}{2}$: 
\begin{eqnarray}
C_{(c^\dagger_\sigma c_\sigma)^2}^M(\tau) &=& -\langle c^\dagger_\sigma(\tau+\tfrac{\beta\hbar}{2}) c_\sigma(\tfrac{\beta\hbar}{2}) c^\dagger_\sigma(\tau) c_\sigma(0) \rangle, \label{fermion_otoc}\\
C_{(n_\sigma n_\sigma)^2}^M(\tau) &=& -\langle \hat n_\sigma(\tau+\tfrac{\beta\hbar}{2})\hat n_\sigma(\tfrac{\beta\hbar}{2})\hat n_\sigma(\tau)\hat n_\sigma(0) \rangle, \label{otoc_nsns}\\
C_{(nn)^2}^M(\tau) &=& -\langle \hat n(\tau+\tfrac{\beta\hbar}{2})\hat n(\tfrac{\beta\hbar}{2})\hat n(\tau)\hat n(0) \rangle,\label{otoc_nn}
\end{eqnarray}
where $c_\sigma^\dagger$ ($c_\sigma$) are the fermionic creation (annihilation) operators for spin $\sigma$ (and orbital $\alpha=1$ in the multi-orbital case), $\hat n_\sigma=c^\dagger_\sigma c_\sigma$ is the corresponding spin-dependent density operator, and $\hat n=\hat n_\uparrow + \hat n_\downarrow$ the total density operator. Note that in all the three cases above we have $\hat B=\hat A^\dagger$. 
We measure these local correlation functions in the impurity model using a hybridization expansion continuous-time Monte Carlo algorithm (CT-HYB) \cite{Werner2006}. In this algorithm, the 
two-particle
Green's functions of the type (\ref{fermion_otoc}) can be measured by removing two hybridization lines, i.e., from the elements of the inverse hybridization matrix, while the density-density correlation functions can be easily measured either by insertion of density operators (matrix formalism) \cite{Werner2006Kondo} or by reading off the occupation of the orbitals at the four time points in the segment implementation \cite{Werner2006}. We use the latter algorithm since we consider only density-density interactions. 
The values at the end-points $\tau=0$ and $\tau=\frac{\beta\hbar}{2}$ reduce to standard density-density correlation functions, which in the case of Eq.~(\ref{fermion_otoc}) are measured separately. 

We perform the analytic continuation to the real-frequency axis using the Maximum Entropy method \cite{Jarrell1996,Lewin_maxent} with a bosonic kernel. This yields the imaginary part of the retarded correlator 
$-\frac{1}{\pi}\text{Im}\, C^R_{(AA^\dagger)^2}(\omega)=\mathscr A_{(AA^\dagger)^2}(\omega)$.  
When $\hat B=\hat A^\dagger=\hat A$ (which is the case for $\hat A=\hat n_\sigma, \hat n$), we have $C_{(AA)^2}^R(\omega)^\ast=C_{(AA)^2}^R(-\omega)$. At half filling, the particle-hole symmetry furthermore ensures $C_{c_\sigma^\dagger c_\sigma}^R(\omega)^\ast=C_{c_\sigma^\dagger c_\sigma}^R(-\omega)$.
Thus, in all the cases considered in this study, the retarded OTOC has the symmetry property $C_{(AA^\dagger)^2}^R(\omega)^\ast=C_{(AA^\dagger)^2}^R(-\omega)$. Using this, as well as the out-of-time-order fluctuation-dissipation
theorem discussed in the previous section, we can obtain the real-time OTOC
$\langle \hat A(t)\hat A^\dagger(0), \hat A(t)\hat A^\dagger(0)\rangle$ from the following inverse Fourier transformation, 
\begin{align}
{\rm Re}\langle \hat A(t)\hat A^\dagger(0), \hat A(t)\hat A^\dagger(0)\rangle
&=
{\rm Re}
\int_0^\infty d\omega\, e^{-i\omega t}\coth\left(\frac{\beta\hbar\omega}{4}\right)
\left(-\frac{1}{\pi}{\rm Im}\, C_{(AA^\dagger)^2}^R(\omega)\right),
\\
{\rm Im}\langle \hat A(t)\hat A^\dagger(0), \hat A(t)\hat A^\dagger(0)\rangle
&=
{\rm Im}\int_0^{\infty} d\omega\, e^{-i\omega t} \left(-\frac{1}{\pi} {\rm Im}\, C_{(AA^\dagger)^2}^R(\omega)\right),
\end{align}
for $\hat A=c_\sigma^\dagger, \hat n_\sigma$, and $\hat n$.

\begin{figure}[b]
\begin{center}
\includegraphics[angle=-90, width=0.48\columnwidth]{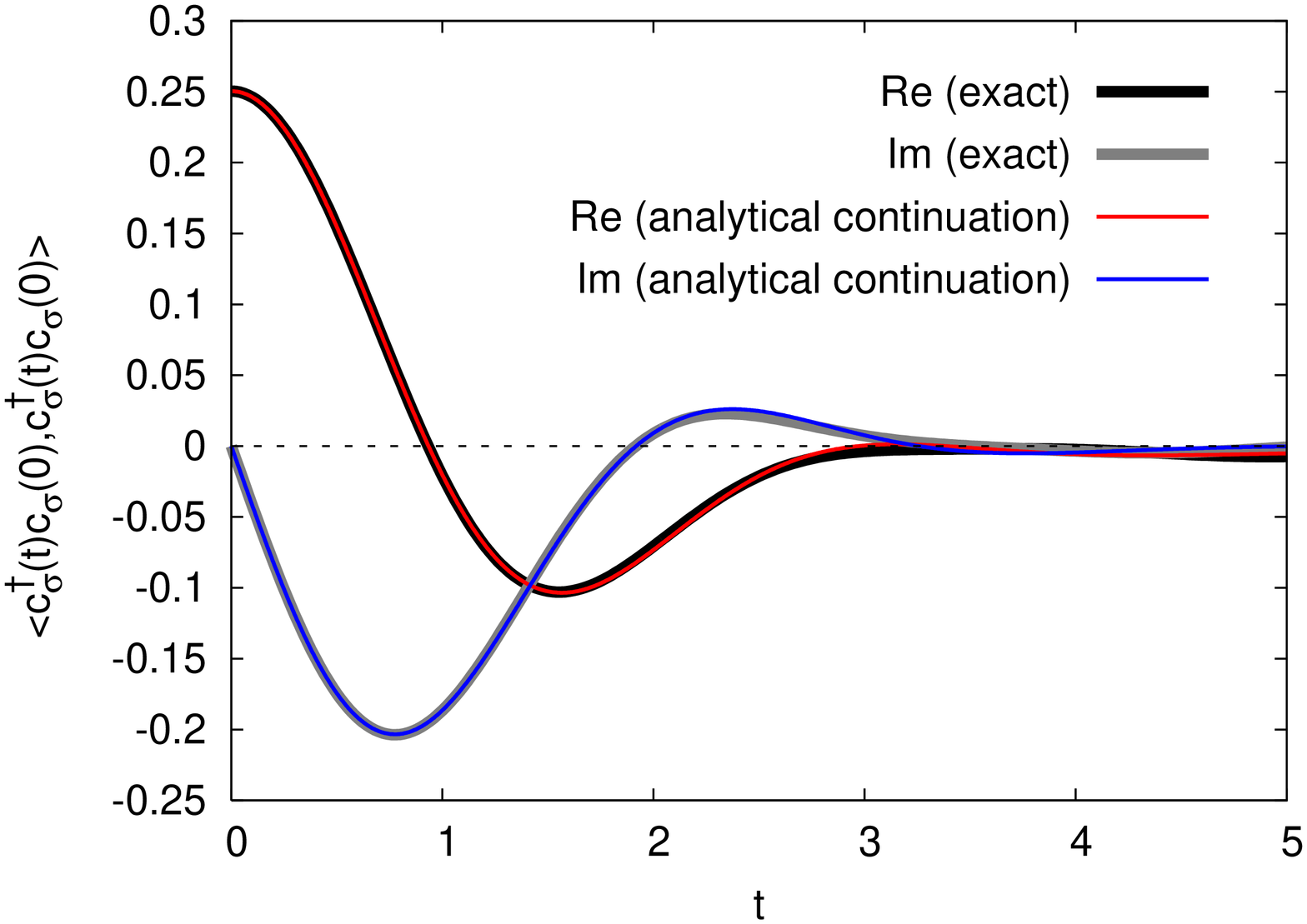}\hfill
\includegraphics[angle=-90, width=0.48\columnwidth]{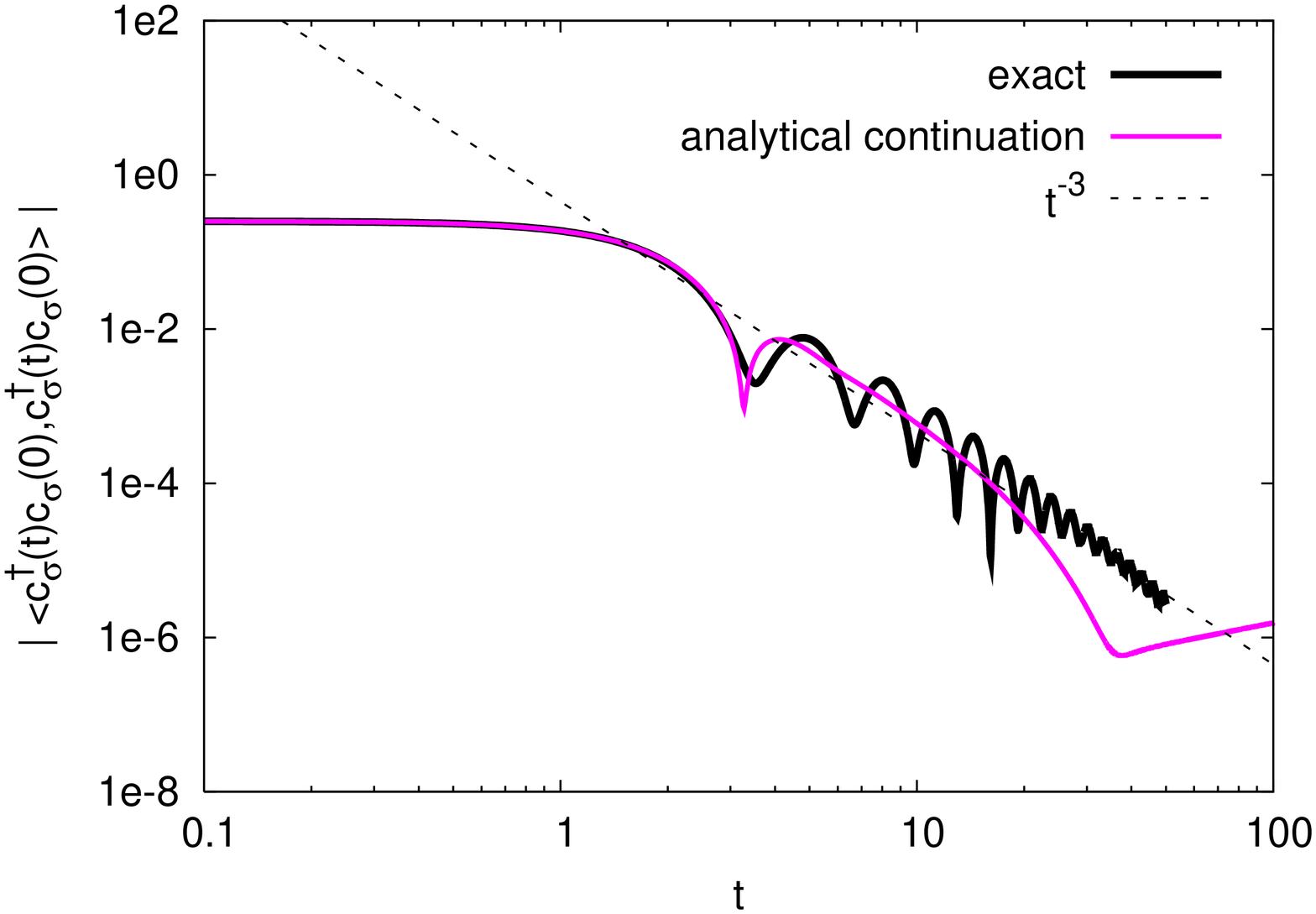}
\caption{
Left panel: Comparison between the exact solution and the result obtained by analytical continuation
for the real and imaginary parts of the OTOC $\langle c_\sigma^\dagger(t)c_\sigma(0), c_\sigma^\dagger(t)c_\sigma(0)\rangle$ 
in the noninteracting case ($U=0, \beta\hbar=50$) of the single-orbital Hubbard model.
Right panel: Comparison of the long-time behavior for the modulus $|\langle c_\sigma^\dagger(t)c_\sigma(0), c_\sigma^\dagger(t)c_\sigma(0)\rangle|$. 
The exact result shows a power-law decay ($\sim 1/t^3$). 
}
\label{fig:test_u0}
\end{center}
\end{figure}

As a check of our procedure, we first compare the results for the noninteracting model,
for which an exact solution of the OTOC $\langle c_\sigma^\dagger(t)c_\sigma(0), c_\sigma^\dagger(t)c_\sigma(0)\rangle$ 
is available \cite{TsujiWernerUeda2017}.
It is a nontrivial task to measure the noninteracting two-particle Green's function in CT-HYB, 
and to analytically continue 
$C_{(c^\dagger_\sigma c_\sigma)^2}^M(\tau)$ 
by the Maximum Entropy method.  
The results for the inverse temperature $\beta\hbar=50$ are plotted in Fig.~\ref{fig:test_u0}. 
One can see that the OTOC oscillates for about one cycle, and quickly decays to zero.
The analytically continued data show a good agreement with the exact solution. 
In particular, for short times (up to about two inverse hoppings) the dynamics is accurately reproduced, while deviations appear at longer times. 
It is known that for the noninteracting fermion system the OTOC decays as a power law at long time ($\sim 1/t^3$) \cite{TsujiWernerUeda2017}.
The analytical continuation method cannot reproduce the details of the oscillations at intermediate or long times, but it roughly captures the long-time decay, which is controlled by low-frequency spectral features. 
Usually, analytic continuation is unreliable for high frequency components, because high-frequency information is suppressed 
by the kernel $K(\tau,\omega)$ in the transformation from the spectral function $A_O(\omega)$ to the (bosonic) Matsubara correlation function $O(\tau)$ [$O(\tau)=\int_{-\infty}^{\infty} d\omega\, K(\omega,\tau) A_O(\omega)$ with $K(\tau,\omega)=e^{-\tau\omega}/(1-e^{-\beta\omega})$]. On the other hand, the low frequency components can be rather accurately determined. As a result, in Fig. 4 we more or less recover the low-frequency features of the time-dependent correlation function, while the details of rapid oscillations are not captured. The accurate result at short times can be understood from the fact that this region is close to the imaginary time axis in the complex time plane, where the original QMC data are available.
At higher temperatures, the Maximum Entropy method becomes less reliable, so that the resulting OTOCs are expected to be less accurate.

\subsection{Single-orbital Hubbard model}
\label{sec:hubbard}

In this section, we calculate the OTOCs (\ref{fermion_otoc})-(\ref{otoc_nn}) for the interacting single-orbital Hubbard model with the Hamiltonian 
\begin{align}
H
&=
-v\sum_{\langle i,j\rangle\sigma} (c^\dagger_{i\sigma}c_{j\sigma}+ \text{h.c.})
-\mu\sum_i \hat n_i+U\sum_i \hat n_{i\uparrow}\hat n_{i\downarrow}, 
\end{align}
where $v$ is the hopping amplitude, $\langle i,j\rangle$ represents the nearest-neighbor site pairs, 
$\mu$ is the chemical potential, and $U$ is the on-site interaction.
In this section, we focus on the half-filled case (i.e., $\mu=U/2$). The DMFT solution for the simplified self-consistency (\ref{semicirc}) is the exact solution for an infinitely connected Bethe lattice \cite{GeorgesKotliarKrauthRozenberg1996}.
Let us briefly recall the paramagnetic phase diagram for the single-orbital Hubbard model on this lattice \cite{Bluemer_thesis}. At half-filling, there is a metal-insulator crossover at high temperature and a first-order Mott transition below a temperature corresponding to $\beta\hbar\approx 17$ with a coexistence region between $U_{c1}$ and $U_{c2}$. The finite-temperature critical endpoint is at $U\approx 4.7$, while the zero-temperature Mott transition occurs at $U=U_{c2}\approx 5.6$. 
At low temperature, the self-energy shows the Fermi liquid behavior [${\rm Im}\,\Sigma(\omega)\sim \omega^2$] 
in the weakly correlated metallic phase, 
while it shows an insulating behavior [${\rm Im}\,\Sigma(\omega)\sim \delta(\omega)$] in the Mott phase
\cite{GeorgesKotliarKrauthRozenberg1996}. 
In the correlated metallic phase, as the temperature is increased, deviations from the Fermi liquid behavior become apparent, 
but an ${\rm Im}\,\Sigma(\omega)\sim \sqrt{\omega}$ scaling as in the SYK model is not observed. 
In the following, we will start the DMFT iterations from the noninteracting solution, which means that the finite-temperature Mott transition occurs at $U_{c2}$.
From now on, we set $\hbar=1$. 

\begin{figure}[t]
\begin{center}
\includegraphics[angle=-90, width=0.49\columnwidth]{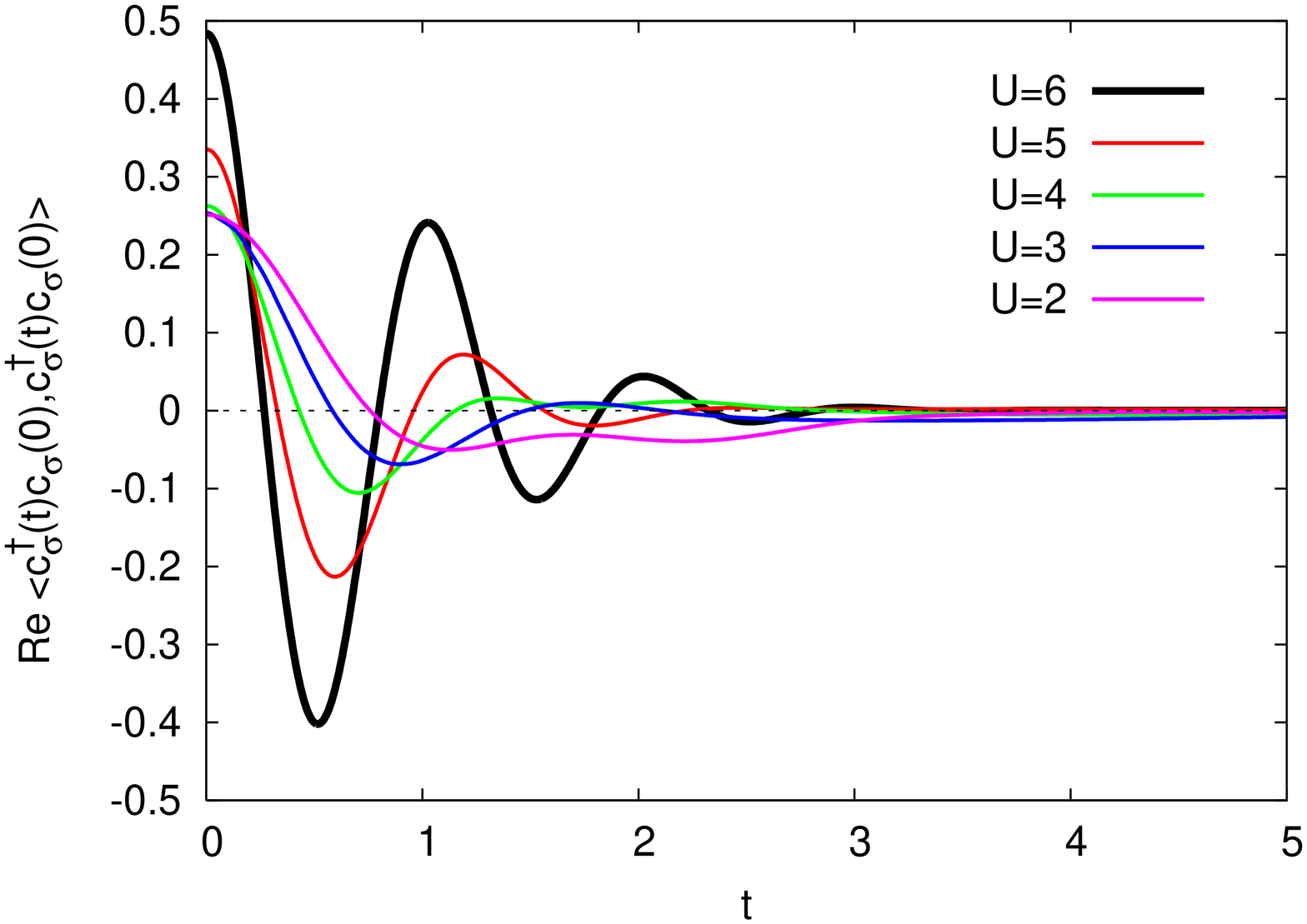}
\includegraphics[angle=-90, width=0.49\columnwidth]{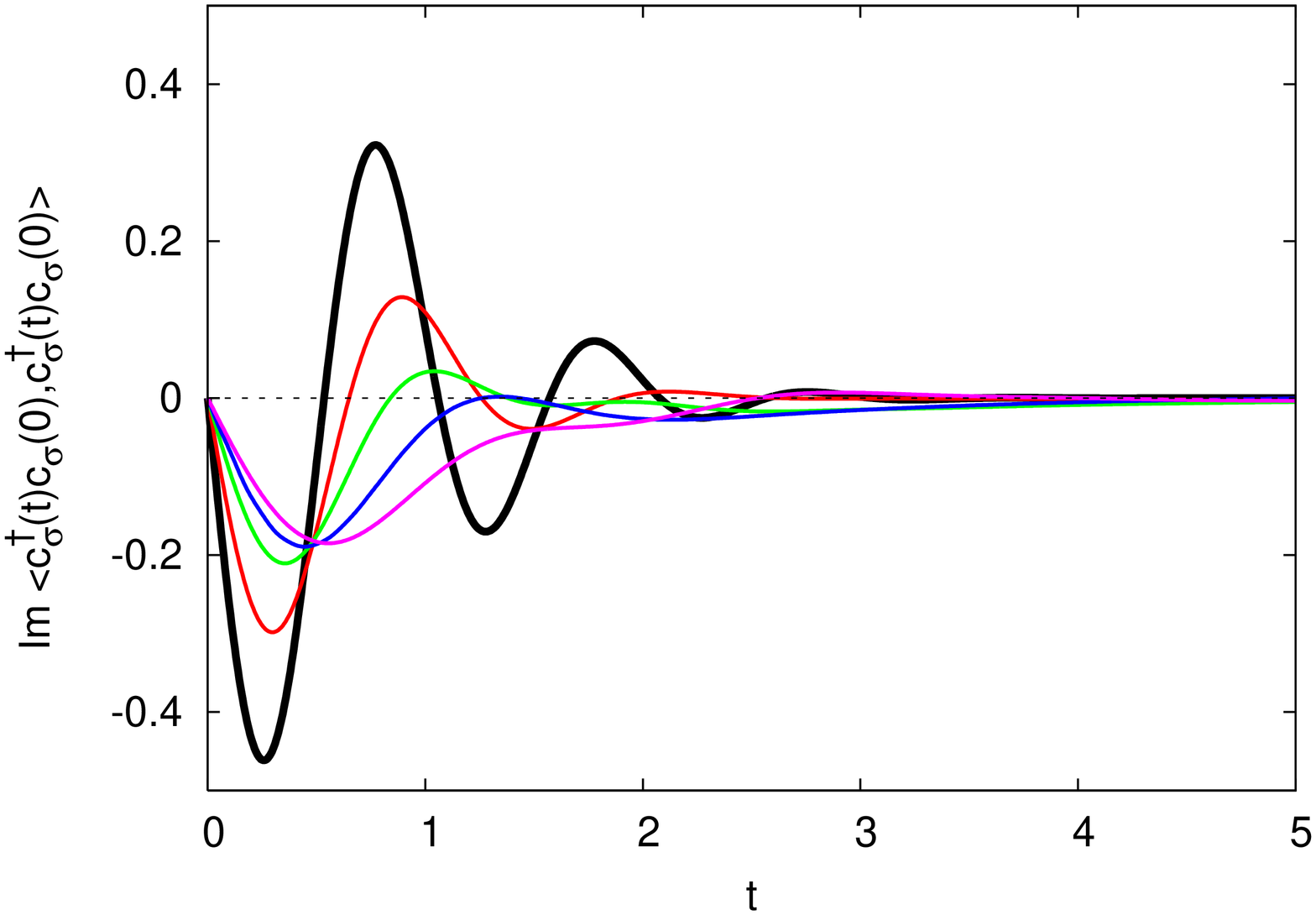}
\includegraphics[angle=-90, width=0.49\columnwidth]{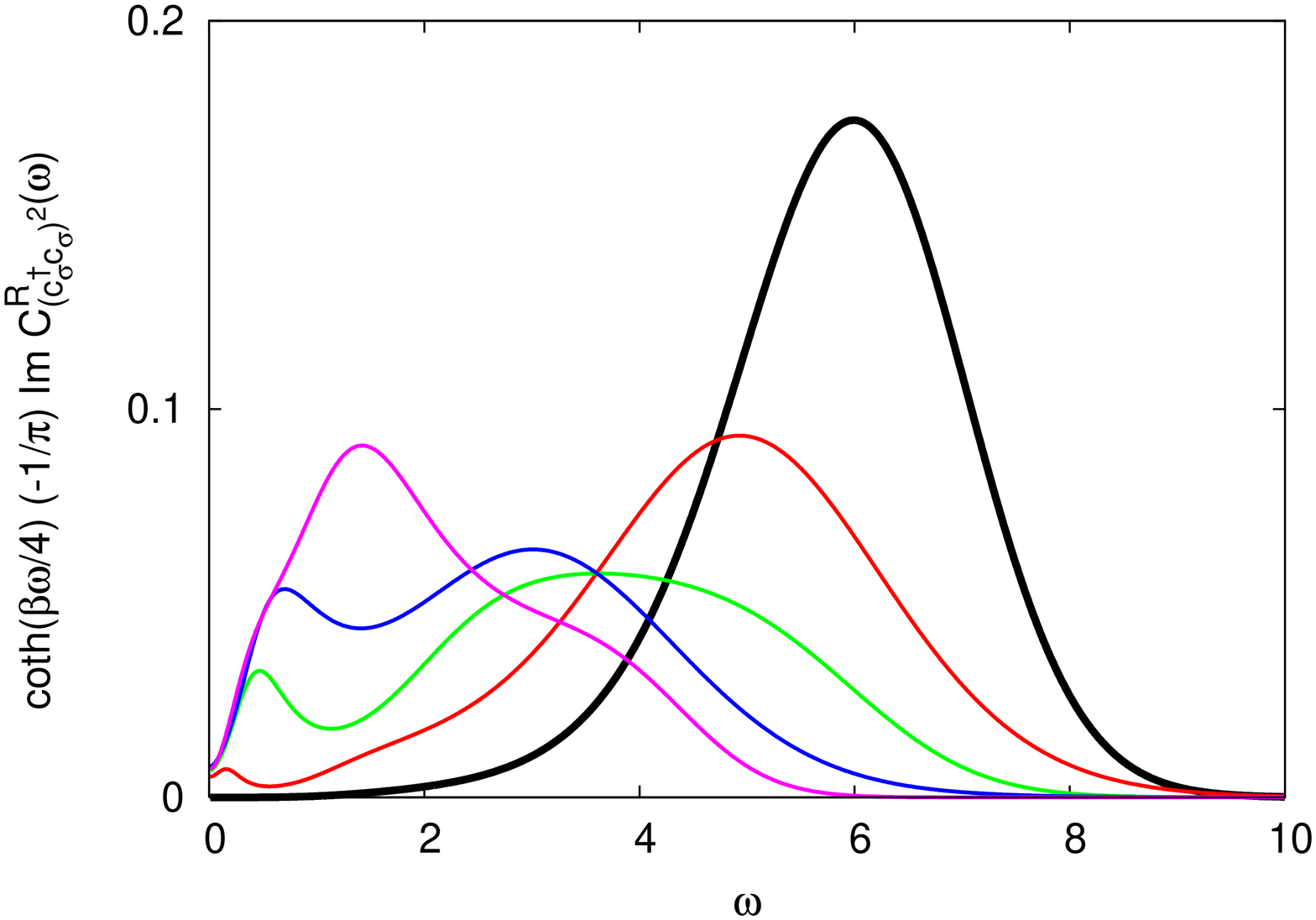}
\includegraphics[angle=-90, width=0.49\columnwidth]{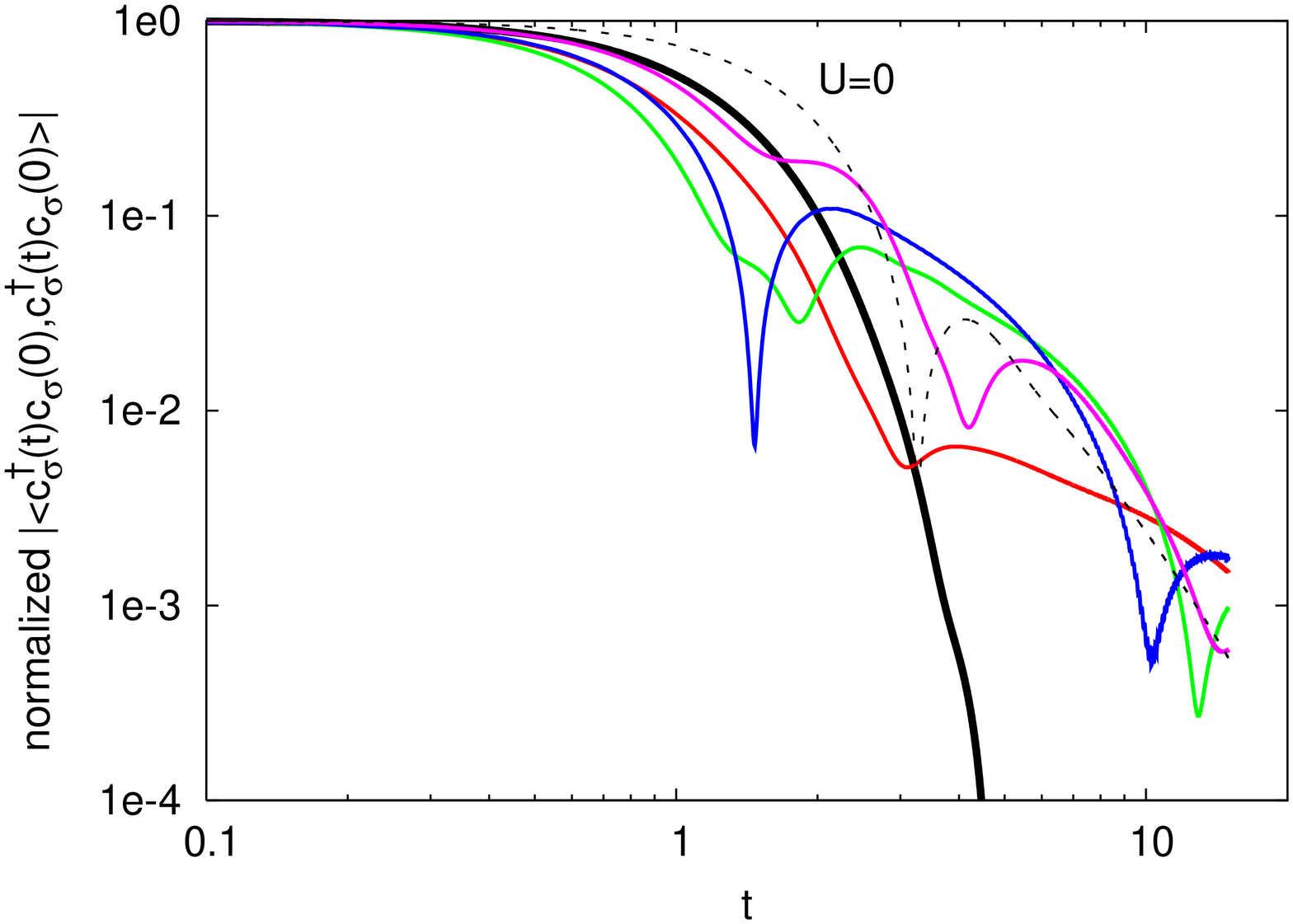}
\caption{
Top panels: Real and imaginary parts of the OTOC $\langle c^\dagger_\sigma(t)c_\sigma(0), c^\dagger_\sigma(t)c_\sigma(0)\rangle$ for the single-orbital Hubbard model with $\beta=50$ at half filling. The thin lines correspond to metallic solutions, while the thick line ($U=6$) corresponds to a Mott insulating solution. 
Bottom left panel: Spectral function $\mathscr A_{(c_\sigma^\dagger c_\sigma)^2}(\omega)=-\frac{1}{\pi}{\rm Im}\,C_{(c^\dagger_\sigma c_\sigma)^2}^R(\omega)$ multiplied by $\coth\big(\frac{\beta\omega}{4}\big)$. 
Bottom right panel: Modulus of the OTOC $|\langle c^\dagger_\sigma(t)c_\sigma(0), c^\dagger_\sigma(t)c_\sigma(0)\rangle|$ (normalized at $t=0$) on a log-log scale. The dashed line is the result for $U=0$.
}
\label{fig:results_fermion}
\end{center}
\end{figure}

In Fig.~\ref{fig:results_fermion}, we show the results of $\langle c^\dagger_\sigma(t)c_\sigma(0), c^\dagger_\sigma(t)c_\sigma(0)\rangle$ for the interacting single-orbital Hubbard model at $\beta=50$. 
The top panels present the real and imaginary parts of the OTOC. One can see that the oscillations become more pronounced as one increases the interaction. In the bottom left panel of Fig.~\ref{fig:results_fermion},
we show the corresponding spectral function $\mathscr A_{(c_\sigma^\dagger c_\sigma)^2}(\omega)=-\frac{1}{\pi}{\rm Im}\,C_{(c^\dagger_\sigma c_\sigma)^2}^R(\omega)$ multiplied by $\coth\big(\frac{\beta\omega}{4}\big)$.
The Mott transition, which occurs between $U=5$ and $U=6$ at $\beta=50$, manifests itself by the opening of a gap in the spectral function $\mathscr A_{(c_\sigma^\dagger c_\sigma)^2(\omega)}$. This translates into more weakly damped oscillations in the real-time evolution in the Mott phase.  
The bottom right panel of Fig.~\ref{fig:results_fermion} plots the modulus of the OTOC (normalized at time $t=0$, i.e., $|\langle c^\dagger_\sigma(t)c_\sigma(0), c^\dagger_\sigma(t)c_\sigma(0)\rangle|/|\langle c^\dagger_\sigma(0)c_\sigma(0), c^\dagger_\sigma(0)c_\sigma(0)\rangle|$) on a log-log scale. 
Due to the oscillations and the limited accuracy of the analytic continuation procedure, it is hard to clearly identify the nature of the long-time decay of the OTOC,
but the results indicate that the OTOC decays much faster (possibly exponentially) in the Mott phase than in the metallic phase. This is consistent with the results for the spectral functions of the OTOC. The comparison to the noninteracting case (dashed curve in the bottom right panel of Fig.~\ref{fig:results_fermion}) suggests that the correlated metallic phase has a similar (possibly power-law) decay behavior of the OTOC as in the noninteracting case.

\begin{figure}[t]
\begin{center}
\includegraphics[angle=-90, width=0.49\columnwidth]{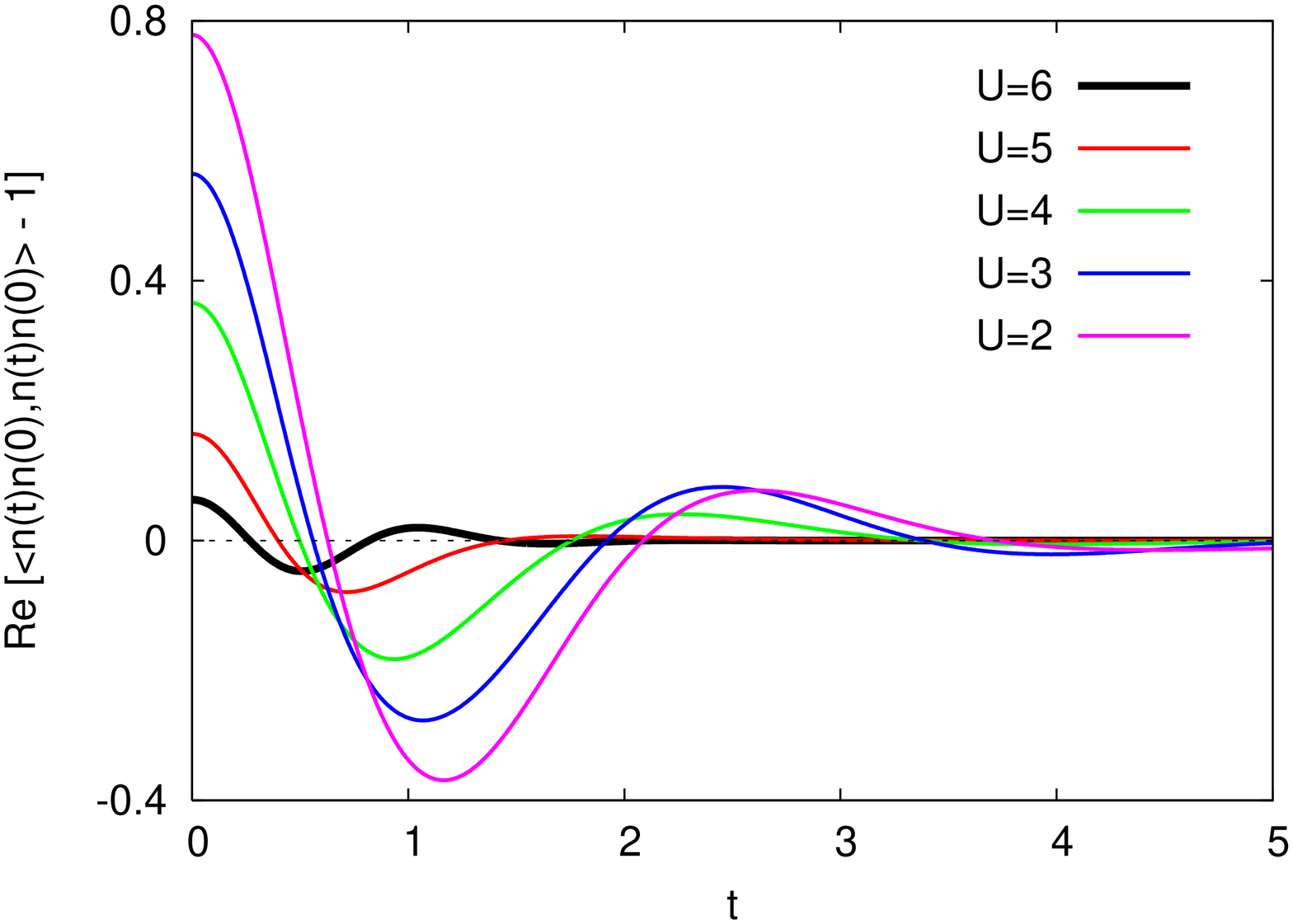}
\includegraphics[angle=-90, width=0.49\columnwidth]{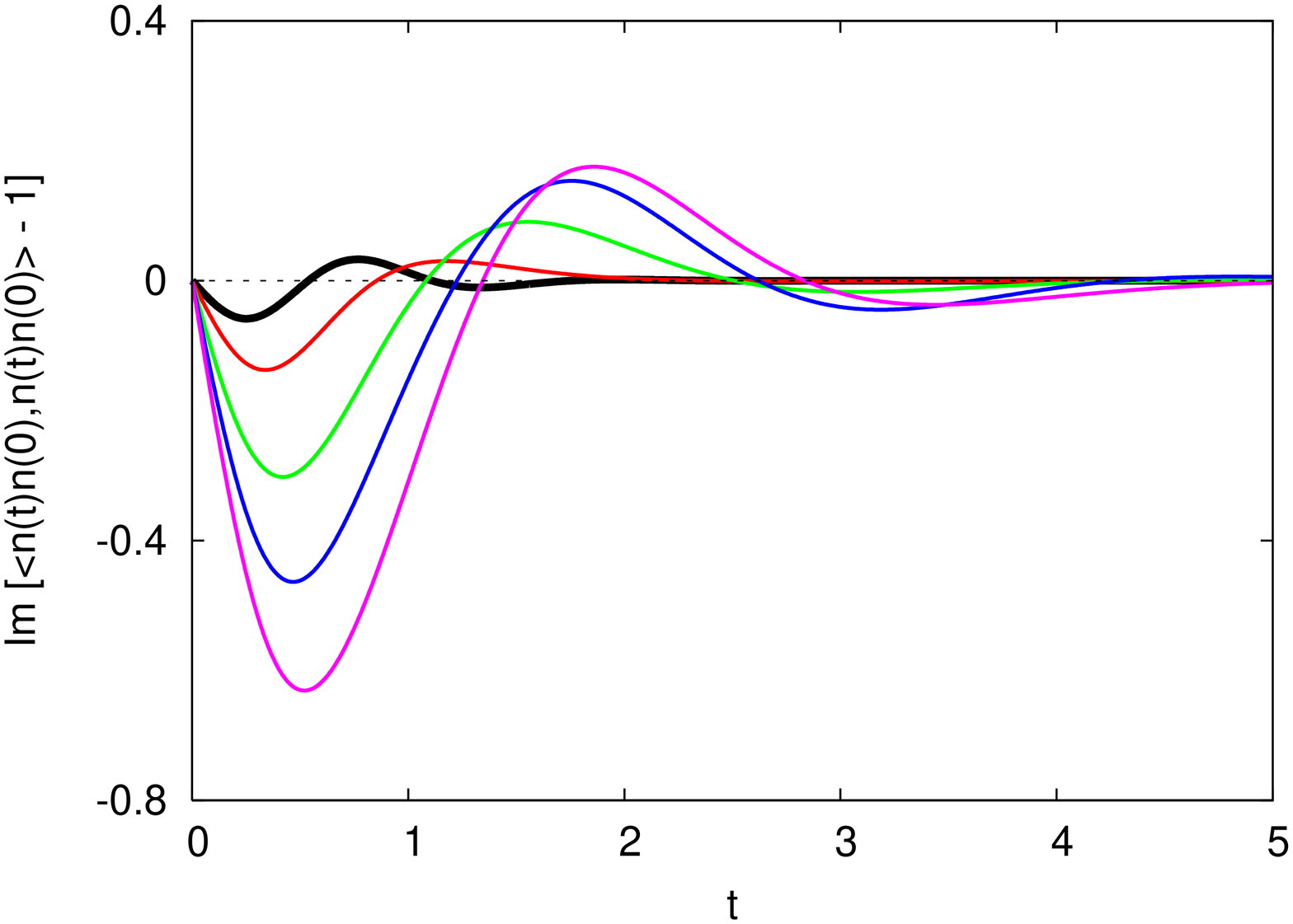}
\includegraphics[angle=-90, width=0.49\columnwidth]{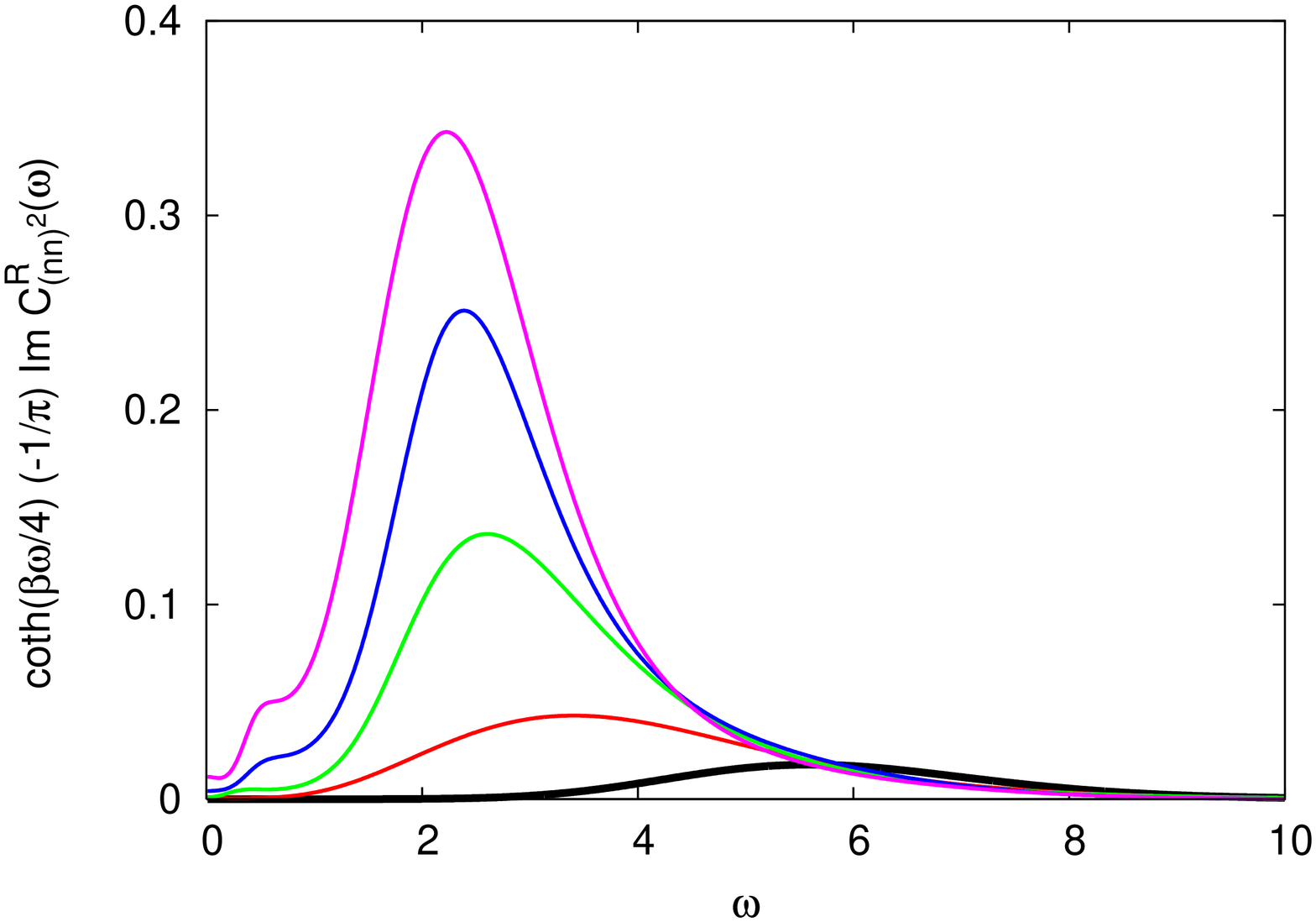}
\includegraphics[angle=-90, width=0.49\columnwidth]{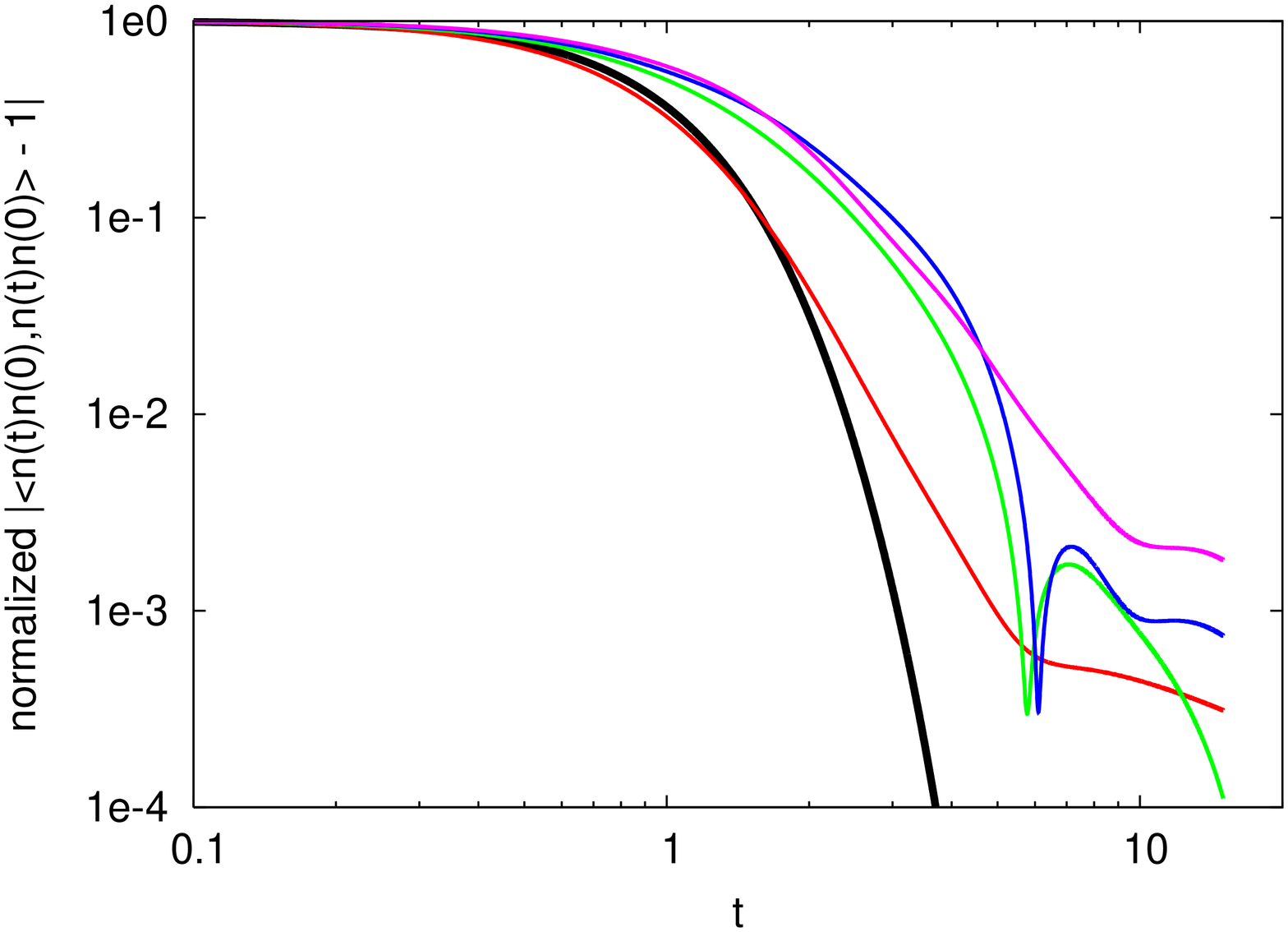}
\caption{
Top panels: Real and imaginary parts of the OTOC $\langle \hat n(t)\hat n(0), \hat n(t)\hat n(0)\rangle - 1$ for the single-orbital Hubbard model with $\beta=50$ at half filling. The thin lines correspond to metallic solutions, 
while the thick line ($U=6$) corresponds to a Mott insulating solution. 
Bottom left panel: Spectral function $\mathscr A_{(nn)^2}(\omega)=-\frac{1}{\pi}{\rm Im}\,C_{(n n)^2}^R(\omega)$ multiplied by $\coth\big(\frac{\beta\omega}{4}\big)$.
Bottom right panel: Modulus of the OTOC $|\langle \hat n(t)\hat n(0), \hat n(t)\hat n(0)\rangle - 1|$ (normalized at $t=0$) on a log-log scale. 
}
\label{fig:results_nn}
\end{center}
\end{figure}

In Fig.~\ref{fig:results_nn}, we plot the OTOC $\langle \hat n(t)\hat n(0), \hat n(t)\hat n(0)\rangle$
for the single-orbital Hubbard model in a similar manner as in Fig.~\ref{fig:results_fermion}.
Since at half filling this correlation function approaches $\langle \hat n\rangle^4=1$ at long times
and at sufficiently low temperature,
we perform the analytical continuation procedure for $\langle \hat n(t)\hat n(0), \hat n(t)\hat n(0)\rangle-1$. 
The top panels of Fig.~\ref{fig:results_nn} show the real and imaginary parts of this shifted OTOC. 
Contrary to the case of $\langle c_\sigma^\dagger(t)c_\sigma(0), c_\sigma^\dagger(t)c_\sigma(0)\rangle$, 
the amplitude of the OTOC $\langle \hat n(t)\hat n(0), \hat n(t)\hat n(0)\rangle-1$ is suppressed as one increases the interaction, which reflects the reduced charge fluctuations in the Mott insulator.
The bottom left panel in Fig.~\ref{fig:results_nn} shows the analytically continued spectral function $\mathscr A_{(nn)^2}(\omega)=-\frac{1}{\pi}{\rm Im}\, C_{(nn)^2}^R(\omega)$ multiplied by $\coth\big(\frac{\beta\omega}{4}\big)$. 
Again we notice the opening of a gap in the Mott phase, and a corresponding shift of the spectral weight to higher energies. This results in more rapid oscillations of the OTOC in the Mott state.
The bottom right panel in Fig.~\ref{fig:results_nn} plots the modulus of the OTOC $|\langle \hat n(t)\hat n(0), \hat n(t)\hat n(0)\rangle-1|$ (normalized at $t=0$) on a log-log scale.
The long-time behavior of the OTOC is consistent with a power-law decay in the metallic phase, and an exponential decay in the Mott insulating phase.

\begin{figure}[t]
\begin{center}
\includegraphics[angle=-90, width=0.49\columnwidth]{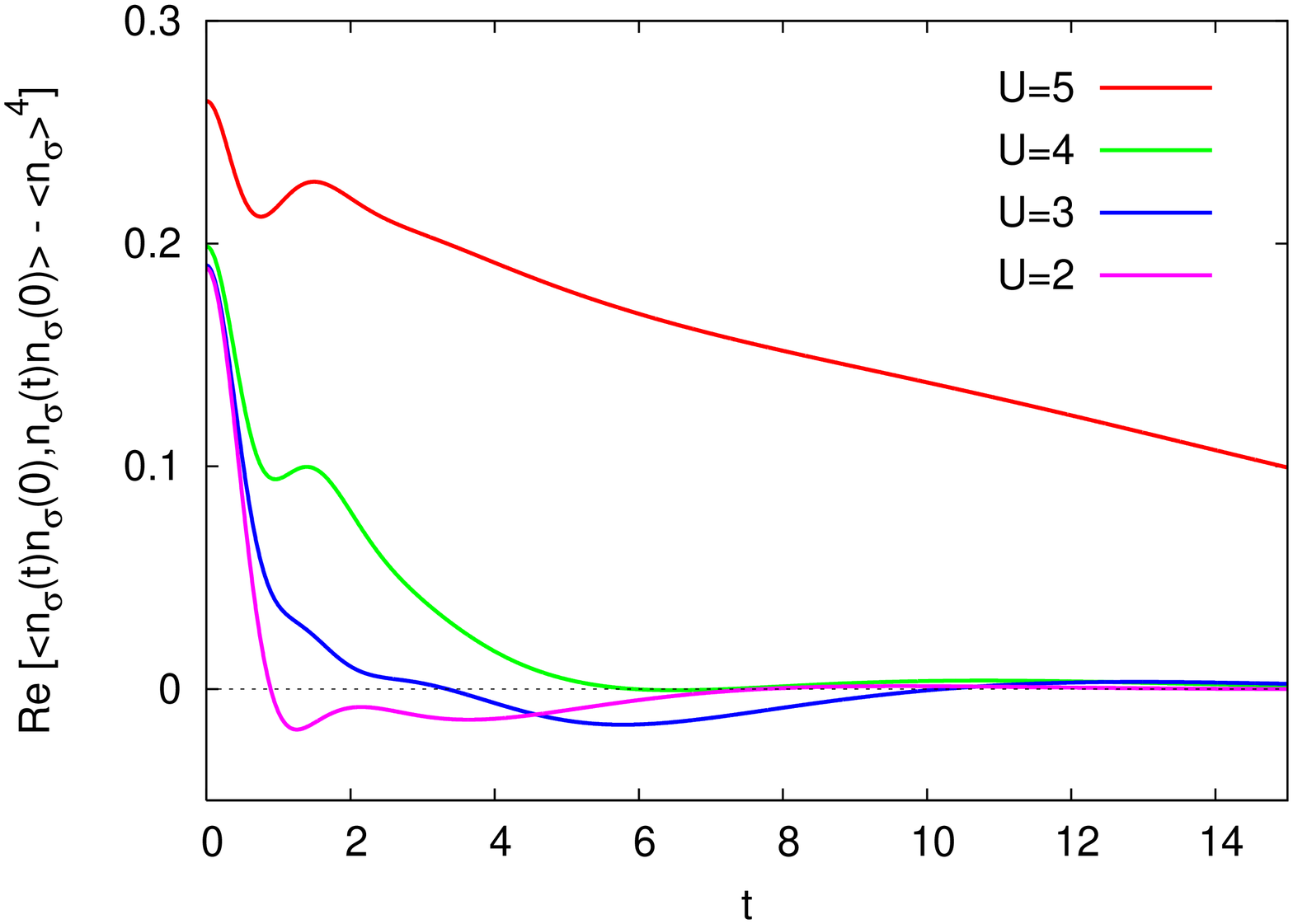}
\includegraphics[angle=-90, width=0.49\columnwidth]{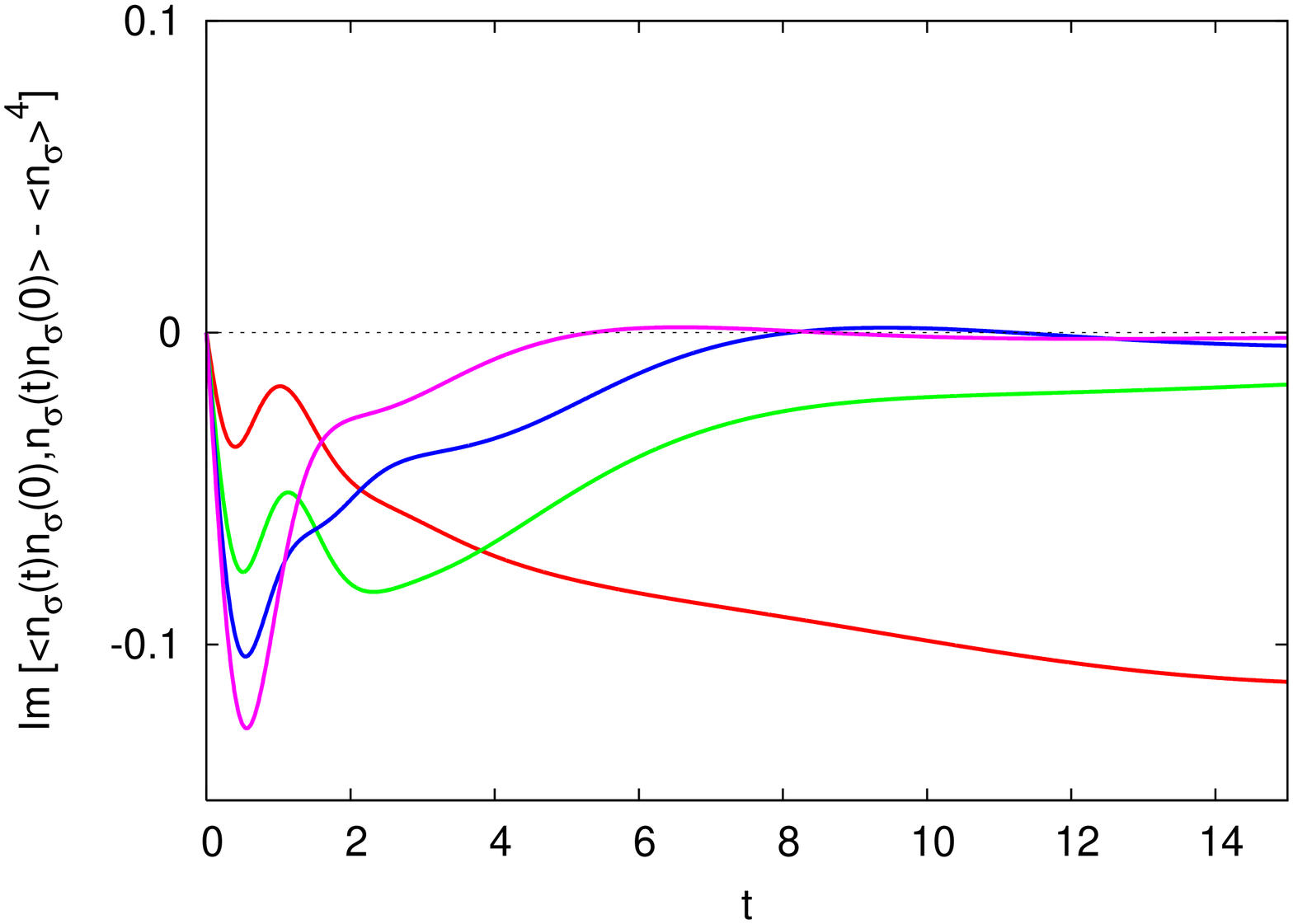}
\includegraphics[angle=-90, width=0.49\columnwidth]{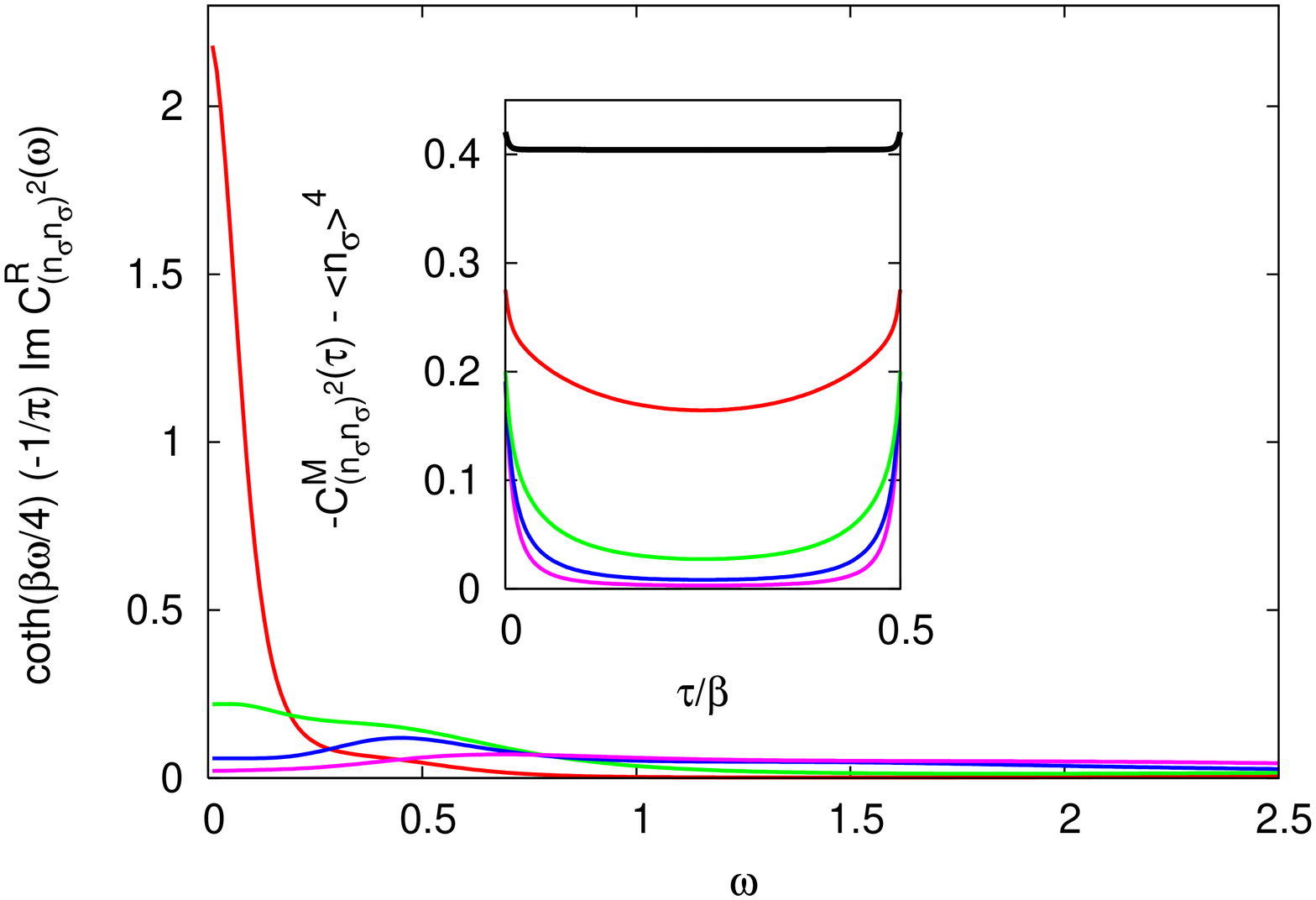}
\includegraphics[angle=-90, width=0.49\columnwidth]{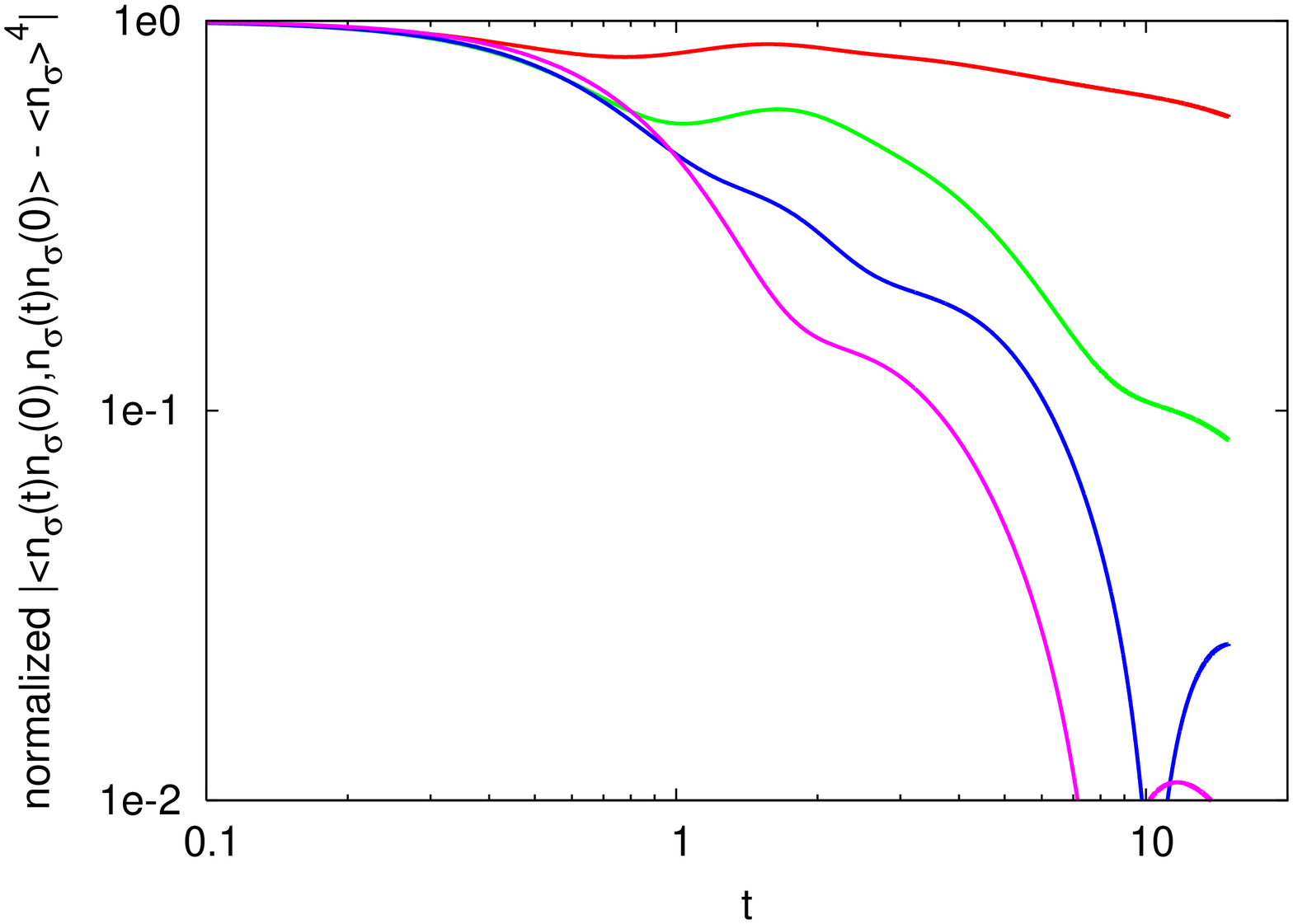}
\caption{
Top panels: Real and imaginary parts of the OTOC $\langle \hat n_\sigma(t)\hat n_\sigma(0), \hat n_\sigma(t)\hat n_\sigma(0)\rangle  - \langle \hat n_\sigma\rangle^4$ for the single-orbital Hubbard model with $\beta=50$ at half filling. All the colored lines correspond to metallic solutions. 
Bottom left panel: Spectral function $\mathscr A_{(n_\sigma n_\sigma)^2}(\omega)=-\frac{1}{\pi}{\rm Im}\,C_{(n_\sigma n_\sigma)^2}^R(\omega)$ multiplied by $\coth\big(\frac{\beta\omega}{4}\big)$. In the inset, we plot the imaginary-time four-point function $\langle \hat n_\sigma(\tau)\hat n_\sigma(0), \hat n_\sigma(\tau)\hat n_\sigma(0)\rangle-\langle \hat n_\sigma\rangle^4$, with the black line corresponding to the Mott insulating solution at $U=6$.  
Bottom right panel: Modulus of the OTOC $|\langle \hat n_\sigma(t)\hat n_\sigma(0), \hat n_\sigma(t)\hat n_\sigma(0)\rangle  - \langle \hat n_\sigma\rangle^4|$ (normalized at $t=0$) on a log-log scale.
}
\label{fig:results_nsns}
\end{center}
\end{figure}

Figure~\ref{fig:results_nsns} 
shows the results of the OTOC $\langle \hat n_\sigma(t)\hat n_\sigma(0), \hat n_\sigma(t)\hat n_\sigma(0)\rangle$
for the single-orbital Hubbard model. At half filling, this correlation function approaches $\langle \hat n_\sigma\rangle^4=\frac{1}{16}$ in the long-time limit and at sufficiently low temperature, 
so that we perform the analytical continuation procedure for $\langle \hat n_\sigma(t)\hat n_\sigma(0), \hat n_\sigma(t)\hat n_\sigma(0)\rangle-\langle \hat n_\sigma\rangle^4$. 
The top panels of Fig.~\ref{fig:results_nsns} show the real and imaginary parts of 
this shifted OTOC. 
Here, we only present results for the metallic phase, because resolving the very sharp low-energy feature of the OTOC spectrum in the half-filled Mott state is challenging. One can see that coherent oscillations are not observed for this type of OTOC, and that the incoherent part becomes dominant compared to the other two OTOCs shown in Figs.~\ref{fig:results_fermion} and \ref{fig:results_nn}. The amplitude of the OTOC $\langle \hat n_\sigma(t)\hat n_\sigma(0), \hat n_\sigma(t)\hat n_\sigma(0)\rangle-\langle \hat n_\sigma\rangle^4$ is enhanced as $U$ increases. As we argue below, this is related to the enhancement of spin fluctuations 
and local moment formation close to the Mott transition.

The bottom left panel in Fig.~\ref{fig:results_nsns} plots the analytically continued spectral function $\mathscr A_{(n_\sigma n_\sigma)^2}(\omega)=-\frac{1}{\pi}{\rm Im}\, C_{(n_\sigma n_\sigma)^2}^R(\omega)$ multiplied by $\coth\big(\frac{\beta\omega}{4}\big)$. 
We can see that the spectral weight is concentrated in the low-energy region as we approach the Mott transition point.
The inset in the bottom left panel of Fig.~\ref{fig:results_nsns} shows the imaginary-time four-point function $\langle \hat n_\sigma(\tau)\hat n_\sigma(0), \hat n_\sigma(\tau)\hat n_\sigma(0)\rangle-\langle \hat n_\sigma\rangle^4$, with the black line corresponding to a Mott insulating solution ($U=6$).  
For $U\gtrsim 4$, the imaginary-time four-point function does not decay to zero but remains relatively large near $\tau=\frac{\beta}{4}$. 
This behavior is reminiscent of the spin-freezing physics seen in the correlated Hund metal phase of multi-orbital Hubbard models, where the dynamical spin correlation function $\langle \hat S_z(\tau)\hat S_z(0)\rangle$ is trapped at a finite value at long $\tau$ \cite{Werner2008}. 
In fact, the OTOC $\langle \hat n_\sigma(\tau)\hat n_\sigma(0), \hat n_\sigma(\tau)\hat n_\sigma(0)\rangle$
measures spin correlations (in addition to charge correlations) through $\hat n_\sigma=\frac{1}{2}\hat n+\sigma\hat S_z$.
Physically, the trapping of $\langle \hat S_z(\tau)\hat S_z(0)\rangle$ signals the formation of frozen local magnetic moments.
In the metal-insulator crossover region, the scattering induced by these magnetic moments results in incoherent metal states. Intuitively, one might expect fast scrambling in the corresponding regions of the phase diagram. 
Although there is some resemblance with the spin freezing crossover,
the self-energy does not exhibit a square-root frequency dependence in the single-orbital case, and a spin-freezing crossover in the sense of Ref.~\cite{Werner2008} does not exist.  

The bottom right panel of Fig.~\ref{fig:results_nsns} plots the modulus of the OTOC $|\langle \hat n_\sigma(t)\hat n_\sigma(0), \hat n_\sigma(t)\hat n_\sigma(0)\rangle-\langle \hat n_\sigma\rangle^4|$ (normalized at $t=0$) on a log-log scale. 
The dynamics of $\langle \hat n_\sigma(t)\hat n_\sigma(0), \hat n_\sigma(t)\hat n_\sigma(0)\rangle$ is quite distinct from that of $\langle \hat n(t)\hat n(0), \hat n(t)\hat n(0)\rangle$. As the Mott transition is approached, we observe a slow-down in the decay of the modulus, 
which indicates the presence of slowly fluctuating local moments. 
The long-time behavior may be consistent with a power-law, although slow oscillations make it difficult to determine the long-time asymptotic form from the numerics.

\begin{figure}[t]
\begin{center}
\includegraphics[angle=-90, width=0.49\columnwidth]{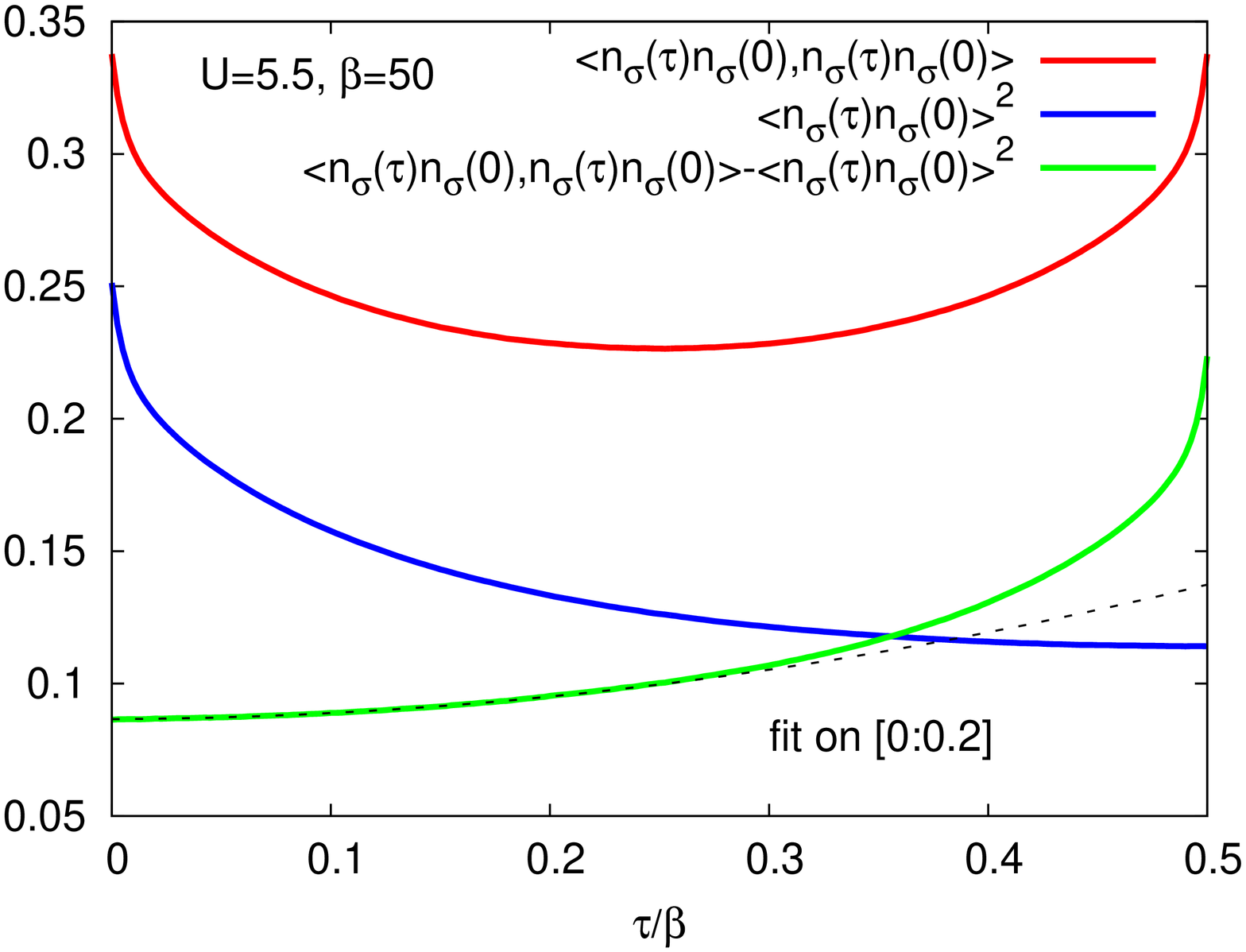}
\includegraphics[angle=-90, width=0.49\columnwidth]{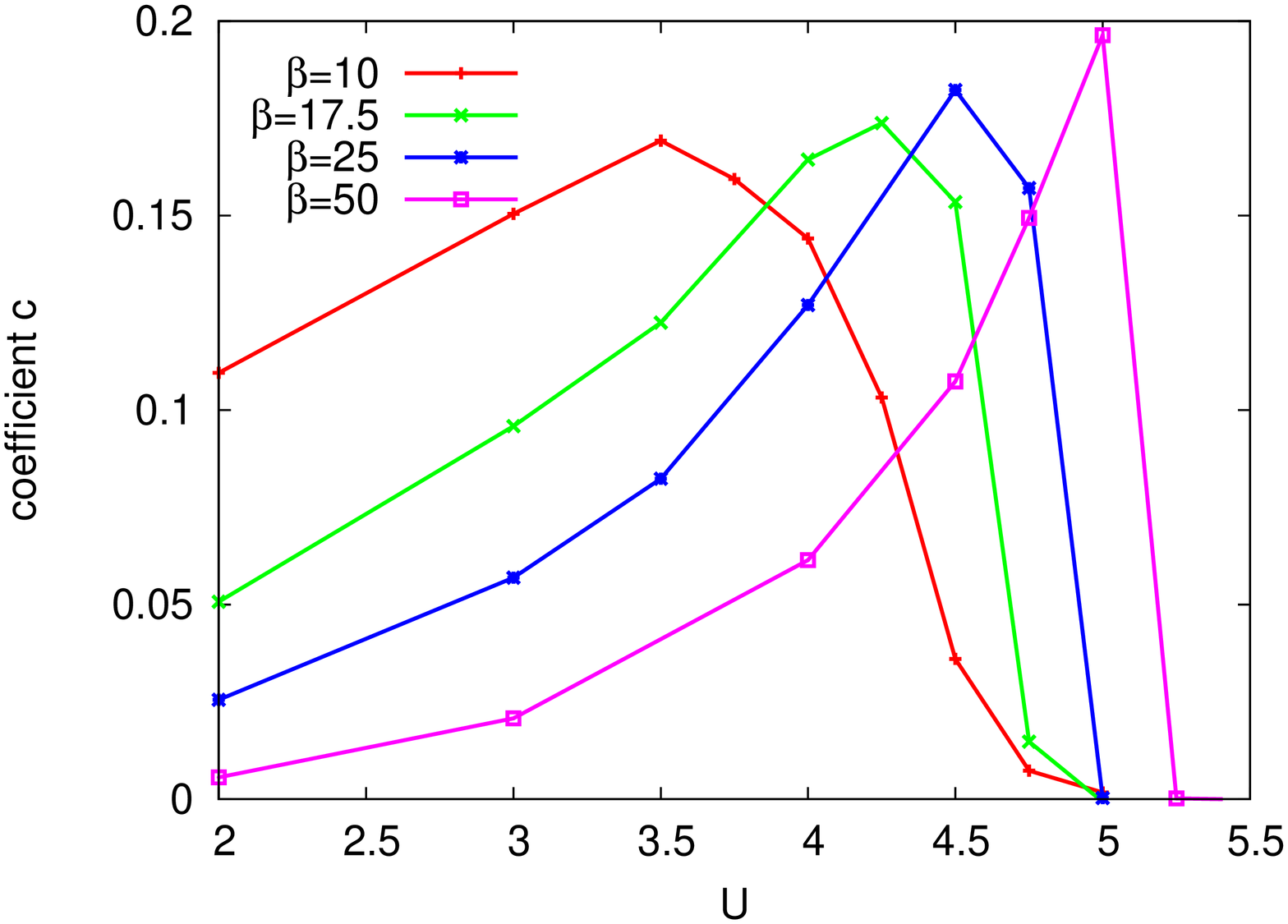}
\caption{
Left panel: 
The imaginary-time four-point function $\langle \hat n_\sigma(\tau)\hat n_\sigma(0), \hat n_\sigma(\tau)\hat n_\sigma(0)\rangle$, the squared density-density correlation function $\langle \hat n_\sigma(\tau)\hat n_\sigma(0)\rangle^2$,
and their difference for the single-orbital Hubbard model with $U=5.5$ and $\beta=50$.
The dashed curve shows a fit of the difference with the function $f_\text{fit}(\tau)=a+b\tau+c\tau^2$ in the interval $\tau/\beta \in [0: 0.2]$.
Right panel: Fitting coefficient $c$ as a function of $U$ for indicated values of $\beta$.
}
\label{fig:coeffs}
\end{center}
\end{figure}

For large enough $\beta$ and small enough $\tau$, the OTOC (\ref{otoc_nsns}) factorizes into $\langle \hat n_\sigma(\tau)\hat n_\sigma(0), \hat n_\sigma(\tau)\hat n_\sigma(0)\rangle \approx \langle \hat n_\sigma(\tau)\hat n_\sigma(0)\rangle^2$. 
On the real-time axis, this corresponds to the decoupling $\langle \hat n_\sigma(t)\hat n_\sigma(0), \hat n_\sigma(t)\hat n_\sigma(0)\rangle \approx \langle \hat n_\sigma(t)\hat n_\sigma(0)\rangle^2$, that is, 
the OTOC is reduced to a product of ordinary density-density correlation functions.
In order to see whether nontrivial correlations are captured beyond the decoupled form by the OTOC function, 
we consider the difference $\langle \hat n_\sigma(\tau)\hat n_\sigma(0), \hat n_\sigma(\tau)\hat n_\sigma(0)\rangle-\langle \hat n_\sigma(\tau)\hat n_\sigma(0)\rangle^2$ and fit the short-time behavior of this function to $f_\text{fit}(\tau)=a+b\tau+c\tau^2$, as illustrated in the left panel of Fig.~\ref{fig:coeffs}. The coefficients $b$ and $c$ serve as  indicators of the nontrivial correlations that cannot be attributed to the decoupled form of the OTOC.
The coefficient $c$ is plotted as a function of $U$ and for different $\beta$ in the right hand panel. The coefficient $b$ (not shown) exhibits a similar trend. We notice that the OTOC picks up nontrival correlations near the metal-insulator transition, and in particular at 
intermediate
temperatures, while the Mott phase shows no such correlations. The crossover region  
associated with the emergence of local moments 
corresponds, roughly, to the interaction range where the nontrivial correlations start to become significant.

\subsection{Two-orbital Hubbard model}
\label{sec:two-orbital hubbard}

In this section, we investigate OTOCs in the two-orbital Hubbard model with Hund coupling $J>0$ 
and density-density interactions. (Three-orbital results are presented in Appendix~\ref{sec:3orbital}.) 
The Hamiltonian is given by
\begin{align}
H
&=
-v\sum_{\langle i,j\rangle\alpha\sigma} (c^\dagger_{i\alpha\sigma}c_{j\alpha\sigma}+ \text{h.c.})
-\mu\sum_{i\alpha\sigma} \hat n_{i\alpha\sigma}
+U\sum_{i\alpha} \hat n_{i\alpha\uparrow}\hat n_{i\alpha\downarrow}
+(U-2J)\sum_{i\sigma} \hat n_{i1\sigma}\hat n_{i2\bar\sigma}
\notag
\\
&\quad
+(U-3J)\sum_{i\sigma} \hat n_{i1\sigma}\hat n_{i2\sigma},
\end{align}
with $U$ being the intra-orbital interaction $U$, $U-2J$ the inter-orbital antiparallel-spin interaction, and $U-3J$ the inter-orbital parallel-spin interaction. This is the simplest model that shows a crossover from a spin-frozen metal to a Fermi-liquid metal as one dopes the half-filled Mott insulator, with the self-energy scaling as $\text{Im}\Sigma(\omega_n)\sim\sqrt{\omega_n}$ over a significant energy range in the crossover regime \cite{Hafermann2012}.
The sketch of the phase diagram is shown in Fig.~\ref{fig:illustration}. 
As we have seen in the previous section, the spin-related OTOC of the type (\ref{otoc_nsns}) is relevant for the analysis of  the spin-freezing crossover, so that we concentrate on this OTOC here.
In the following calculations, we choose $U=8$, $J=U/4$, and compute $\langle \hat n_\sigma(t)\hat n_\sigma(0), \hat n_\sigma(t)\hat n_\sigma(0)\rangle - \langle n_\sigma \rangle^4$ 
as a function of filling at $\beta=50$ (dashed lines in Fig.~\ref{fig:illustration}).

\begin{figure}[t]
\begin{center}
\includegraphics[angle=-90, width=0.49\columnwidth]{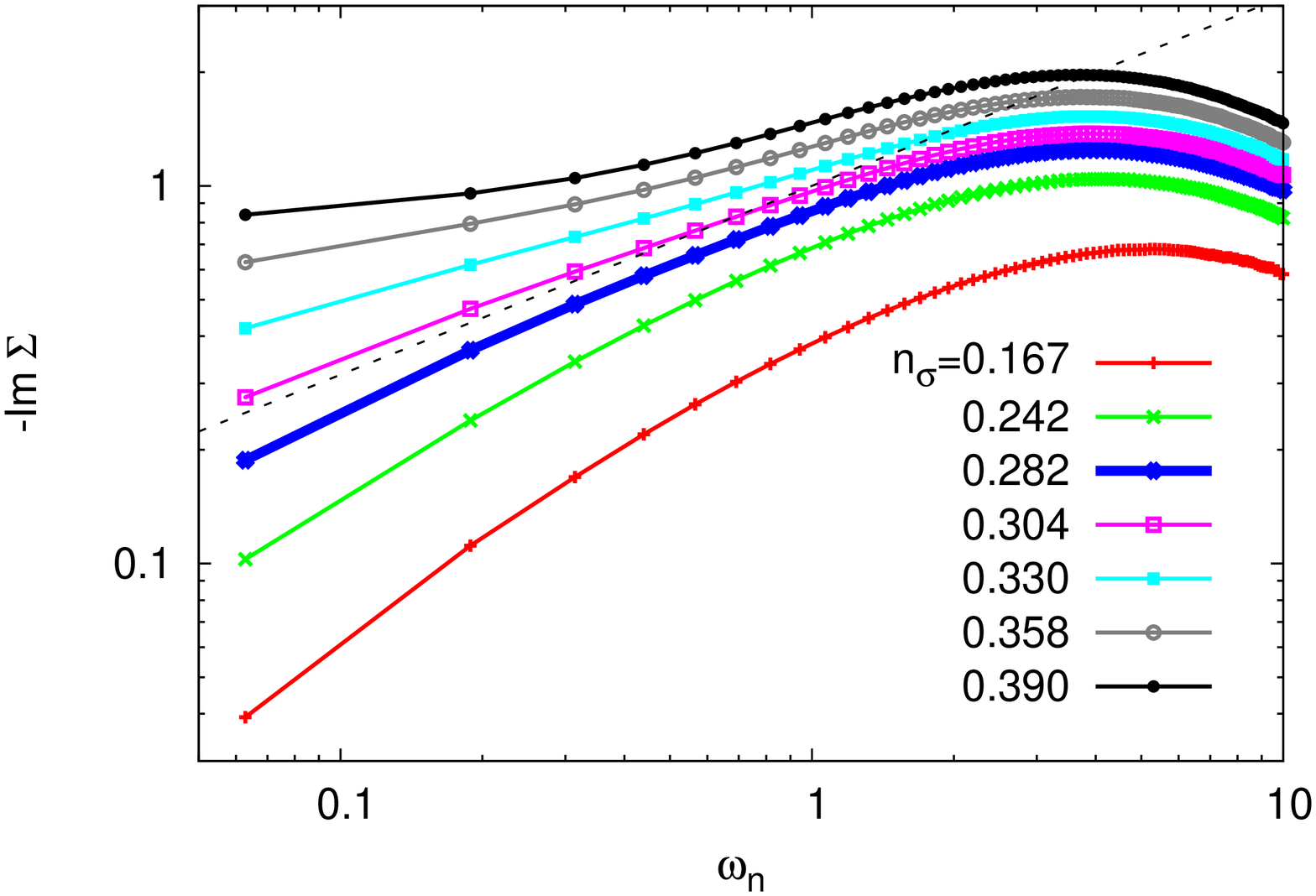}\hfill
\includegraphics[angle=-90, width=0.49\columnwidth]{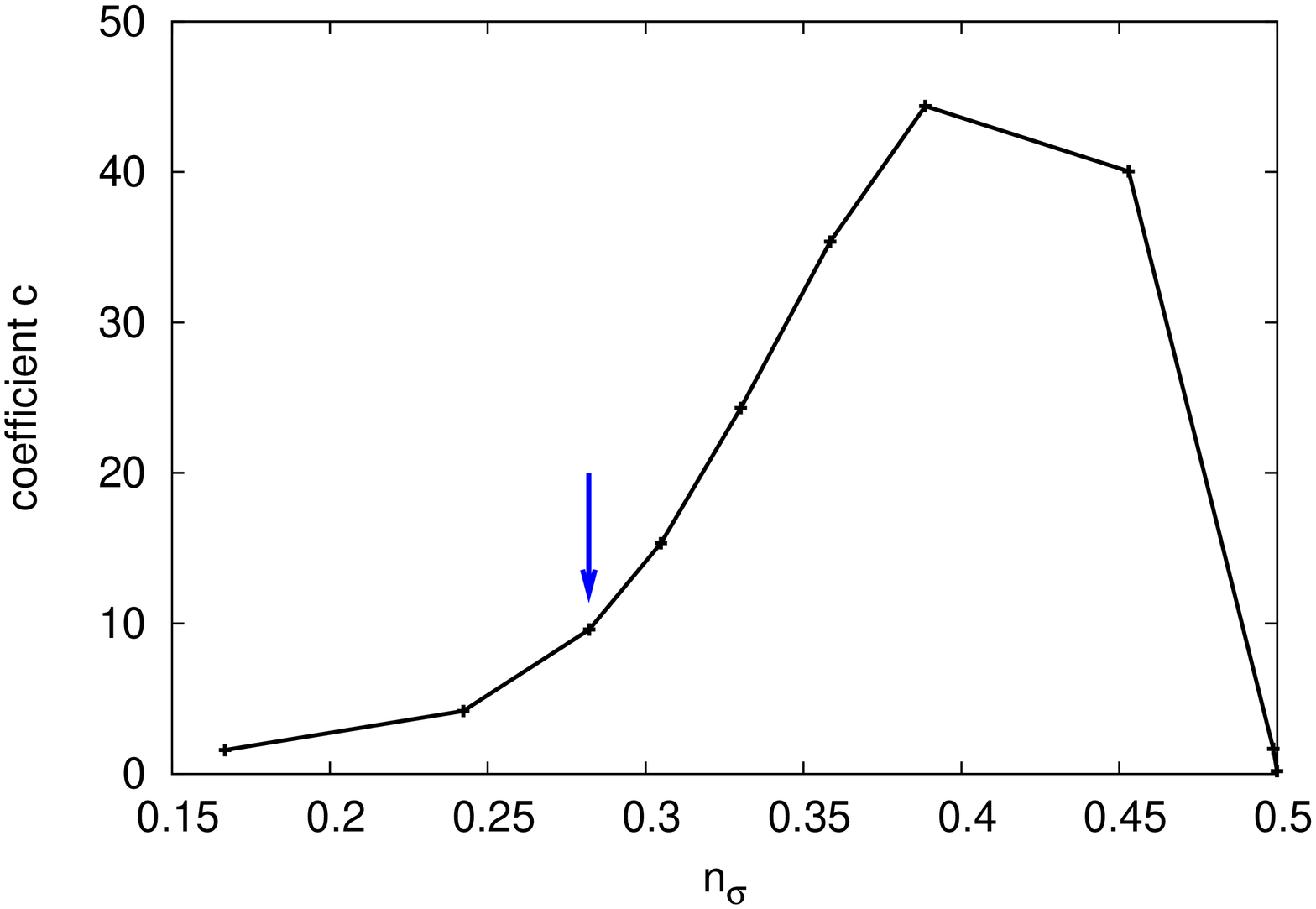}
\caption{
Left panel: Imaginary part of the self-energy as a function of Matsubara frequency for the two-orbital Hubbard model with $U=8$, $J=U/4$, and $\beta=50$ on a log-log scale, with the dashed line indicating $\sqrt{\omega_n}$ behavior. 
Right panel: The coefficient $c$ extracted from
fitting $\langle \hat n_\sigma(\tau)\hat n_\sigma(0), \hat n_\sigma(\tau)\hat n_\sigma(0)\rangle-\langle \hat n_\sigma(\tau)\hat n_\sigma(0)\rangle^2$ with a function $f_\text{fit}(\tau)=a+b\tau+c\tau^2$
in the interval $\tau/\beta\in [0: 0.05]$ for the two-orbital Hubbard model. 
The spin-freezing crossover occurs roughly at the filling $n_\sigma\approx 0.28$ (blue curve in the left panel and blue arrow in the right panel). 
}
\label{fig:self}
\end{center}
\end{figure}

The left panel of Fig.~\ref{fig:self} plots the imaginary part of the self-energy as a function of Matsubara frequency in order to identify the filling corresponding to the spin-freezing crossover \cite{Werner2008}. An approximate square-root scaling is observed over a wide range of frequencies near $n_\sigma\approx 0.28$ (blue line). In the right panel of Fig.~\ref{fig:self}, we present the coefficient $c$ obtained from a similar analysis as presented in Fig.~\ref{fig:coeffs}, but with a fitting range $\tau/\beta\in [0 : 0.05]$. Again, we find that the OTOC $\langle \hat n_\sigma(t)\hat n_\sigma(0), \hat n_\sigma(t)\hat n_\sigma(0)\rangle - \langle \hat n_\sigma\rangle^4$ detects nontrivial correlations in the spin-freezing crossover (blue arrow) and spin-frozen metal regime, but not in the Fermi liquid or the Mott insulating phase.

\begin{figure}[t]
\begin{center}
\includegraphics[angle=-90, width=0.49\columnwidth]{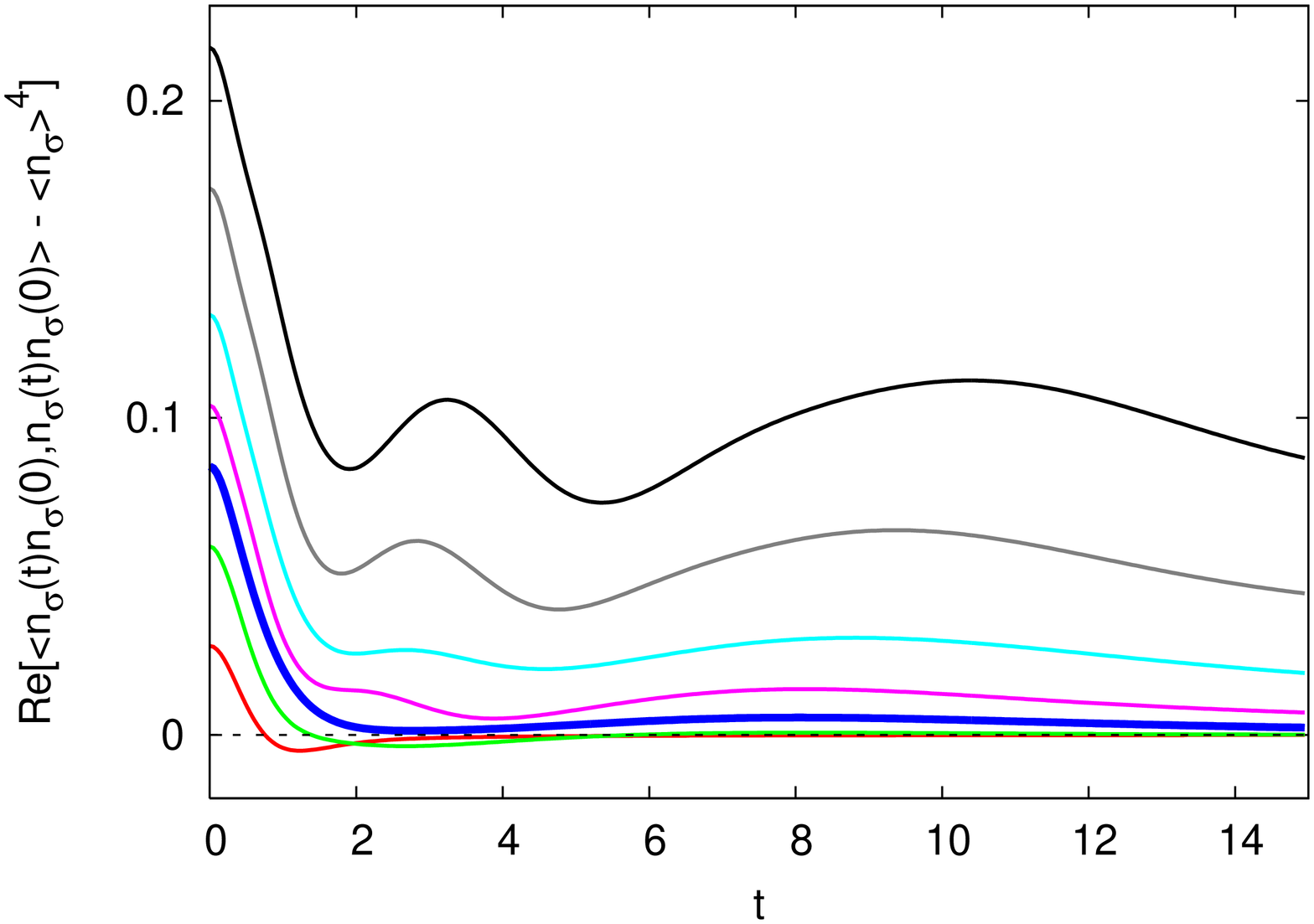}
\includegraphics[angle=-90, width=0.486\columnwidth]{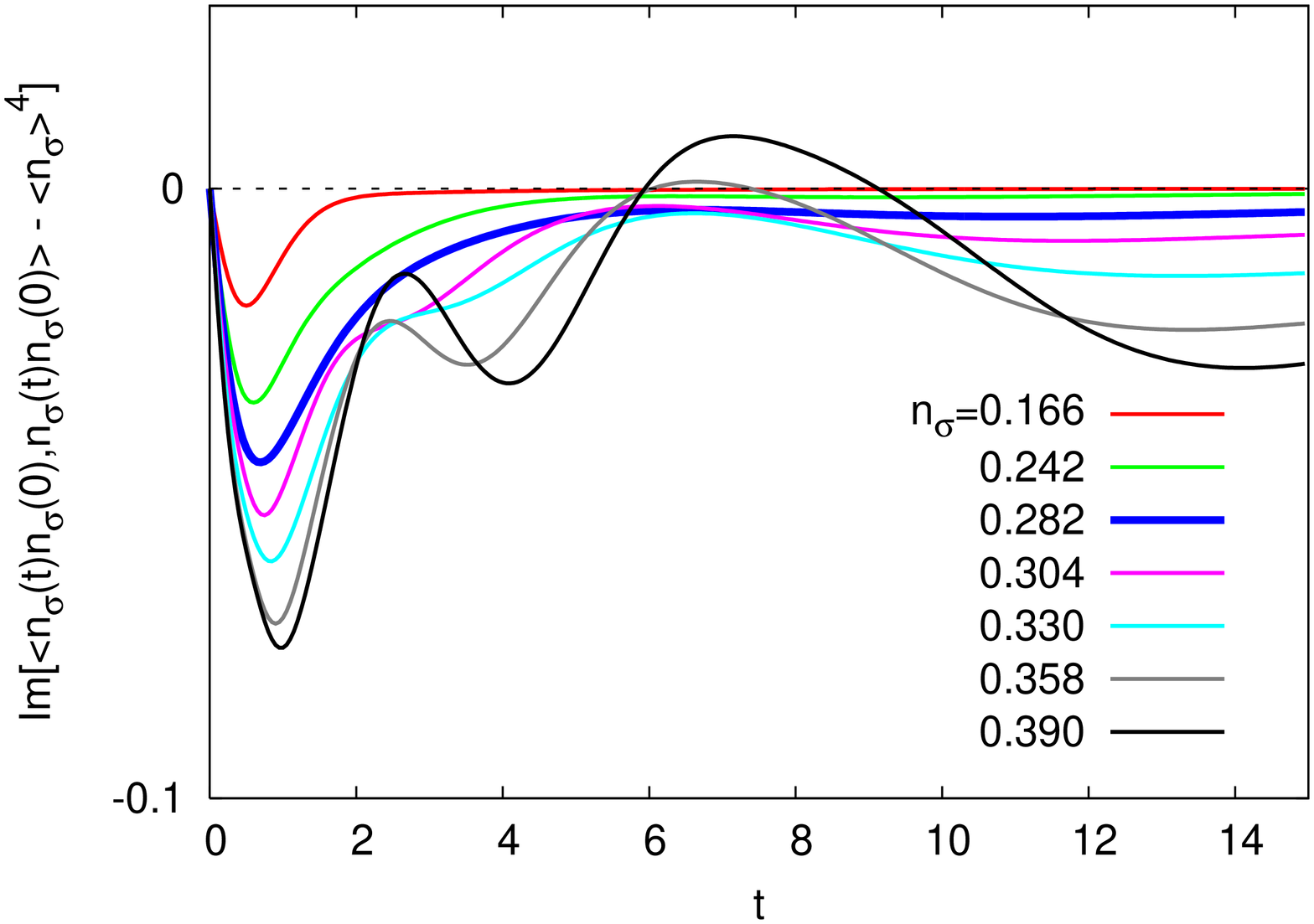}
\includegraphics[angle=-90, width=0.484\columnwidth]{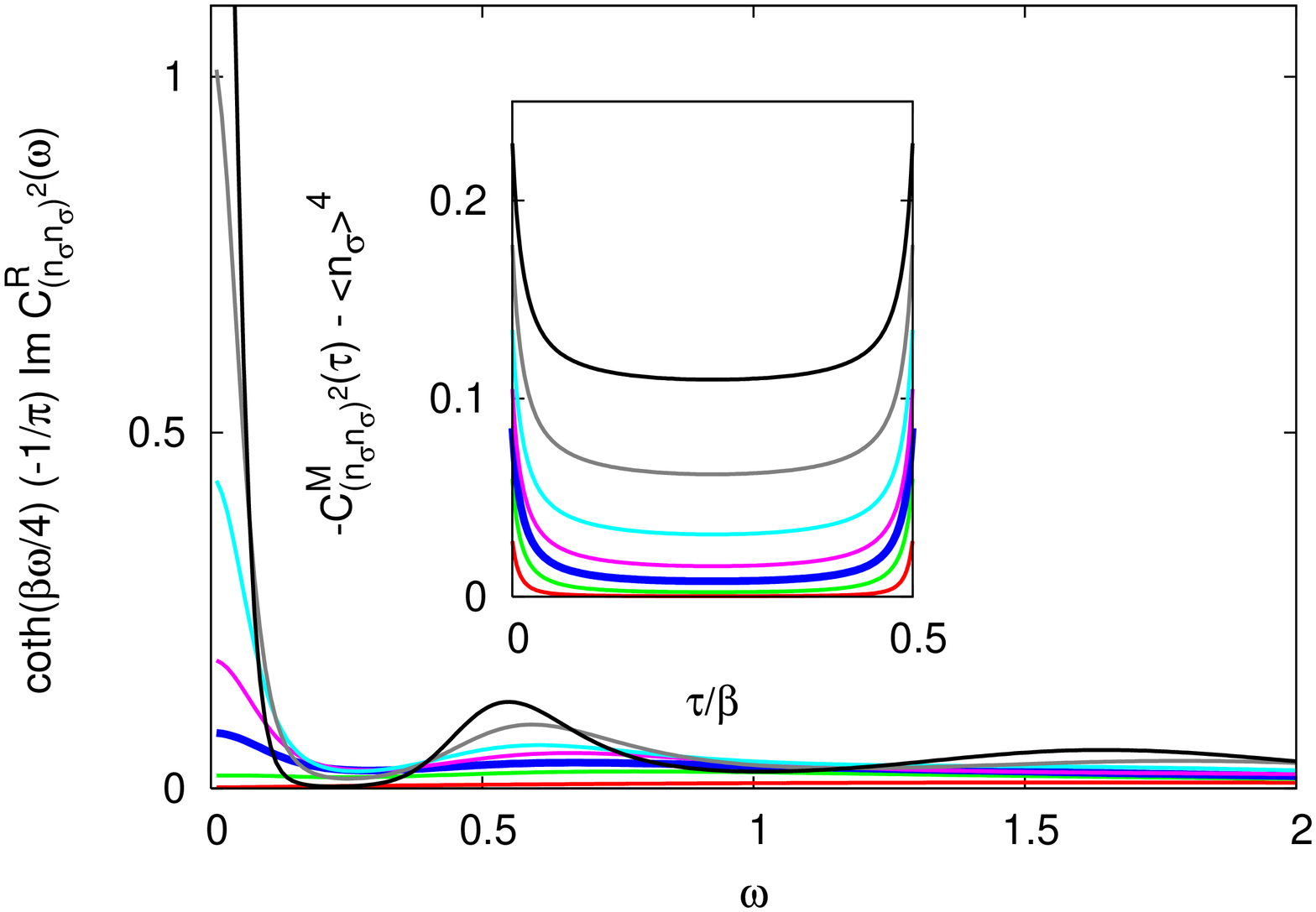}
\includegraphics[angle=-90, width=0.486\columnwidth]{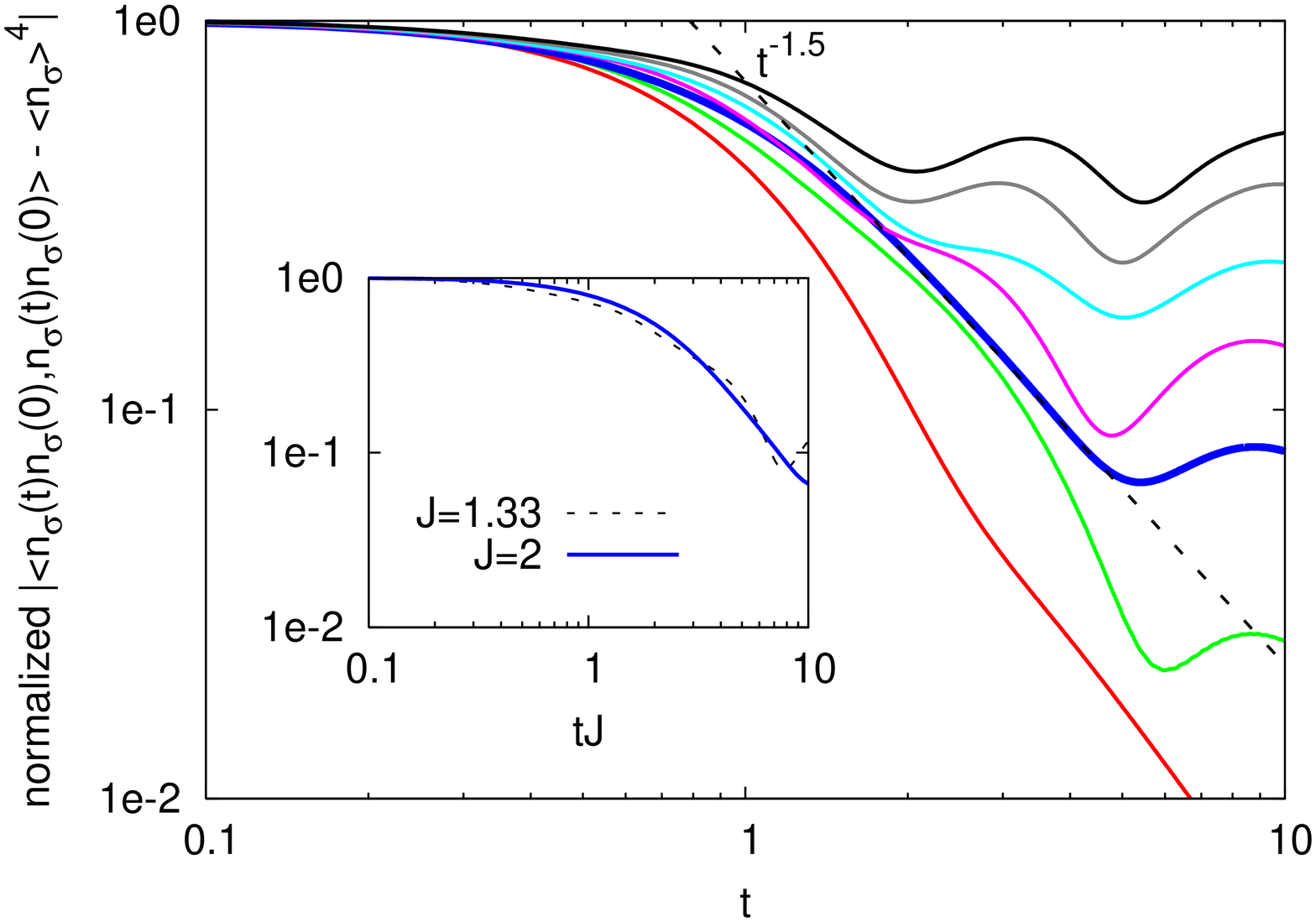}
\caption{
Top panels: Real and imaginary parts of the OTOC $\langle \hat n_\sigma(t)\hat n_\sigma(0), \hat n_\sigma(t)\hat n_\sigma(0)\rangle  - \langle \hat n_\sigma\rangle^4$ for the two-orbital Hubbard model with $U=8$, $J=U/4$, $\beta=50$ and indicated fillings. 
Bottom left panel: Spectral function $\mathscr A_{(n_\sigma n_\sigma)^2}(\omega)=-\frac{1}{\pi}{\rm Im}\,C_{(n_\sigma n_\sigma)^2}^R(\omega)$ multiplied by $\coth\big(\frac{\beta\omega}{4}\big)$. In the inset, we plot the imaginary-time four-point function $\langle \hat n_\sigma(\tau)\hat n_\sigma(0), \hat n_\sigma(\tau)\hat n_\sigma(0)\rangle-\langle \hat n_\sigma\rangle^4$. 
Bottom right panel: Modulus of the OTOC $|\langle \hat n_\sigma(t)\hat n_\sigma(0), \hat n_\sigma(t)\hat n_\sigma(0)\rangle  - \langle \hat n_\sigma\rangle^4|$ (normalized at $t=0$) on a log-log scale.
In the inset of the bottom right panel, we illustrate the approximate data collapse obtained by considering the spin-freezing crossover regimes for different values of $J$. 
In all the panels, the thick blue curves represent the results for the spin-freezing crossover region. 
}
\label{fig:2orbital}
\end{center}
\end{figure}

The top panels of Fig.~\ref{fig:2orbital} show the real and imaginary parts of the OTOC
$\langle \hat n_\sigma(t)\hat n_\sigma(0), \hat n_\sigma(t)\hat n_\sigma(0)\rangle-\langle \hat n_\sigma\rangle^4$
for the two-orbital Hubbard model. In the spin-frozen phase ($n_\sigma \gtrsim 0.3$), both components exhibit 
highly incoherent oscillations, which get suppressed as the filling is reduced. As one moves across the crossover point ($n_\sigma \sim 0.28$),
the oscillations completely disappear, and the real part of the OTOC starts to overshoot to the negative side
in the Fermi liquid regime.
The bottom left panel in Fig.~\ref{fig:2orbital} presents the analytically continued spectral function $\mathscr A_{(n_\sigma n_\sigma)^2}(\omega)=-\frac{1}{\pi}{\rm Im}\,C_{(n_\sigma n_\sigma)^2}^R(\omega)$ multiplied by $\coth\big(\frac{\beta\omega}{4}\big)$. 
In the spin-frozen regime ($n_\sigma \gtrsim 0.3$) a sharp peak appears at low energy, which originates from the saturation of the imaginary-time four point function $\langle \hat n_\sigma(\tau)\hat n_\sigma(0), \hat n_\sigma(\tau)\hat n_\sigma(0)\rangle$ at large $\tau$ (see the inset of the bottom left panel of Fig.~\ref{fig:2orbital}). This low-energy peak represents the slow dynamics of the frozen spins, and translates into a slow long-time decay of $|\langle \hat n_\sigma(t)\hat n_\sigma(0), \hat n_\sigma(t)\hat n_\sigma(0)\rangle-\langle \hat n_\sigma\rangle^4|$,
as shown in the bottom right panel of Fig.~\ref{fig:2orbital}
on a log-log scale. Again, it is difficult to clearly resolve the long-time behavior of the OTOC due to the limited
accuracy of the analytic continuation, but in the spin-freezing crossover regime (thick blue curves)
the decay of the OTOC is consistent with a power law $\sim 1/t^{1.5}$ at least for $2\lesssim t \lesssim 5$. 
The shorter-time behavior ($0.5\lesssim t \lesssim 2$) is instead well fitted by an exponential decay $\sim e^{-\alpha t}$ with a decay constant $\alpha=0.78$. 
This exponent exhibits
a strong dependence on the Hund coupling. 
In fact, the OTOCs in the spin-freezing regime for $J=U/4$ and $J=U/6$ can be approximately collapsed by plotting them as a function of $tJ$ (see the inset in the bottom right panel of Fig.~\ref{fig:2orbital}), which indicates that the Hund coupling is the parameter which controls the dynamics in this regime. 
This is distinct from the Lyapunov behavior in which the exponential decay scales with $t/\beta$ \cite{Bagrets2017}.
A possible reason is that we are still far from the large $N$ limit (where $N$ corresponds to the number of orbitals times the number of spin degrees of freedom) so that the large $N$ behavior 
such as the Lyapunov growth is not observed in our calculations. 
In fact, the strong dependence of the time scale on $J$ and weak dependence on $\beta$ is consistent
with the behavior of the finite $N$ SYK model in Ref.~\cite{FuSachdev2016}.
In the spin-frozen regime ($n_\sigma\gtrsim 0.3$) the decay is much slower than near the crossover point, while in the Fermi liquid regime the decay is accelerated, approaching the free fermion behavior of $t^{-3}$
as shown in Fig.~\ref{analytic continuation}. 
Qualitatively similar results for this OTOC are obtained 
for the three-orbital Hubbard model (see Appendix~\ref{sec:3orbital}).

The time dependence of the modulus of 
$\langle \hat n_\sigma(\tau)\hat n_\sigma(0), \hat n_\sigma(\tau)\hat n_\sigma(0)\rangle-\langle \hat n_\sigma\rangle^4$ 
in the spin-freezing crossover regime of multi-orbital Hubbard models (exponential decay crossing over into a power-law) bears a close resemblance to a (different) OTOC for the SYK model discussed in Ref.~\cite{Bagrets2017}. 
Also, a recent study of yet another type of OTOCs for the finite-$N$ SYK model found a qualitatively similar decay as observed here in the spin-freezing crossover regime \cite{FuSachdev2016}. In the following section, we will make the connection to the SYK dynamics more quantitative.

\subsection{Comparison to the SKY model}
\label{sec:SYK}

In this section, we compare the results of the multi-orbital Hubbard models to the SYK model.
We consider the particle-hole symmetric SYK model of complex fermions \cite{FuSachdev2016},
whose Hamiltonian is given by
\begin{align}
H
&=
\frac{1}{(2N)^{3/2}} \sum_{i,j,k,l=1}^N 
J_{ij;kl} (c_i^\dagger c_j^\dagger c_k c_l
+\delta_{ik}n c_j^\dagger c_l
-\delta_{il}n c_j^\dagger c_k
-\delta_{jk}n c_i^\dagger c_l
+\delta_{jl}n c_i^\dagger c_k).
\end{align}
The coupling constant $J_{ij;kl}$ is a gaussian random variable, satisfying $J_{ij;kl}=-J_{ji;kl}=-J_{ij;lk}=J_{kl;ij}^\ast$, and
\begin{align}
\overline{({\rm Re}\, J_{ij;kl})^2}
&=
\begin{cases}
J_{\rm SYK}^2/2 & (i,j)\neq (k,l) \\
J_{\rm SYK}^2 & (i,j)=(k,l)
\end{cases},
\\
\overline{({\rm Im}\, J_{ij;kl})^2}
&=
\begin{cases}
J_{\rm SYK}^2/2 & (i,j)\neq (k,l) \\
0 & (i,j)=(k,l)
\end{cases},
\end{align}
where the overline represents an average over each realization of $J_{ij;kl}$. The parameter 
$J_{\rm SYK}$ ($J_{\rm SYK}^{-1}$) defines the unit of energy (time) in the following calculations. 
The system is particle-hole symmetric when
$n=0.5$, which fixes the total number of fermions as $N^{-1}\sum_{i=1}^N \langle c_i^\dagger c_i\rangle=0.5$.
We take the particle-hole symmetric form of the SYK model because it has been well studied previously \cite{FuSachdev2016} and avoids the effect of the filling drift.
We numerically solve the model by exact diagonalization for finite $N(\le 12)$, and evaluate the density-density OTOC
$\overline{\langle \hat n_i(t)\hat n_i(0), \hat n_i(t)\hat n_i(0)\rangle-\langle \hat n_i\rangle^4}$
and its imaginary-time counterpart.

\begin{figure}[t]
\begin{center}
\includegraphics[angle=-90,width=0.49\columnwidth]{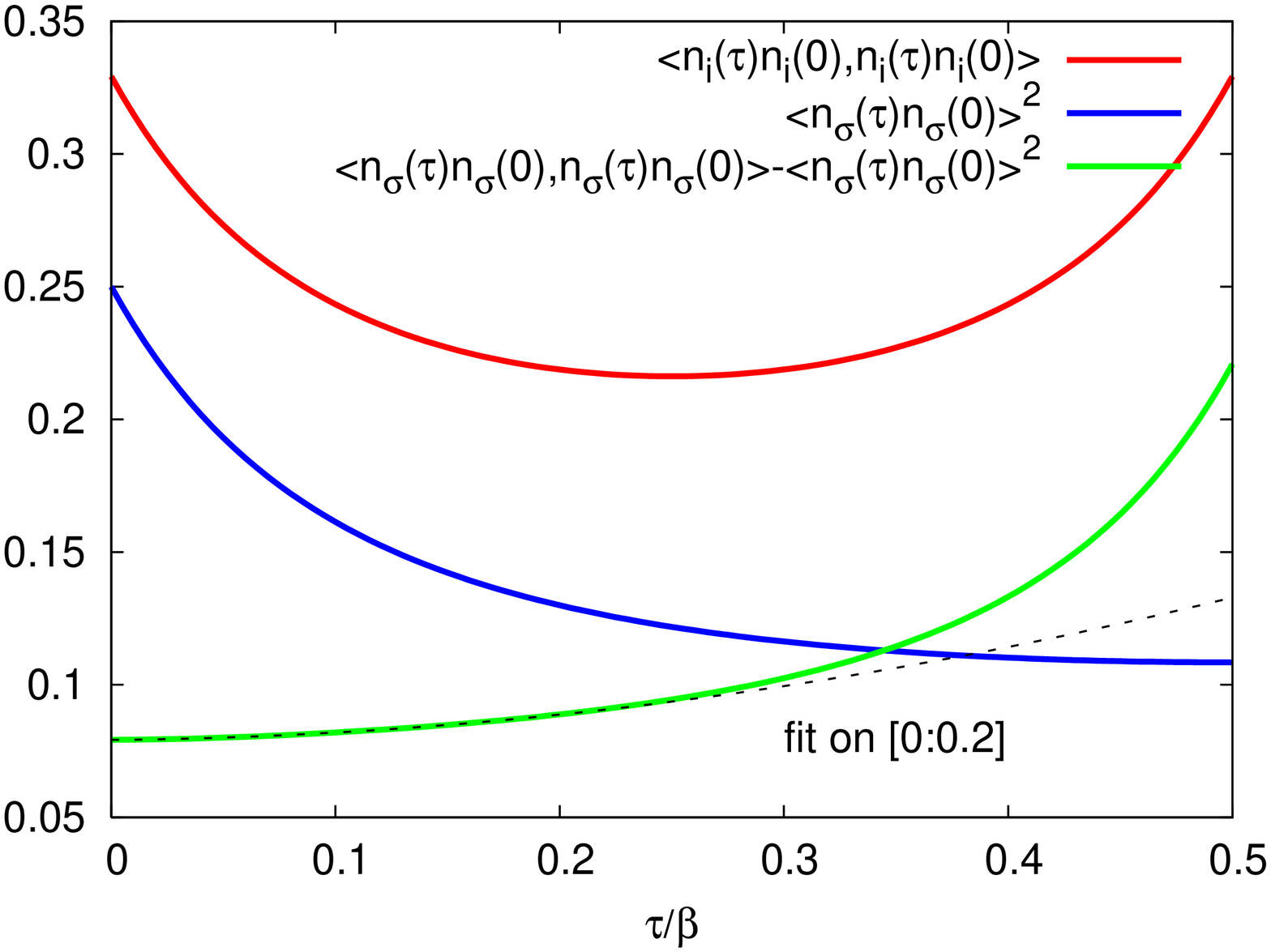}
\includegraphics[angle=-90,width=0.49\columnwidth]{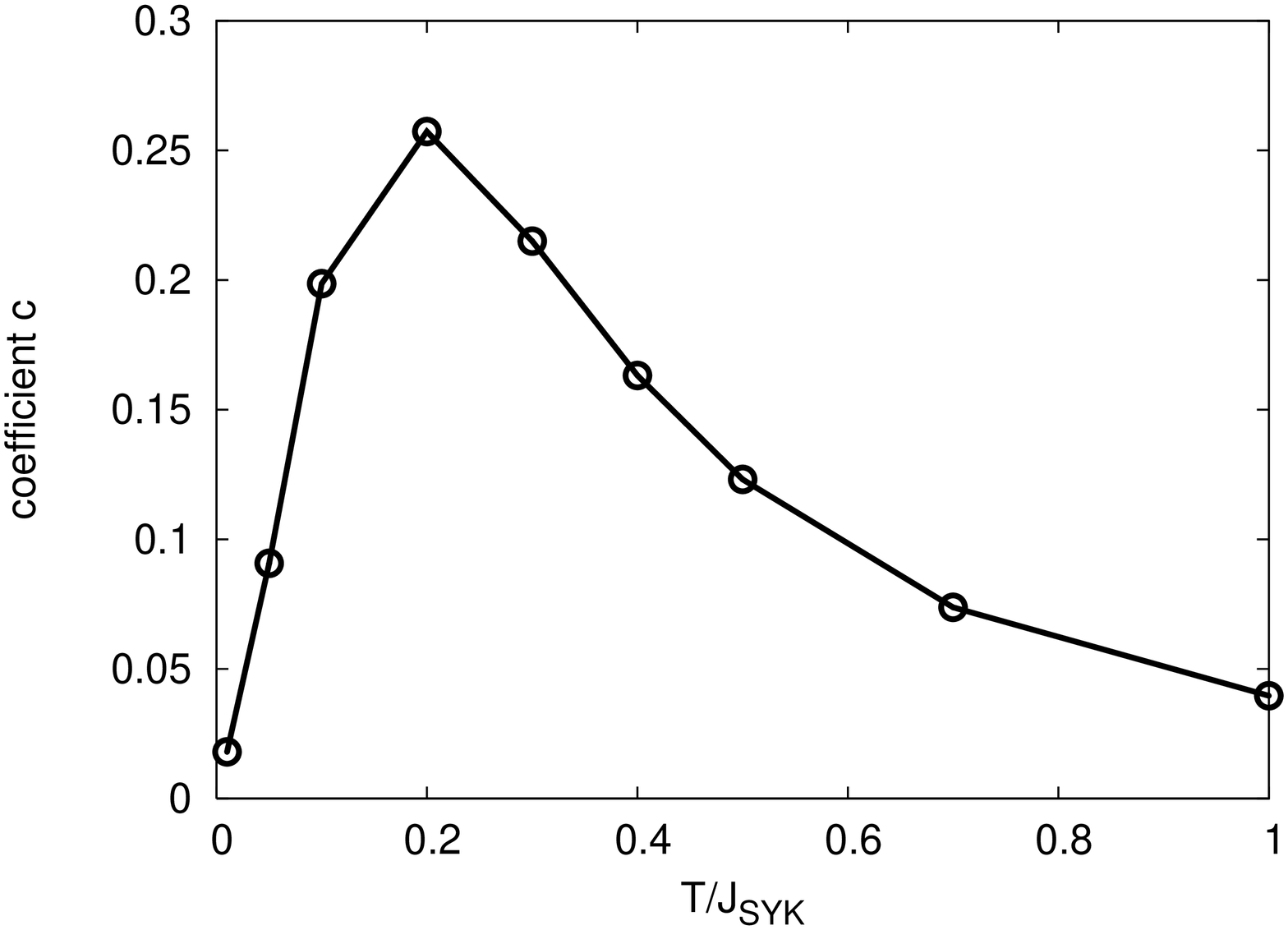}
\caption{
Left panel: 
The imaginary-time four-point function $\overline{\langle \hat n_i(\tau)\hat n_i(0), \hat n_i(\tau)\hat n_i(0)\rangle}$, the density-density correlation function $\overline{\langle \hat n_i(\tau)\hat n_i(0)\rangle^2}$,
and their difference for the SYK model with $N=12$ and $\beta J_{\rm SYK}=10$.
The dashed curve shows a fit of the difference with a function $f_\text{fit}(\tau)=a+b\tau+c\tau^2$ in the interval $\tau/\beta \in [0: 0.2]$. 
Right panel: Fitting coefficient $c$ as a function of $T/J_{\rm SYK}$.
}
\label{fig:SYK fit}
\end{center}
\end{figure}

In the left panel of Fig.~\ref{fig:SYK fit}, we plot the imaginary-time four-point function 
$\overline{\langle \hat n_i(\tau)\hat n_i(0), \hat n_i(\tau)\hat n_i(0)\rangle}$
for the SYK model with $N=12$ and $\beta J_{\rm SYK}=10$. 
At low temperature, we expect that $\overline{\langle \hat n_i(\tau)\hat n_i(0), \hat n_i(\tau)\hat n_i(0)\rangle}$ is decoupled into $\overline{\langle \hat n_i(\tau)\hat n_i(0)\rangle^2}$.
To see whether nontrivial correlations beyond the decoupled form are present, we fit the difference
$\overline{\langle \hat n_i(\tau)\hat n_i(0), \hat n_i(\tau)\hat n_i(0)\rangle}
-\overline{\langle \hat n_i(\tau)\hat n_i(0)\rangle^2}$ with a function $f_{\rm fit}(\tau)=a+b\tau+c\tau^2$
in the interval $\tau/\beta\in[0:0.2]$, in the same manner as for the Hubbard models.
The obtained coefficient $c$ is plotted as a function of the temperature $T$ in the right panel of
Fig.~\ref{fig:SYK fit}. One can see that nontrivial correlations exist in a wide range of temperature.
Especially, they are enhanced in the intermediate temperature regime. This is because in the zero-temperature limit
the four-point function is decoupled as 
$\langle \hat n_i(\tau)\hat n_i(0), \hat n_i(\tau)\hat n_i(0)\rangle \approx \langle \hat n_i(\tau)\hat n_i(0)\rangle^2$, while in the high-temperature limit ($\beta\to 0$) the imaginary-time dependence is washed out.

\begin{figure}[t]
\begin{center}
\includegraphics[angle=-90, width=0.49\columnwidth]{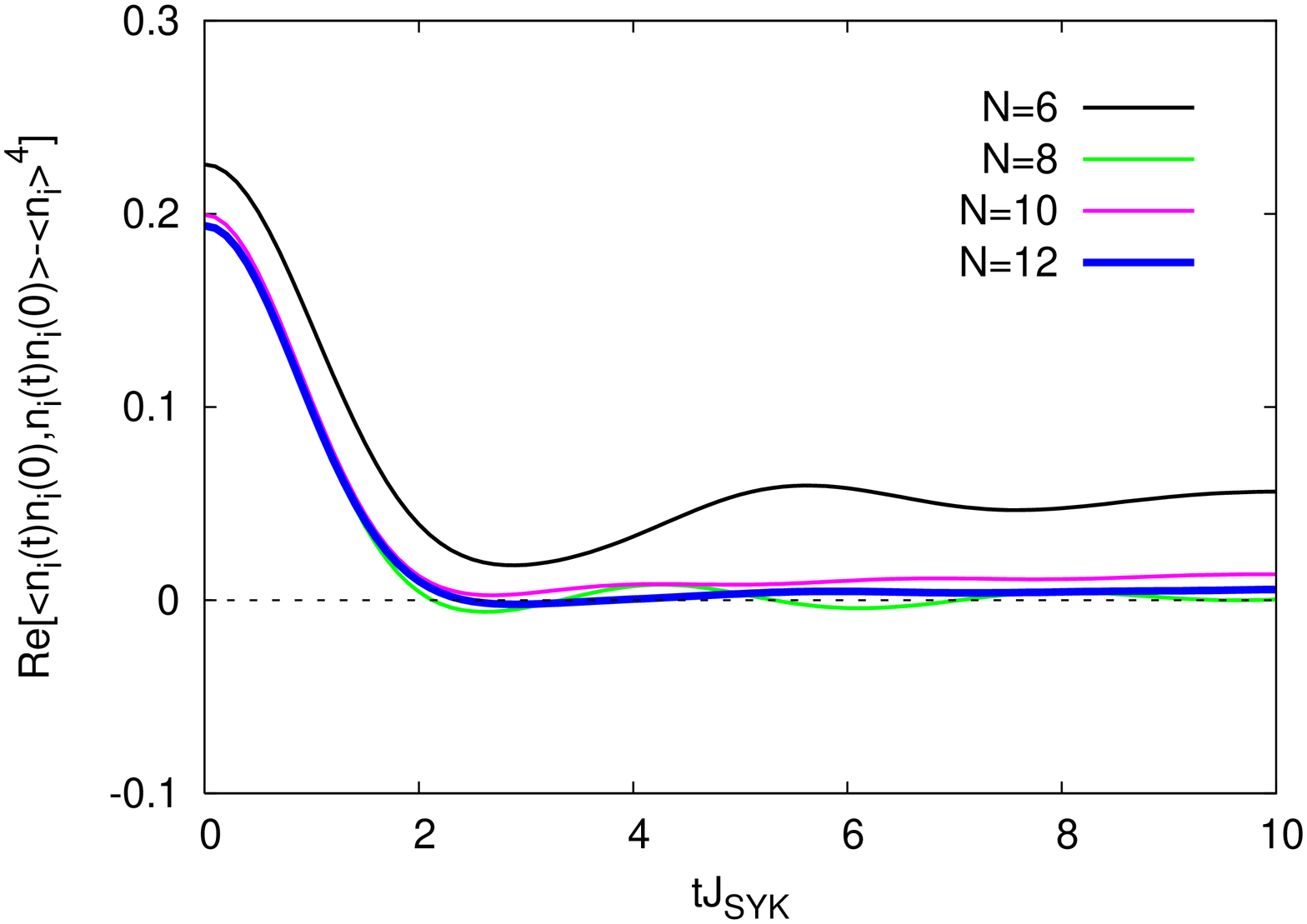}
\includegraphics[angle=-90, width=0.49\columnwidth]{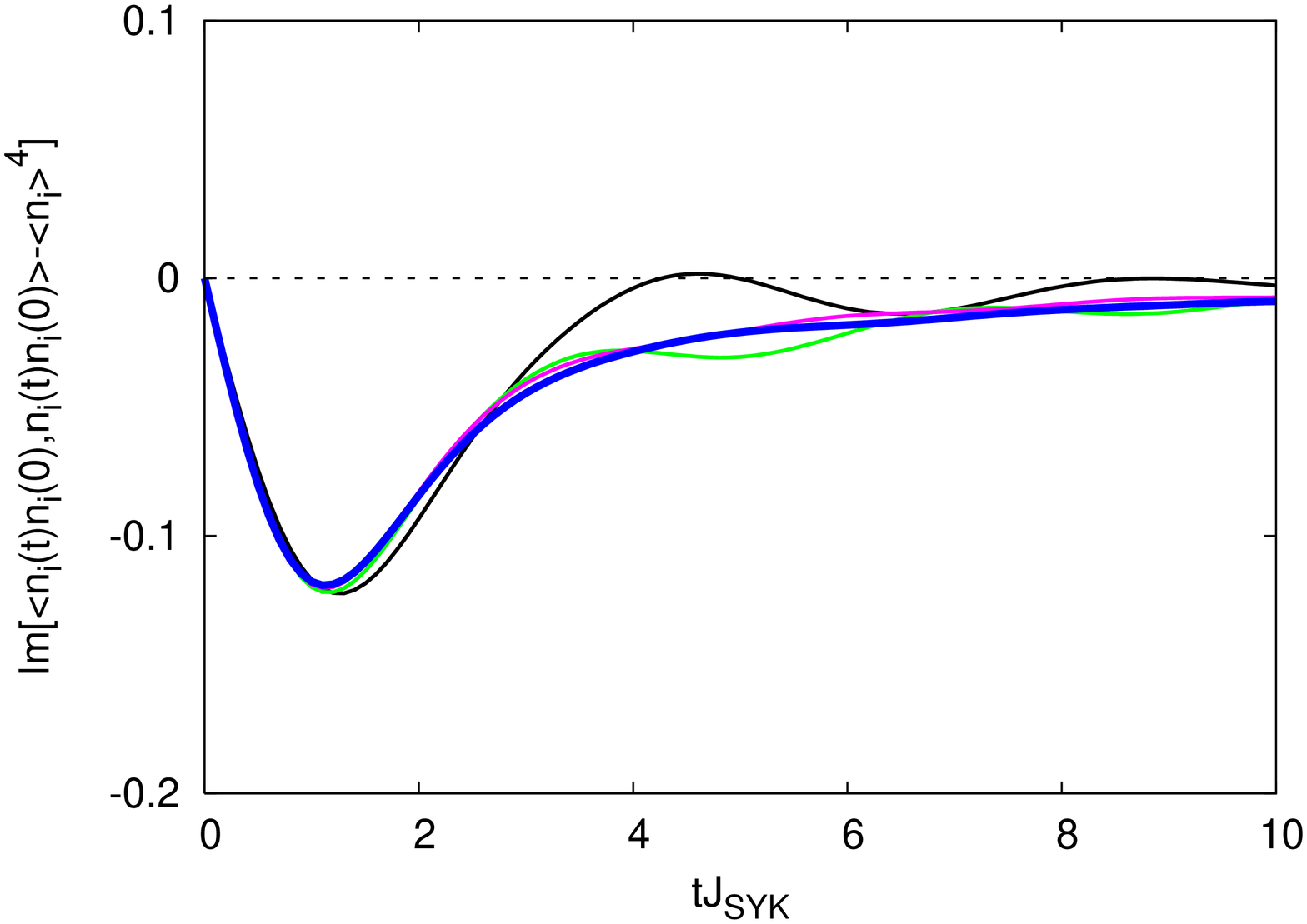}
\includegraphics[angle=-90, width=0.49\columnwidth]{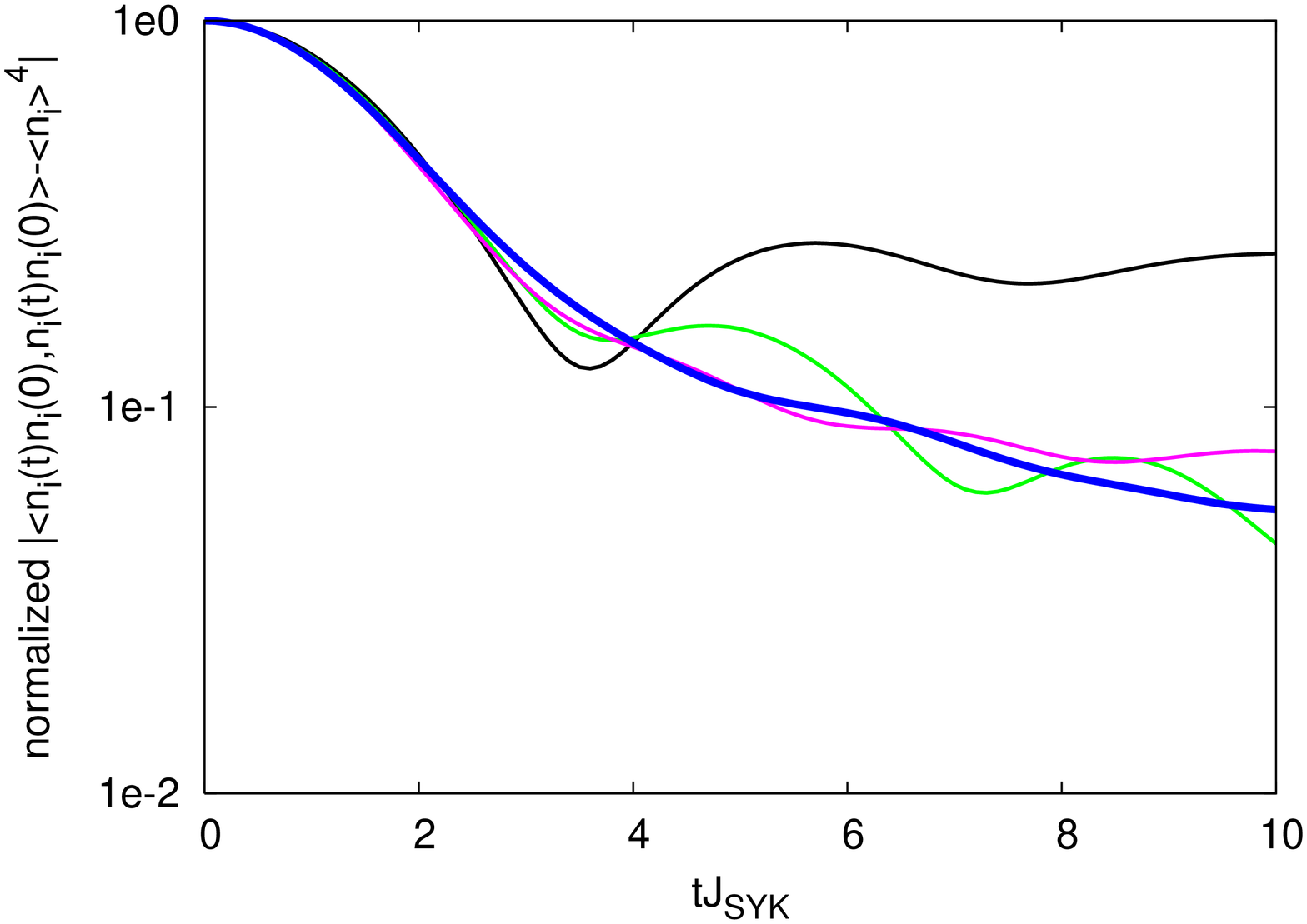}
\includegraphics[angle=-90, width=0.49\columnwidth]{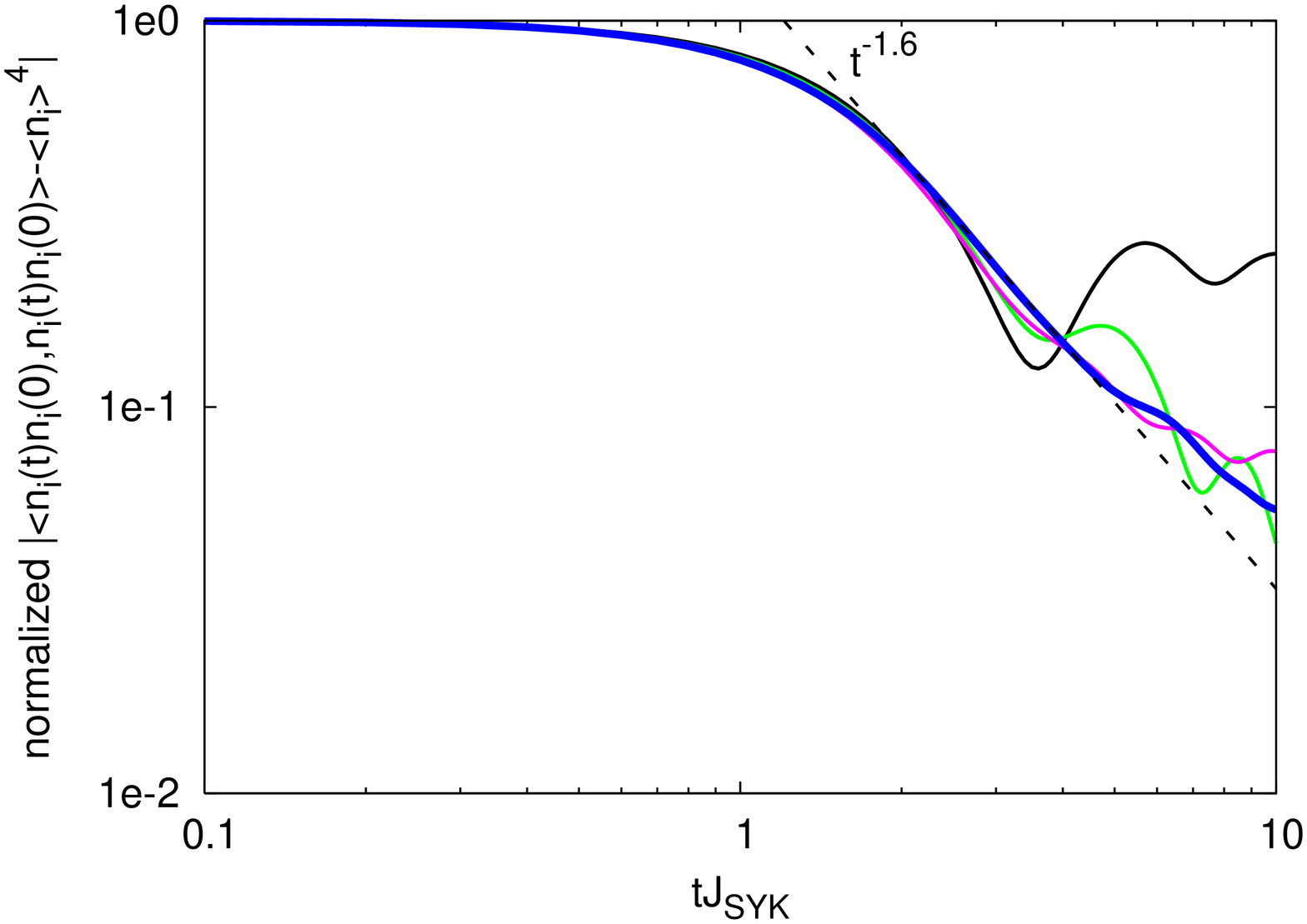}
\caption{
The density-density OTOC $\overline{\langle \hat n_i(t)\hat n_i(0), \hat n_i(t)\hat n_i(0)\rangle-\langle \hat n_i\rangle^4}$ for the particle-hole symmetric SYK model averaged over $10^2$ samples with $\beta J_{\rm SYK}=100$ and various $N$. 
The top left and right panels show the real and imaginary parts of the OTOC,
while the bottom left and right panels show the log and log-log plots of the modulus of the OTOC (normalized at $t=0$).
The dashed line shows a power-law decay $\propto t^{-1.6}$ for comparison.}
\label{fig:SYK}
\end{center}
\end{figure}

\begin{figure}[t]
\begin{center}
\includegraphics[angle=-90, width=0.49\columnwidth]{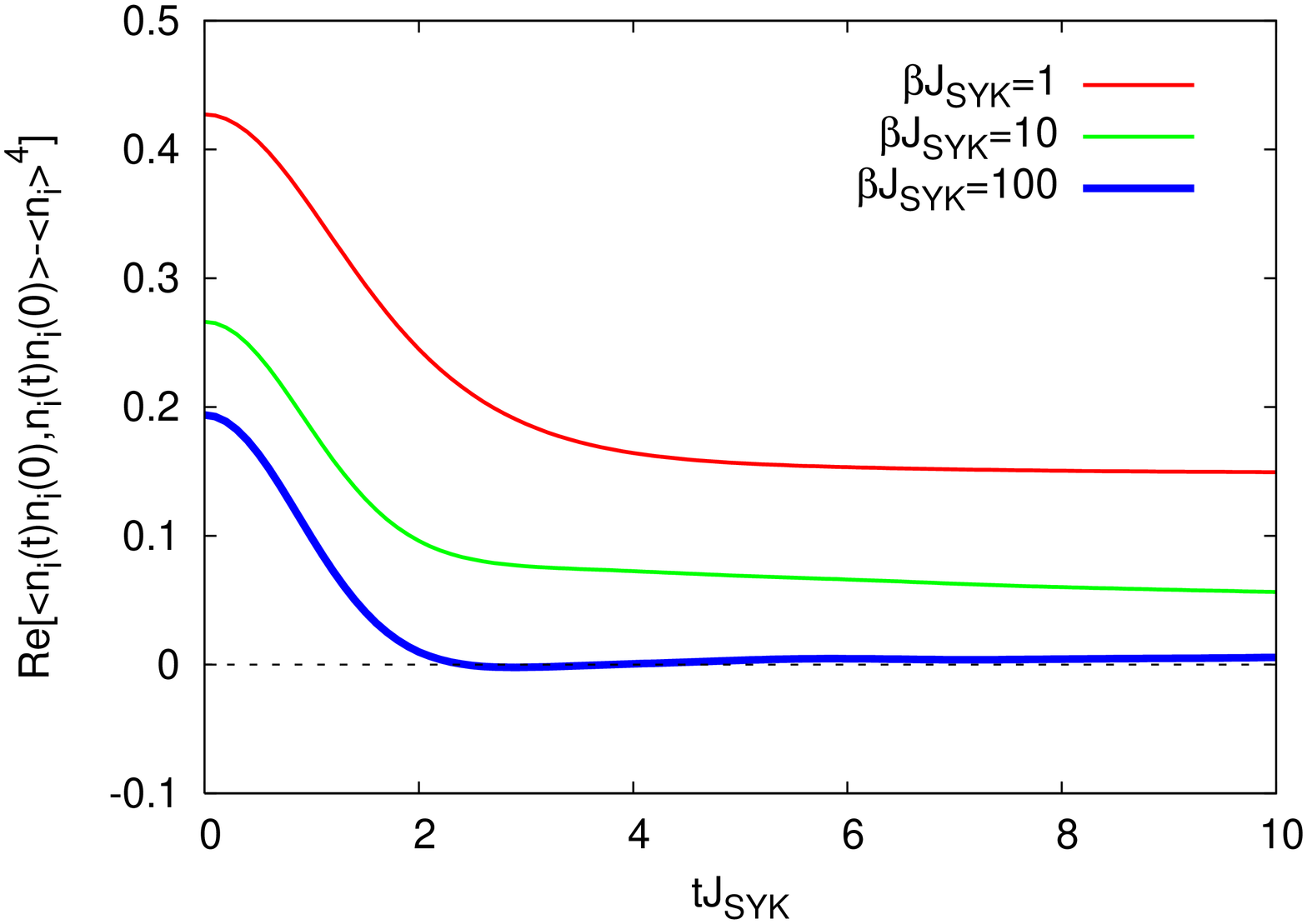}
\includegraphics[angle=-90, width=0.49\columnwidth]{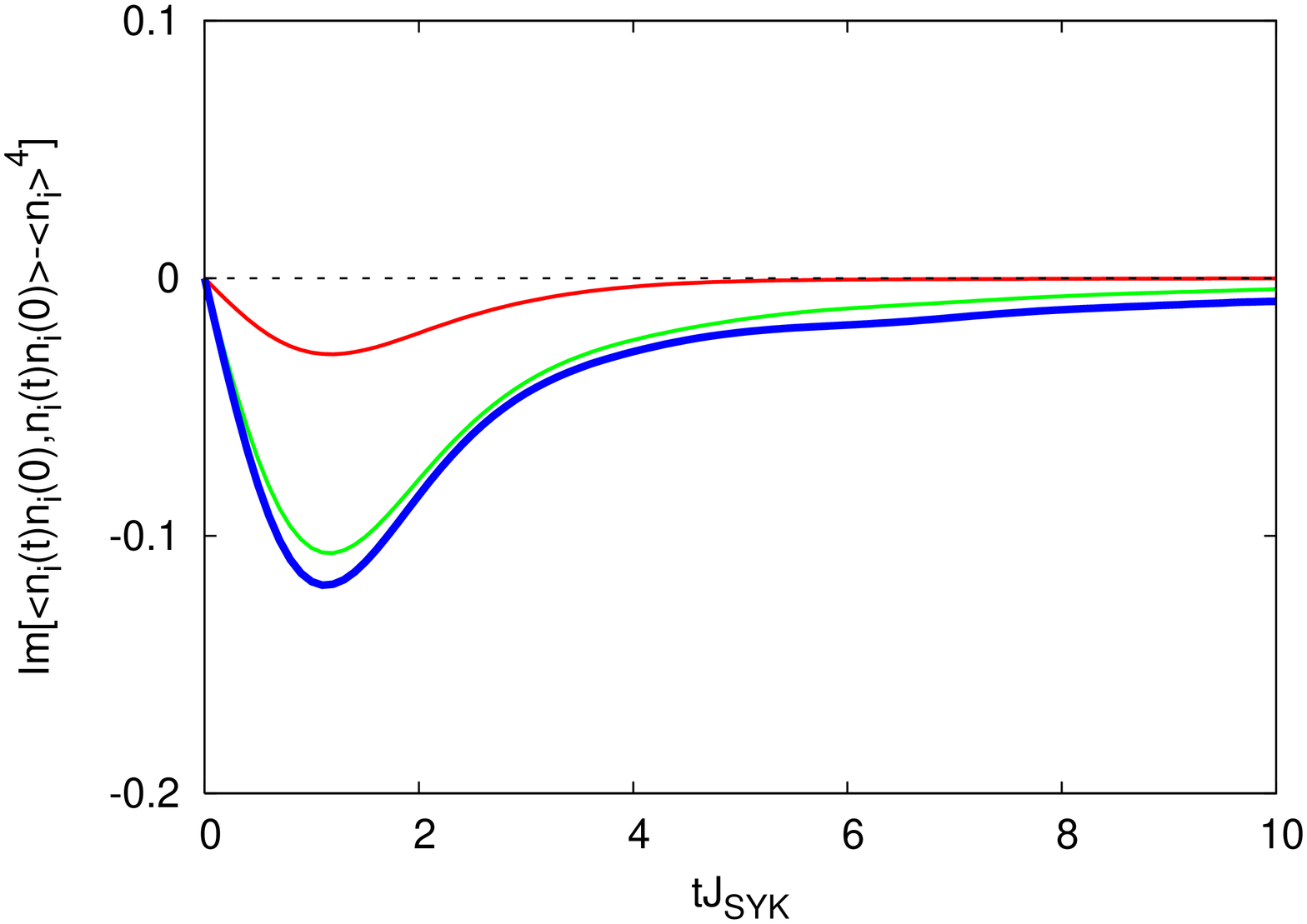}
\includegraphics[angle=-90, width=0.49\columnwidth]{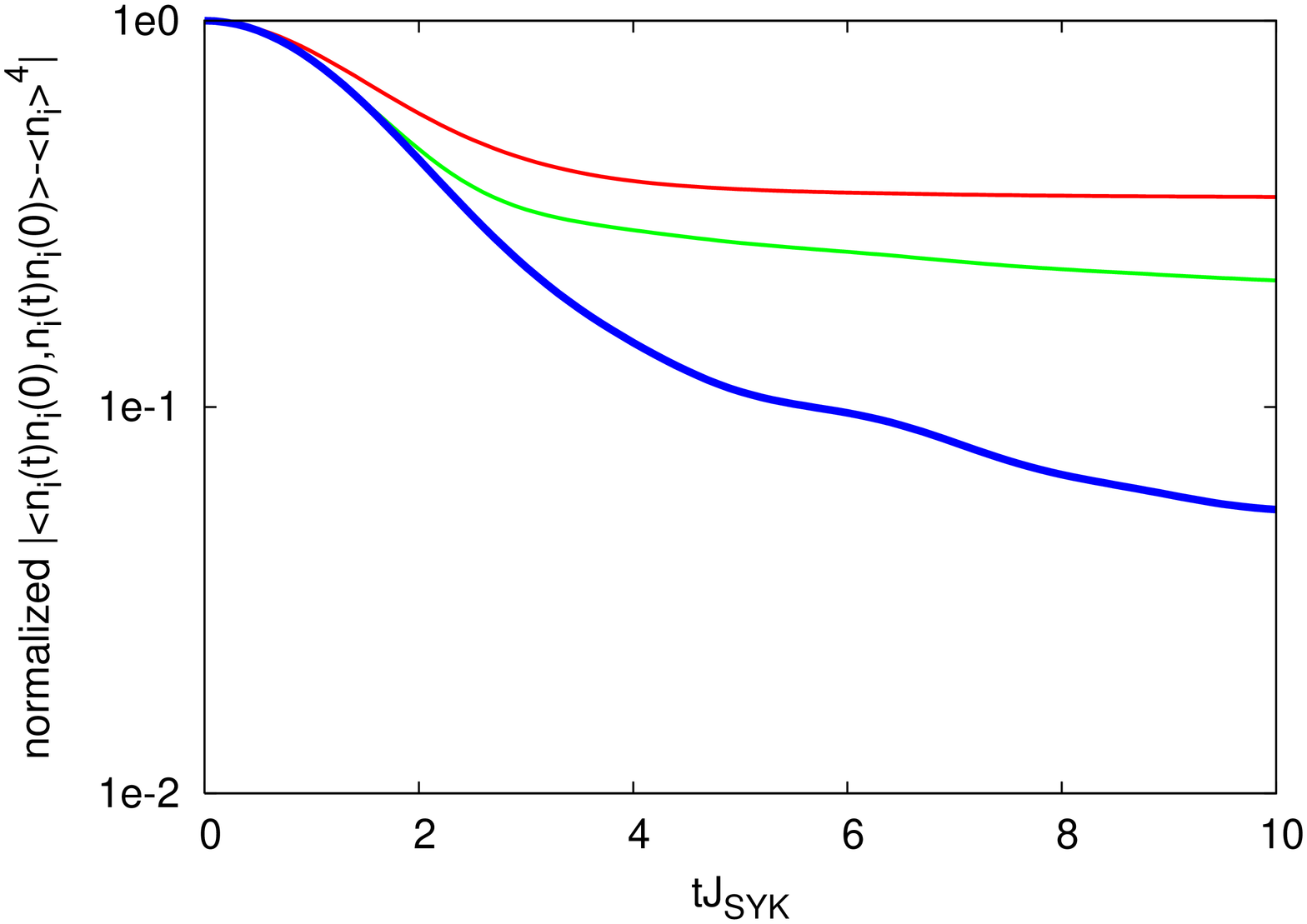}
\includegraphics[angle=-90, width=0.49\columnwidth]{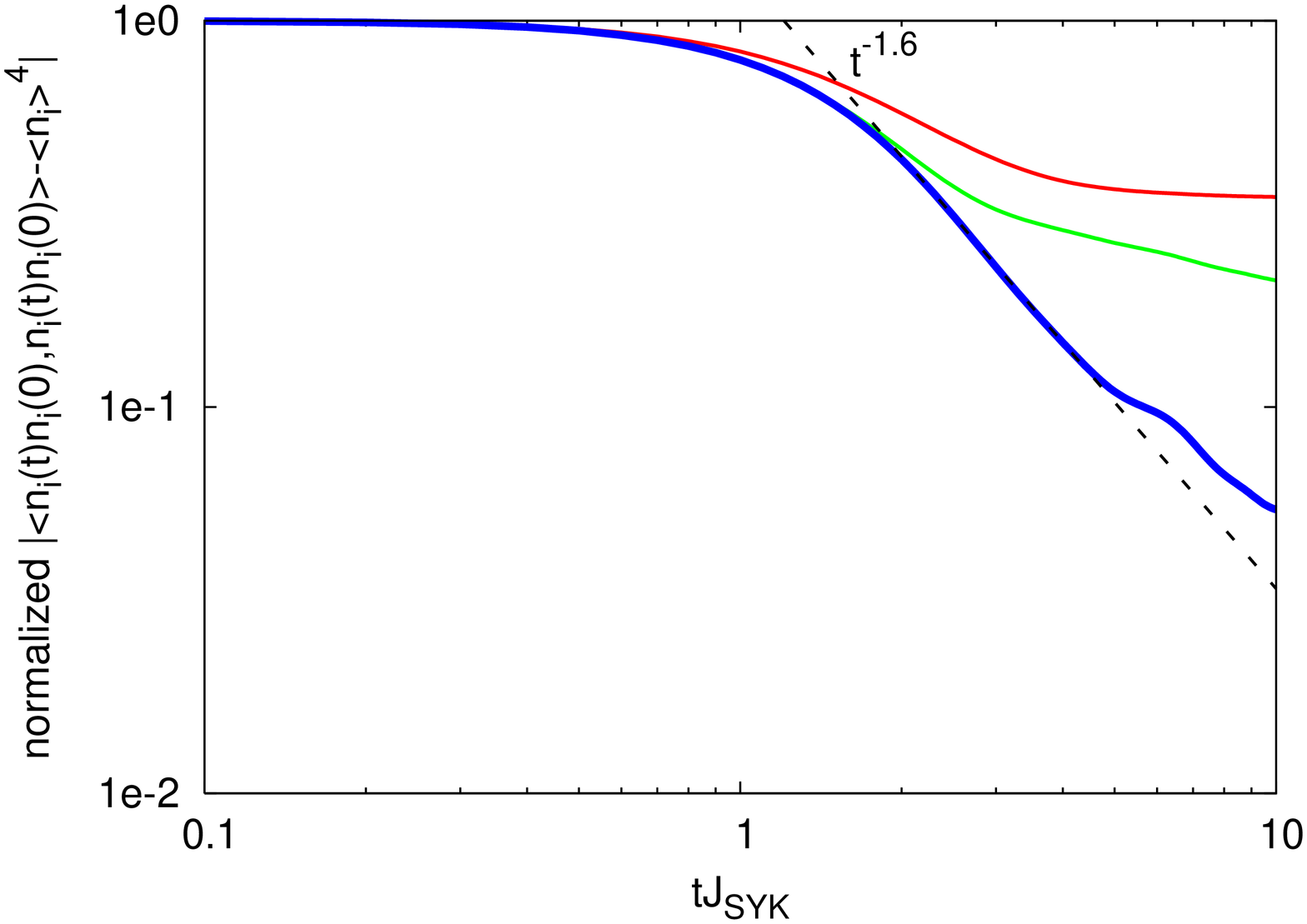}
\caption{
Temperature dependence of the density-density OTOC
$\overline{\langle \hat n_i(t)\hat n_i(0), \hat n_i(t)\hat n_i(0)\rangle-\langle \hat n_i\rangle^4}$ for the particle-hole symmetric SYK model averaged over $10^2$ samples with $N=12$.
The top left and right panels show the real and imaginary parts of the OTOC,
while the bottom left and right panels show the log and log-log plots of the modulus of the OTOC.
The dashed line shows a power-law decay $\propto t^{-1.6}$ for comparison.}
\label{fig:SYK-T}
\end{center}
\end{figure}

In Fig.~\ref{fig:SYK}, we show the numerical results of the density-density OTOC for $\beta J_{\rm SYK}=100$.
The real part of the OTOC rapidly drops from the initial value within the time scale of $tJ_{\rm SYK}\sim 2$,
and slowly decays to zero in the long time limit.
As one increases $N$, the oscillations that appear at longer time tend to be suppressed.
One can see that the result for $N=12$ in the top panels of Fig.~\ref{fig:SYK} closely resembles the OTOC in 
the spin-freezing crossover regime of the two-orbital Hubbard model shown
by the blue curve in the top panels of Fig.~\ref{fig:2orbital},
while it differs from the single-orbital result shown in the top panels of Fig.~\ref{fig:results_nsns}.
Due to the limitation of our calculations to small system sizes ($N\le 12$), we do not clearly observe
an exponential growth ($\sim c_0-c_1 e^{\lambda t}$)  at short times or an exponential decay ($\sim e^{-\alpha t}$) 
at intermediate times. 
However, we find that for $tJ_{\rm SYK}\gtrsim 2$ the OTOC decays
approximately in a power law $\propto t^{-\gamma}$ with $\gamma=1.6$ (see bottom right panel in Fig.~\ref{fig:SYK}). 
The time interval exhibiting the power-law decay becomes longer as we increase $N$.  
The temperature dependence of the OTOC for the SYK model is shown in Fig.~\ref{fig:SYK-T}.
The time scale of the initial drop is more or less independent of the temperature, while 
the power-law-like decay is only visible in the low-temperature regime.

In Ref.~\onlinecite{Werner2018}, it has been argued that the spin-freezing crossover regime of multi-orbital Hubbard models is effectively described by the SYK model in which $J_{\rm SYK}$ is replaced by the Hund coupling $J$. The results in Fig.~\ref{fig:2orbital} are for the two-orbital Hubbard model with $U=8$, $J/U=1/4$ and $\beta=50$, corresponding to $\beta J=100$, which means that it is meaningful to directly compare the blue lines in Fig.~\ref{fig:2orbital} with the blue lines in Figs.~\ref{fig:SYK} and \ref{fig:SYK-T}. Not only is the qualitative agreement remarkable, even the power-law exponent measured for the two-orbital Hubbard model ($\gamma=1.5$) is very close to the exponent extracted from the finite-$N$ SYK calculation. 
The power-law decay in the three-orbital case is approximately $1/t^{1.75}$ (see Appendix~\ref{sec:3orbital}), and thus also in semi-quantitative agreement with the SYK model behavior. 
This nontrivial result provides further support for the identification of the spin-freezing crossover regime of multi-orbital Hubbard systems with an SYK strange metal. 

In the large-$N$ and low-$T$ limit of the SYK model, we expect that the power-law behavior approaches $1/t^2$, since in this limit the OTOC is decoupled as $\langle \hat n_i(t)\hat n_i(0), \hat n_i(t)\hat n_i(0)\rangle\approx \langle \hat n_i(t)\hat n_i(0)\rangle^2$ and the decay of $\langle \hat n_i(t)\hat n_i(0)\rangle$ is dominated by what corresponds to the slow spin relaxation $\langle \hat S_z(t)\hat S_z(0)\rangle\sim 1/t$ in the Sachdev-Ye model.

\section{Conclusions}
\label{sec:conclusions}

We have studied out-of-time-ordered correlators for single- and multi-orbital Hubbard models on the infinitely connected Bethe lattice using DMFT simulations combined with an analytical continuation procedure. 
The basis for this approach is the fact that real-time OTOCs and suitably defined imaginary-time four-point correlation functions are analytically connected through the spectral representation and the out-of-time-order fluctuation-dissipation theorem. 
We showed that it allows to accurately compute the short-time dynamics of OTOCs (up to a few inverse hopping times), as well as the rough decay at long times, which is controlled by low-energy spectral features. The details of oscillations at intermediate and long times can, however, not be accurately captured by the analytical continuation method.

We have used this novel procedure to explore OTOCs in the correlated metal phase, with a particular focus on the spin-related OTOC $\langle \hat n_\sigma(t)\hat n_\sigma(0), \hat n_\sigma(t)\hat n_\sigma(0)\rangle$ which detects the formation of local moments. In the context of the recently proposed connection between the spin-freezing crossover regime of multi-orbital Hubbard models and the SYK model, it is particularly interesting to investigate the doped Mott regime of multi-orbital systems with nonzero Hund coupling. We found that the spin-freezing crossover regime, which is characterized by a $\sqrt{\omega}$ behavior of the self-energy and a $1/\tau$ decay of the spin correlations, is  reflected in a qualitative change of the short-time decay of the spin-related OTOC. More specifically, the real part of this OTOC changes from a slowly decaying function in the weakly doped bad metal phase with frozen magnetic moments to a fast decaying and overshooting function in the strongly doped Fermi liquid phase. In the spin-freezing crossover region, the modulus of the OTOC exhibits an approximately exponential short-time decay, which crosses over into a power-law at $tJ\approx 3$, and the results for different $J$ can be approximately collapsed by plotting the OTOC as a function of $tJ$. 
The power-law exponents are similar for the two- and three-orbital models ($\gamma=1.5$ and 1.75, respectively), and in almost quantitative agreement with the result for the finite-$N$ SYK model ($\gamma=1.6$). 

Our results provide further evidence that the spin-freezing crossover regime of multi-orbital Hubbard models is effectively described by the SYK model, and that $J_\text{SYK}$ in this context can be identified with the Hund coupling parameter,  
which differentiates between the energies of same-spin and opposite spin inter-orbital interactions. The SYK model is thus relevant for the description of an important class of correlated materials, the so-called Hund metals \cite{Georges2013}, which includes a broad range of unconventional superconductors, such as Sr$_2$RuO$_4$ and iron-based superconductors. Via a suitable mapping, it also becomes relevant for the description of optimally doped cuprates \cite{Werner2016}.

In a broader context, our work demonstrates a general strategy for calculating real-time OTOCs of interacting lattice models in the thermodynamic limit. This opens the door to systematic explorations of strange metal physics and information scrambling based on OTOCs in models that are relevant for the description of condensed matter systems. 
   
\acknowledgements

The DMFT calculations were performed on the Beo04 cluster at the University of Fribourg, using a code based on ALPS \cite{Albuquerque2007}. N.T. is supported by JSPS KAKENHI Grant No. JP16K17729.
P.W. thanks the Aspen Center for Physics for its hospitality. Most of the numerical work was carried out during the 2017 summer workshop on ``Correlations and Entanglement in and out of Equilibrium:
from Cold Atoms to Electrons" and the 2018 workshop on ``Topological Phases and Excitations of Quantum Matter."

\appendix

\section{Results for the three-orbital Hubbard model}
\label{sec:3orbital}

\begin{figure}[t]
\begin{center}
\includegraphics[angle=-90, width=0.505\columnwidth]{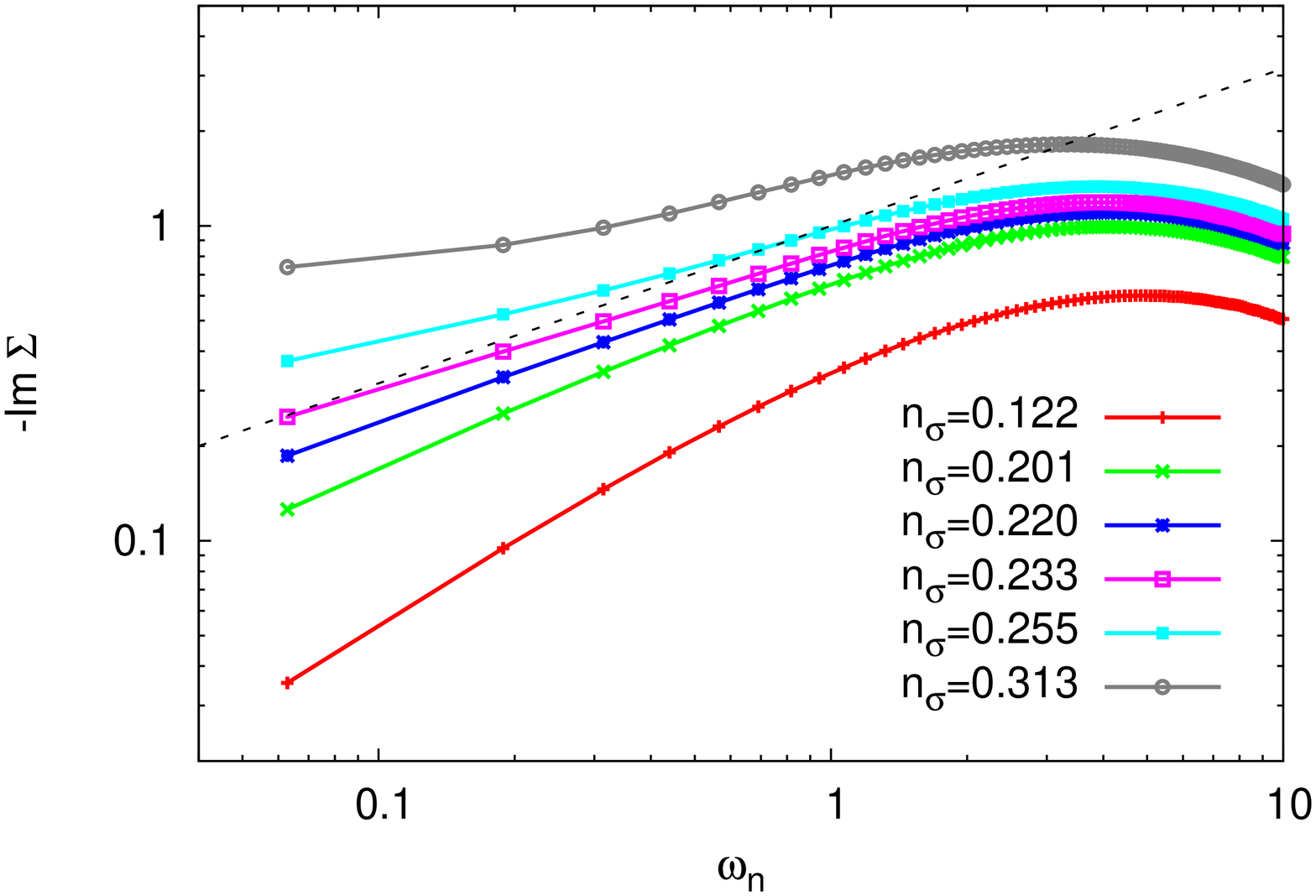}\hfill
\includegraphics[angle=-90, width=0.48\columnwidth]{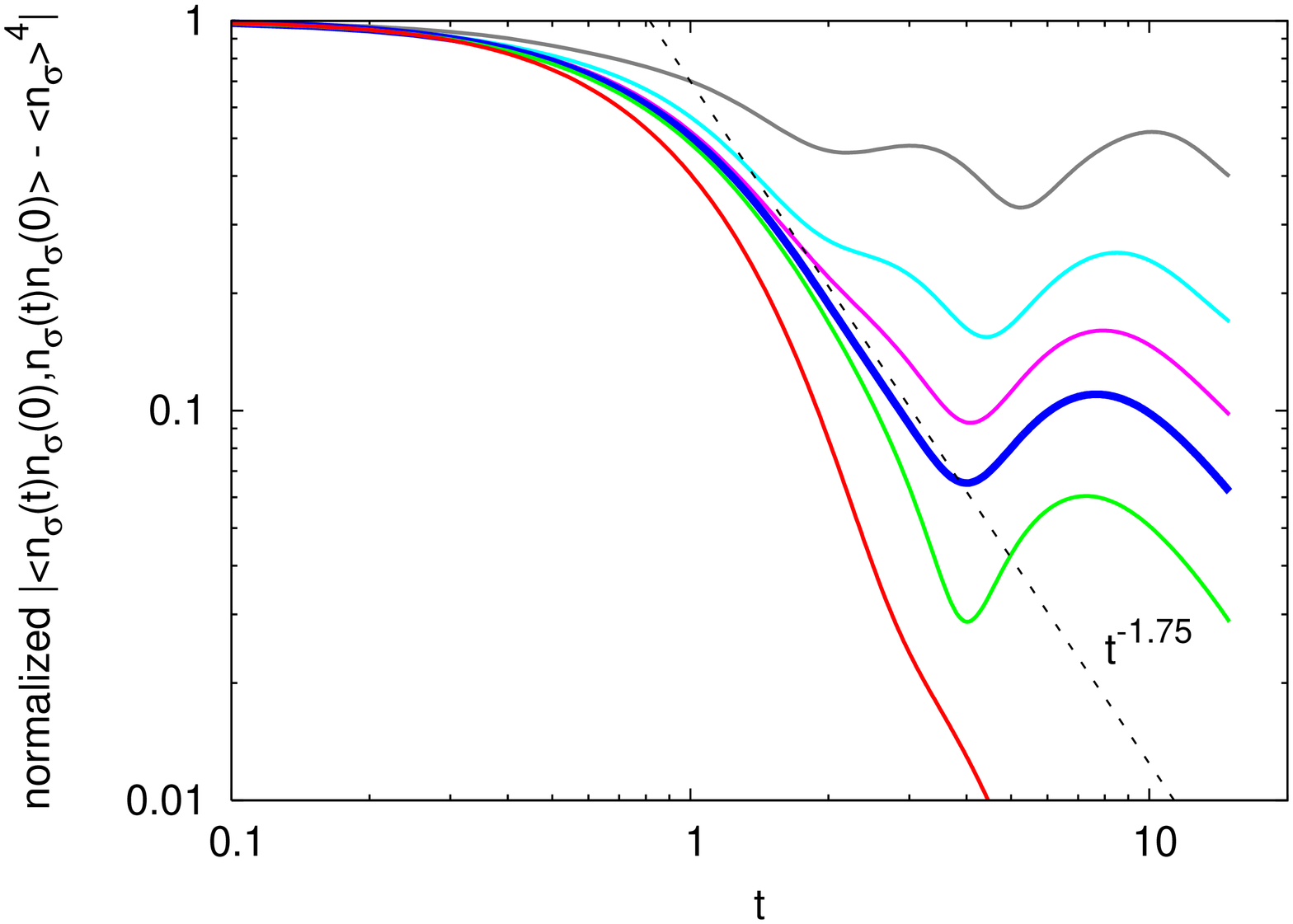}
\caption{
Results for the three-orbital Hubbard model with $U=8$, $J=U/4$, and $\beta=50$. 
Left panel: Imaginary part of the self-energy as a function of Matsubara frequency $\omega_n$.
The dashed line corresponds to $\sqrt{\omega_n}$. 
The spin-freezing crossover is located near $n_\sigma=0.220$.  
Right panel: The modulus of the OTOC $|\langle \hat n_\sigma(t)\hat n_\sigma(0), \hat n_\sigma(t)\hat n_\sigma(0)\rangle  - \langle \hat n_\sigma\rangle^4|$ (normalized at $t=0$) on a log-log scale. In the spin-freezing crossover region, there is an approximate power-law behavior $\sim 1/t^{1.75}$ indicated by the dashed line.
}
\label{fig:3orbital}
\end{center}
\end{figure}

We show the results for the self-energy and the modulus of the OTOC $|\langle \hat n_\sigma(t)\hat n_\sigma(0), \hat n_\sigma(t)\hat n_\sigma(0)\rangle  - \langle \hat n_\sigma\rangle^4|$ for the three-orbital Hubbard model
with $U=8$, $J=U/4$, and $\beta=50$ in Fig.~\ref{fig:3orbital}.  
The spin-freezing crossover roughly occurs at the filling $n_\sigma=0.220$ (blue lines), where we observe an approximate power-law decay $1/t^\gamma$ of the modulus of the OTOC (\ref{otoc_nsns}) at longer times, with $\gamma=1.75$.

\bibliography{ref}

\end{document}